\documentclass[11pt]{report}

\usepackage{amssymb}
\usepackage{amsfonts}
\usepackage{graphics}
\begin{document}

\begin{titlepage}
\begin{center}
\huge On a Dynamical Origin for Fermion Generations

\vspace{2cm}
\Large Jim Bashford

Ph.D. Thesis

Department of Physics and Mathematical Physics

University of Adelaide

July 2003
\end{center}
\end{titlepage}
\tableofcontents
\vspace{7cm}
\abstract
We investigate a proposal to address several outstanding shortcomings of the 
perturbative Standard Model (SM) of particle physics, specifically a common, 
dynamical origin 
for the number of fermion generations, the spectrum of fermion
masses and for Charge-Parity (CP) violating processes. 
The appeal of this proposal is that these features are a 
manifestation of the non-perturbative sector of the SM, requiring no 
assumptions about new physics beyond presently
attainable experimental limits.

In this thesis we apply non-perturbative techniques,  
which have been used to investigate dynamical symmetry breakdown in other 
quantum field theories, to two complementary models: a toy 4-fermion model 
containing explicit chiral symmetry-breaking terms (of an anomalous origin) 
and the quenched  hypercharge gauge interaction.
The key difference from ``conventional'' studies of dynamical breaking of
chiral symmetry breakdown in field theories 
is the possibility that scalar fermion-pairing terms are a necessary but 
not sufficient requirement for dynamical mass generation, 
analogous to the pseudogap phenomenon observed in systems of 
strongly-correlated electrons.

Understanding of how the mass, generations and CP-violation
might arise are first investigated in the toy 4-fermion model.
It is shown that different scale-invariant 4-fermion operators
are present for the three subspaces of the full theory, enabling
self-consistent introduction of three fermion generations.

The second part of the thesis is concerned with dynamical fermion
mass generation in the quenched hypercharge interaction. In particular we 
follow the successful procedure developed for QED, developing a
1-loop renormalisable vertex {\it ansatz} for solution of the fermion self-energy
Dyson-Schwinger equation. 
In the absence of dynamical fermion-antifermion bound states it is found there
exist two mass ``gaps'', potentially corresponding to two types of scalar 4-fermion pairing. These
``gaps'' cannot, however, be interpreted as physical fermion mass. It is 
suggested that only after the incorporation of the composite scalars 
does the self-energy equation admit multiple (physical) solutions.
An alternative possibility, that of a rearrangement of fermionic degrees 
of freedom analogous to spin-charge separation (SCS) in condensed matter physics,
is also briefly outlined.

\vspace{10cm}

\pagebreak

\section{Statement}
 
This work contains no material which has been accepted for the award
of any other degree or diploma in any university or other tertiary 
institution and, to the best of my knowledge and belief, contains
no material previously published or written by another person, except where 
due reference has been made in the text.

\noindent
I give consent to this copy of my thesis, when deposited in the 
University Library, being available for loan and photocopying.

\vspace{2cm}

Jim Bashford
\vspace{8cm}
\pagebreak 

\section{Acknowledgements}
$\bullet$ First and foremost, I wish thank to my supervisor Tony Thomas, 
for his unfailing support and enormous patience. \\

\noindent
$\bullet$ To my wonderful partner Virginia and also to Hayden and Emily 
for their continuing insights into the equation-free world at large.
And to Lydia for providing the necessary impetus to finally finish. \\

\noindent
$\bullet$ To my parents and sisters for their moral support. \\

\noindent
$\bullet$ To Stuart Corney for friendship and bravely trusting me with the 
run of his machine at the University of Tasmania when facilities in Adelaide 
were incapacitated.\\

\noindent
$\bullet$ To Peter Jarvis, for encouragement, friendship and several helpful 
discussions. \\

\noindent
$\bullet$ To my friends Natalie, Michael and Sam, who made life in
Adelaide enjoyable.\\

\noindent
$\bullet$ I wish to acknowledge financial support from the 
University of Adelaide through an APRA scholarship, and a generous extension 
from the Centre for the Subatomic Structure of Matter. \\

\chapter{Introduction}
The Standard Model (SM) of particle physics has had remarkable success
describing the behaviour of subatomic matter (for a review see \cite{Pich94}).
The scalar sector of the SM, however is constructed to accommodate, rather 
than explain, certain phenomenological features which, therefore, must be 
regarded as poorly  understood.
Chief among these are the Charge-Parity (CP)-violating electroweak processes, 
the origin and magnitude of mass and the number of fundamental fermion species.

In this thesis we investigate a proposal \cite{BassThomas1996} which
suggested a common, dynamical origin for the mass and fermionic generations 
without the requirement of additional gauge unification. 
The nature of electroweak flavour mixing - and more specifically, CP-violating 
processes - is known to be sensitive to both the ratios of quark mass 
eigenstates and the number of quark flavours. Therefore the origin of
CP violation could also reasonably be expected to lie within the 
hypothesis \cite{BassThomas1996}.

The known fundamental fermions are naturally classified into 3 
families or ``generations'', each consisting of two quarks (colour degrees of 
freedom are largely irrelevant to this thesis and shall be omitted here)
$q_{\uparrow}$, $q_{\downarrow}$, a charged lepton $\ell$ and its associated neutrino $\nu_{\ell}$.
The weak isospin $\tau_{3}$, electromagnetic charge $Q$ and mass $m_{i}$
of each generation are summarised in the columns of Table \ref{qnos}
below.
\begin{table}[htb]
        \centering
        \caption{Fundamental fermion properties. 
Note (estimated) quark masses are in GeV while lepton masses are in MeV. All masses are obtained from \cite{bklt}.}
        \begin{tabular}{cccccc}
        \hline\hline
  & $\tau_{3}$ & $Q$ & $m_{1}$ & $m_{2}$ &$m_{3}$  \\
\hline
$q_{\uparrow}$ & $+\frac{1}{2}$ & $+\frac{2}{3}$ & $0.001-0.005$ & $1.15-1.35$ & $174.3\pm 5.1 $ \\
$q_{\downarrow}$ & $-\frac{1}{2}$ & $-\frac{1}{3}$ & $0.003-0.009$ & $0.075-0.170$ & $4.0-4.4$ \\
$\ell$ & $-\frac{1}{2}$ & $-1$ & $0.511$ & $105.7$ &$1777$  \\
$\nu_{\ell}$ & $+\frac{1}{2}$ & $0$ & $\leq 0.003$ & $\leq 0.19$ & $\leq 18.2$ \\
\hline\hline
\end{tabular}
  \label{qnos}
\end{table}
Within current experimental limits, for example, for lepton
universality \cite{Pich97},\cite{Renton}, the fermions of one generation are 
identical to those of another in terms of gauge interactions, being 
distinguished only by their orders-of-magnitude mass difference.
Essentially the two more massive generations behave as unstable and 
largely redundant copies of the lightest, of which the vast 
majority of stable matter in the observed universe is composed.

An important experimental limit on fermion numbers comes from the LEP 
measurements of partial invisible $Z$-decay width \cite{bklt}, where the
 number of light (with mass $\leq M_{Z}/2$) neutrinos is determined as
\begin{eqnarray*}
n_{\nu}=2.984\pm 0.008,
\end{eqnarray*}
from which it is inferred - if the cancellation of anomalous electroweak 
processes is to occur - that there are only three fermion generations.
The existence of extra half generations of quarks has been considered,
e.g., \cite{FrogNil1996}) however a new confining gauge interaction is 
needed to form composites which mimic the absent leptons.

A number of theories explaining the number of generations exist,
for example, new horizontal flavour gauge interactions.
Variations within this theme include gauge interactions underlying composite 
quarks and leptons \cite{Seiberg},\cite{Adlerpre}  or linking the number of 
generations to the mass hierarchy problem via spontaneously-broken $SU(3)$ 
(see \cite{Volkas} and references contained within).
The simplest horizontal SM extension - an Abelian interaction - taken
to be non-anomalous \cite{Nardi} or anomalous \cite{Ibanez1},\cite{Ibanez2},
\cite{Ramond1}
at the level of the supersymmetric standard model readily reproduces
the required mass hierarchy. Here it should be noted that for
the latter case \cite{Ibanez1},\cite{Ibanez2} anomaly cancellation is
restored at the string-unification scale via the Green-Schwartz mechanism
\cite{GSW}.

While dynamical chiral symmetry breaking (D$\chi$SB) has been used for decades to study the origin of mass in relativistic quantum field theories
such as the Nambu Jona Lasinio (NJL) model \cite{NJL} and strongly-coupled quantum
electrodynamics (QED) 
\cite{MaskawaNakajima} (for  reviews see \cite{Miransky} or
\cite{RobertsWilliams1994}),
the possibility of a dynamical origin for fermion 
generations has a smaller and relatively 
recent literature. 
Some recent theories requiring interesting gauge 
interactions include broken discrete chiral symmetry \cite{Adlerdisc} and 
the $SU(3)$ colour analogue of electromagnetic duality \cite{ChanTsou}.

The possibility that fermion generations could arise dynamically from 
interplay of scalar fields and elementary fermions 
without new gauge interactions has also recently been explored in several 
contexts \cite{Kiselev}, \cite{Visnjic}. 
In the former \cite{Kiselev}, the multiple generations 
are a manifestation of a specially-constrained Higgs potential possessing a 
discrete symmetry of degenerate vacuum minima, while in the latter 
\cite{Visnjic} higher generations effectively appear as a bound state of 
some elementary fermion with excitations of the Higgs scalar. 
However both mechanisms make use of at least one elementary Higgs field and 
in both cases the number of generations is still a parameter to be put in 
by hand. A motivation for the former has been outlined in \cite{K2}.

In contrast, an earlier paper \cite{BassThomas1996} based upon the non-perturbative $U(1)$ hypercharge interaction suggests the existence of 3 families and the 
absence of a fundamental Higgs. The hypothesis, which is the main theme 
of this thesis, is based upon two observations: 

1)
Analysis of strong-coupling QED in a quenched approximation (as in, e.g., cite{MaskawaNakajima}), suggests vector Abelian theories have 2 phases 
separated by a second-order transition associated with breakdown of chiral 
symmetry.
The possibility of a non-trivial ultraviolet (UV)-fixed point is interesting for several
reasons, firstly because the total charge-screening of perturbative QED at large scales fails to eventuate,
facilitating a consistent definition of non-perturbative
QED. Moreover a self-consistent treatment of this non-perturbative QED may dynamically generate fermion mass, abolishing the 
requirement for undetected elementary Higgs scalars.

2) Assuming no further unification physics, the large momentum limit 
of the Standard Model is dominated by the chiral $U(1)$ (hypercharge) sector, 
with interaction Lagrangian term
\begin{equation}
{\cal L}_{I}=-\frac{g_{2}}{\cos\theta_{W}}\left(q\sin^{2}\theta_{W} \bar{\psi}_{R}\gamma^{\mu}\psi_{R} + (-\tau_{3}+q\sin^{2}\theta_{W}) \bar{\psi}_{L}\gamma^{\mu}\psi_{L}\right) Z_{\mu}. \label{Zeq}
\end{equation}
Here $g_{2}$ is the weak coupling constant, and $\theta_{W}$ is the Weinberg 
angle.
The key feature is that, unlike QED, the coupling is dependent on fermion chirality. The hypercharge theory is special 
amongst chiral gauge theories in that the {\it same} gauge
field couples to both chiralities. In particular integration of the gauge-boson field from the lagrangian
Eq.(\ref{Zeq}) leads to effective 4-fermi interactions of the
form (with the perturbative boson propagator)
\begin{eqnarray*}
{\cal L}_{eff}= (c_{L}\psi_{L}\gamma^{\mu}\psi_{L}+c_{R}\psi_{R}\gamma^{\mu}\psi_{R})
(c_{L}\psi_{L}\gamma_{\mu}\psi_{L}+c_{R}\psi_{R}\gamma_{\mu}\psi_{R}).
\end{eqnarray*}
Upon application of Fierz identities (see Appendix \ref{chap:fierz}) the terms $\sim c_{L}^{2}$, $\sim c_{R}^{2}$
are seen to provide the conventional chiral result,
i.e., upon integration over separate left- and right- 
gauge fields, the vector, axial and anomalous 4-fermi
terms are obtained. The ``cross'' terms $\sim c_{L}c_{R}$,
peculiar to the hypercharge theory provide scalar 4-fermi interactions
\begin{eqnarray*}
(\bar{\psi}\psi)^{2}+(\bar{\psi}i\gamma^{5}\psi)^{2},
\end{eqnarray*}
familar to D$\chi$SB studies of QED or the gauged NJL (GNJL) \cite{Bardeen1989} theory.
In principle there are {\it three} running 
couplings (right-right, right-left and left-left) and the theory
could be anticipated to have a more complicated phase structure than QED.

The stability of non-asymptotically free (NAF) $U(1)$ vector theories has 
been studied extensively ( see, e.g. \cite{Beg1989a}, \cite{G1}). While the
question of whether the two-phase structure of strongly-coupled QED 
carries over into the unquenched sector has yet to be fully resolved,
even less is known about the nature of the corresponding chiral theory, 
although a preliminary lattice study \cite{G2} suggests it suffers from the 
same charge-screening problem.
In the absence of any consensus, if one assumes the above phase structure,
the features of fermion mass and generations arise in the qualitative 
picture of \cite{BassThomas1996} as follows:

Fermion masses arise in analogy to the QED case.
In quenched rainbow QED \cite{Bardeen1989}, it was suggested that at large 
momentum scales attractive scalar fermion couplings became strong and
offset the total virtual-pair-screening of a point-like fermion charge
\cite{Reenders}.
In the renormalisation group picture non-renormalisable (at ``low'' momentum) 
4-fermi operators acquire a large anomalous dimension, mixing with the
gauge interaction. At a critical coupling value the vacuum undergoes a 
transition and the fields must be re-quantised with respect to the 
strong-coupling ground state. The difference in zero-point energies of the 
two phases is manifested as a scalar potential in the chirally-symmetric 
phase and thus a fermion mass is dynamically generated \cite{BassThomas1996}.

For the more complicated chiral case, which has three couplings, three such 
phase transitions are suggested \cite{BassThomas1996} to occur, each 
contributing to the scalar potential. 

If the SM couplings evolved to high 
momentum scales are used ($\sin^{2}\theta_{W} \to 1$) the right-right coupling 
is first to become critical. In the neighbourhood of the critical point 
attractive scalar pairing between right-handed fermions comes to dominate
 over the gauge interaction. If there is a vacuum decay, it will be 
associated with a (parity-violating) scalar condensate
of right-handed fermions. The (dynamical) left-handed fermions are highly excited with 
respect to the condensed right sector. The Adler-Bell-Jackiw (ABJ) anomaly 
\cite{Adler}, \cite{BJ} causes a nett 
production of right-handed states (which, being supercritical condense) at the
 expense of left ones, leading to the vacuum decay of the left sector. It is 
in this way that both chiralities of the lightest generation ``freeze out'' at 
the right-right critical coupling, leading to the requisite scalar potential  
at lower scales.
At this stage the right-handed components of the heavier generations exist as 
resonances; 
right-handed bound states of a heavy fermion and light antifermion. These 
generations are
anticipated to condense at the right-left and left-left critical couplings 
respectively. The ABJ
anomaly, once again ensuring that the condensation of left- and right- 
fermions occurs on an equal 
footing. In this picture a single dynamical fermion behaves as three 
``quasi-particles'' of different masses at distinct scales, leading to the 
appearance of generation structure.

Gauge-boson mass is generated by the same dynamical chiral symmetry
breaking via the mechanism outlined in \cite{Gribovgol}, \cite{Li1}:
Massive fermions automatically generate a mass term for
axial fields through the vacuum polarisation \cite{Li1}, which
to be unitary, requires the existence of non-transverse
degrees of freedom \cite{Li2}. The nature of these (composite)
pseudoscalars has been discussed in \cite{Gribovgol}.

The structure of this thesis is as follows.
In chapter \ref{chap:pre} we review the relevant techniques 
required to investigate the phase structure of such a theory, specifically
the effective action and Schwinger-Dyson formalism.
The Schwinger-Dyson equations (SDEs) for the hypercharge interaction are derived
before the renormalisation group (RG) ideas of 
Wilson \cite{Wilson}, \cite{WK} are introduced and a short discussion on
triviality of the vector and chiral NAF theories is presented.

Chapter \ref{chap:mods} consists of two sections, the first dealing with
the incorporation of a scalar sector in the electroweak interactions 
illustrates the conventional mechanisms for including mass and CP violation.
The second reviews the progress made in understanding D$\chi$SB of the gauged NJL model, the motivation being to draw upon 
these techniques for an analysis of our model. 

The following two chapters present different, complementary approaches 
taken towards understanding the problem. In chapter \ref{chap:4fer} we 
present a simple model of fermion pairing interactions, the motivation being 
that the 4-fermi Wilson potential \cite{WK} of the hypercharge theory could contain explicit 
chiral-breaking terms due to the anomaly. 
Specifically fermion pairings of the form
\begin{eqnarray*}
2\tilde{x}\bar{\psi}_{L}\psi_{R}\bar{\psi}_{R}\psi_{L}+\tilde{r}(\bar{\psi}_{L}\psi_{R})^{2}-
\tilde{r}^{*}(\bar{\psi}_{R}\psi_{L})^{2},
\end{eqnarray*}
where $\tilde{x}$, $\tilde{r}$ and $\tilde{r}^{*}$ are dimensionful couplings.
Denoting the renormalisation scale by $\Lambda_{\chi}$ and writing the 
dimensionless couplings $x=\tilde{x}/\Lambda_{\chi}$, $r=\tilde{r}/\Lambda_{\chi}$, their running is investigated 
at 1-loop level by renormalisation group equations. It is found that there are three renormalised 
null-trajectories through different subspaces of the 
$\{x,r,r^{*}\}$ coupling-constant space, depending on whether one or both
of the $x$ and $r$ terms are present,
corresponding to three scalar 
four-fermi operators
\begin{eqnarray*}
O_{1} & = & (\bar{\psi}_{L}\psi_{R})^{2}+(\bar{\psi}_{R}\psi_{L})^{2};
\hspace{0.1cm}x=0, r-r^{*}=0,\\
O_{2} & = & \bar{\psi}_{L}\psi_{R}\bar{\psi}_{R}\psi_{L};
\hspace{0.1cm}r=0=r^{*}, x\neq 0,\\
O_{3} & = & (\bar{\psi}_{L}\psi_{R}+\bar{\psi}_{R}\psi_{L})^{2};
\hspace{0.1cm}x+r+r^{*}=0. \label{rel3}
\end{eqnarray*}
The potential emergence of three scalar condensates 
associated with these operators provides further \cite{K2} 
motivation to consider a dynamical mechanism like that of the special 
3-Higgs potential of \cite{Kiselev}.
Here the Higgs would be composites bound by the underlying attractive
fermion couplings which are required to prevent the total charge-screening scenario, in direct analogy to the GNJL augmentation of QED.
That is, in the language of the renormalisation group,  the above 4-fermi interactions are non-renormalisable
at low momenta, however close to a UV fixed point they are expected to obtain a large anomalous dimension,
becoming strictly renormalisable (marginal) and thus
mixing with the gauge interaction.

In contrast to what is found for chirally-symmetric matter-only theories, the 
auxilary method requires our composite fields to be, in general complex, chiral
objects. With the self-consistent introduction of a ``fundamental'' fermion 
with a component living in each of the three coupling subspaces and
an extended vacuum 
\begin{eqnarray*}
|0,0,0>=|0>_{x}\otimes |0>_{r} \otimes |0>_{\ell},
\end{eqnarray*}
a range of CP-violating terms are possible. It is noted that with the imposition of a discrete symmetry ($Z_{3}\otimes Z_{2}$, in constrast to \cite{Kiselev}, where $Z_{3}$ is used; here $Z_{n}$ is the permutation
group of order $n$) this number is reduced to one and the calculation 
\cite{Kiselev} of the CKM matrix elements to tree-level 
accuracy could be reproduced.

Chiral symmetry breakdown in three potentials defined by the operators 
$O_{1}$-$O_{3}$, presumably corresponding to different critical scales is 
then studied. 
All dimensionful couplings were found to have the critical value 
$4 \pi^{2}/\Lambda$ with the
respective couplings $\tilde{r}$, $\tilde{x}$ and the combination
\begin{eqnarray*}
\tilde{\ell}\equiv \frac{\tilde{x}}{1+\frac{\tilde{r}\tilde{r}^{*}}{\tilde{x}^{2}}}\frac{1}{1-\textrm{Re}(\tilde{r})(1+\tilde{r}\tilde{r}^{*}/\tilde{x}^{2})},
\label{ldef}
\end{eqnarray*}
for $O_{3}$. Here $\Lambda$ is the ultraviolet momentum cutoff for the 
fermion loop. Dimensionless couplings, with the critical value
$4 \pi^{2}\Lambda_{\chi}/\Lambda$ are finite at large scales if $\Lambda_{\chi} \to \infty$ and $\Lambda\to \infty$ in such a way that their ratio is finite.
It turns out moreover, that the null trajectory for the operator $O_{3}$ in 
Eq.(\ref{rel3}), 
flows to a finite (nonzero) IR limit, making this a suitable candidate for 
the effective theory at intermediate or higher scales.

In summary the model is seen to have all the desirable features; 
the dynamical mass generation at different scales, the phase structure
and the existence of three scalar condensates, amenable to a Kiselev-type 
\cite{Kiselev} model of
generations and CP violation. 
It is not clear, however, how these parity- and chiral-breaking scalar terms 
arise in the context of hypercharge: in the Fierz-reordering above they are 
cancelled separately in the right-right and left-left 
terms, and also in the combination of left-right with
right-left terms (see Appendix \ref{chap:fierz}).
The only possibility, therefore, seems to be some dynamical
effect leading to a mismatch between the latter, cross 
terms encountered when unquenching the boson propagator.
Thus in the second part of the chapter we consider the Dyson-Schwinger 
equation for the fermion self-energy in the chirally symmetric, P-violating
 Wilson potential associated with the anomaly-free hypercharge theory.
 
Chapter \ref{chap:fin} discusses the question of criticality in the 
hypercharge theory.  In \cite{BassThomas1996} it was suggested that at the 
lowest (right) critical 
scale the lightest generation ``froze out'' of the theory; fermions
with right chirality formed a condensate $\bar{\psi}_{L}\psi_{R}$.
The left-handed vacuum was highly excited relative to its right counterpart
and via gauge transformations the ABJ anomaly caused a flux of (dynamical) 
left fermions over into the condensed phase. It is for this reason 
that both chiralities are expected to feel the same scalar potential far 
away from the critical point, the manifestation of which is the appearance
of a mass term.

In \ref{sec:rain} in the (non-anomalous) quenched rainbow approximation
we find some indication of the effects of different dynamics of the left and 
right fermion chiralities in that they acquire different form factors $A_{\pm}$, $B_{\pm}$.
As in the case for quenched rainbow QED, we find above a critical coupling
strength that chiral symmetry is dynamically-broken, $B_{\pm} \neq 0$, but 
here seperate left- and right- gaps are found, corresponding to chirality-dependent scalar pairings of fermions.
A physical fermion mass, in the context of Lorentz symmetry and Hermiticity,
is unable to be constructed. 
The effect is found to disappear in the Landau gauge whereupon we require
more a more sophisticated approximation for the vertex appearing in the
fermion SDE. We repeat the analysis with an improved, naively 1-loop 
renormalisable vertex and find the effect, although considerably reduced,
persists in arbitrary gauge.

The anticipated role of composite scalars in enabling physical fermion mass
generation is then investigated. In order to have a low-energy effective
theory with a unitary perturbative limit it is necessary to include the
dynamically-generated Goldstone boson loop.
The effect upon the fermion self energy SDE is found to be that
scalar form factors $B_{\pm}$ are now found to have terms depending upon three 
couplings, suggesting distinct behaviour at three separate scales, 
paving the way for self-consistent introduction of three fermion generations 
as outlined in \cite{BassThomas1996}. For completeness we also attempt
to include the anomalous corrections to the Goldstone vertex, however
these are found to dominate the equations, causing all scalar form factors
to vanish. Possible reasons for failure are discussed, in particular
it is pointed out that correct implementation would involve moving
beyond the quenched approximation.

In chapter \ref{chap:concl} we present our conclusions, and noting
superficial similarities between the problem at hand and strongly correlated
electrons in 1+1 dimensions, outline an alternative mechanism for
an origin of the fermion generations consistent with existing theories.
See, for example, \cite{transmut}.

\chapter{Preliminary} \label{chap:pre}
In this chapter we introduce the formalism for non-perturbative
analyses of a Quantum Field Theory which we shall need in later
chapters. The material is largely standard and in particular we follow 
\cite{IZ} for the Schwinger-Dyson formalism, \cite{Miransky} for the 
discussions of effective action technique and gauge anomalies and 
\cite{Muta} for renormalisation group. 

\section{Chiral symmetry}\label{sec:chi}
Consider the Lagrangian for N species of free fermions in 4 dimensions
\begin{equation}
{\cal L}_{F}=\bar{\psi}\gamma^{\mu}\partial_{\mu}\psi, \label{lag}
\end{equation}
where $\psi = (\psi_{1},\ldots \psi_{N})$. Defining the matrix $\gamma^{5}$ 
via \cite{IZ}
\begin{eqnarray*}
\gamma^{5}=i\gamma^{0}\gamma^{1}\gamma^{2}\gamma^{3}, & \{\gamma^{5},\gamma^{\mu}\}=0,  & (\gamma^{5})^{2}=1,
\end{eqnarray*}
the fermion Lagrangian is invariant under the global $U(N)$ chiral 
transformation
\begin{eqnarray}
\psi \to \exp(i\omega^{a}\lambda_{a} \gamma^{5})\psi, & & \bar{\psi}\to \bar{\psi}\exp(i\omega^{a}\lambda_{a}\gamma^{5}), \label{globchi}
\end{eqnarray}
where $\omega=(\omega_{1},\ldots \omega_{N})$ is an N-dimensional constant vector and $\lambda=(\lambda_{1},\ldots \lambda_{N})$ are the $U(N)$
group generators. Now ${\cal L}_{F}$ is also invariant 
under the global $U(N)$ transformation
\begin{eqnarray}
\psi \to\exp(i\beta^{a}\lambda_{a})\psi, & & \bar{\psi}\to \bar{\psi}\exp(-i\beta^{a}\lambda_{a}) \label{glob},
\end{eqnarray}
and has the corresponding conserved Noether currents
\begin{eqnarray}
J_{5}^{\alpha\mu}& = & \bar{\psi}\gamma^{\mu}\gamma^{5}\lambda^{\alpha}\psi, \label{ax}\\
J^{\alpha\mu}& = & \bar{\psi}\gamma^{\mu}\lambda^{\alpha}\psi.
\end{eqnarray}\label{vec}
Equivalently ${\cal L}_{F}$ can be rewritten in terms of the chiral
spinors
\begin{eqnarray*}
{\cal L}_{F}&=&\bar{\psi}_{L}\gamma^{\mu}\partial_{\mu}\psi_{L} + 
\bar{\psi}_{R}\gamma^{\mu}\partial_{\mu}\psi_{R}, \\
\psi_{a}&=&\chi_{a}\psi,
\end{eqnarray*}
where $\chi_{a}$ are the projection operators
\begin{eqnarray*}
\chi_{L,R}=(1\mp \gamma^{5})/2.
\end{eqnarray*}
That is, left-handed and right-handed components commute and
the theory has chiral symmetry $U(1)_{L}\otimes U(1)_{R}$,
with corresponding Noether currents
\begin{equation}
J^{\alpha\mu}_{L,R} = J^{\alpha\mu}\mp J^{\alpha\mu}_{5}. 
\end{equation}
However a mass term $m\bar{\psi}\psi$ destroys the invariance under the 
transformations Eq.(\ref{globchi}), mixing left and right currents:
\begin{eqnarray*}
m\bar{\psi}\psi \to m\bar{\psi}\exp(2i\omega^{a}\lambda_{a}\gamma^{5})\psi,
\end{eqnarray*}
and the chiral symmetry group is broken to its diagonal subgroup.
\begin{equation}
U(N)_{L}\otimes U(N)_{R} \supset U(N)_{L+R} \equiv U(N)_{V}. \label{dige}
\end{equation}
Here only the vector current is conserved
\begin{eqnarray*}
\partial_{\mu} J^{\alpha\mu} &=& 0,  \label{curr1}\\
\partial_{\mu} J_{5}^{\alpha\mu} &=& 2 i m \bar{\psi}\gamma^{5}\lambda^{\alpha}\psi, \label{curr2}
\end{eqnarray*}
while due to Goldstone's Theorem (see chapter 3 of \cite{Miransky}), the 
global symmetry breaking Eq.(\ref{dige}) is accompanied by the production of 
$N^{2}-1$ massless scalar particles - one for each broken group generator -  or Goldstone 
bosons. 
For the more useful instances of interacting theories, models of scalar 
interactions (such as the NJL model in section 
\ref{sec:NJL}) respecting global chiral symmetry may be constructed.
Moreover vector and axial gauge-interactions
\begin{eqnarray*}
\bar{\psi}\gamma^{\mu}(V_{\mu}+A_{\mu}\gamma^{5})\psi,
\end{eqnarray*}
necessarily preserve global chiral symmetry as they anticommute with 
$\gamma^{5}$. 
\noindent
The case of local chiral symmetry is more subtle, due to a non-vanishing 
surface term in the computation of the associated Noether current.
Chiral gauge anomalies will be dealt with below in section \ref{sec:anom}.

\section{Chiral anomalies}\label{sec:anom}
\begin{figure}[htb]
\centering{
\rotatebox{270}{\resizebox{6cm}{6cm}{\includegraphics{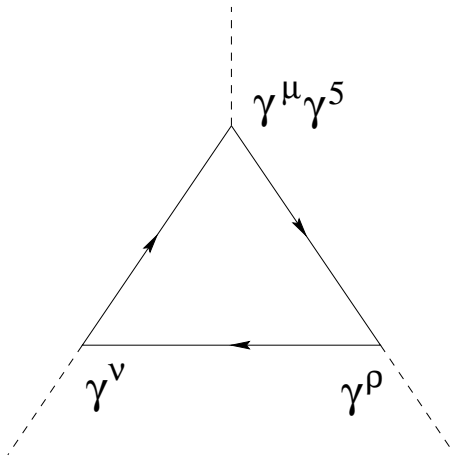}}}
}
 \caption{Anomalous triangle diagram $\Delta^{abc}$.}
 \protect \label{figtri}
 \end{figure}

Gauge anomalies arise in a quantum field theory from the violation of a 
classical symmetry by the second-quantisation procedure.
The chiral gauge anomaly, which is violation of local chiral
symmetry in gauge theories, was originally discovered \cite{Fukuda}
as a lack of gauge-invariance in the computation of the $\pi_{0} \to 2\gamma$ 
decay ``triangle'' diagram $\Delta^{abc}$ with one axial and two vector 
vertices, as shown in Figure \ref{figtri}.
Since the definitive analyses of Adler \cite{Adler}, Bell and Jackiw
\cite{BJ} (ABJ) in 1969 illustrating the lack of a chiral-invariant
regularisation procedure, the anomaly has been interpreted and derived in a 
number of ways. 
\noindent
Perhaps the most intuitive picture is provided by the
Dirac sea picture (\cite{Jackiw},\cite{Gribbk},\cite{Dass}).
Here second-quantisation is achieved by filling the (infinite) 
negative-energy states. On introducing a regularisation, that is, giving
the sea a finite ``depth'', anomalies appear as an exchange
of degrees of freedom (such as charge, chirality or spin) across the
cut-off momentum scale.
In this thesis the path-integral approach of Fujikawa \cite{Fuji1}  
will be the most convenient method of computation. As outlined below,
here the anomaly arises from the failure of the path-integral measure
to be invariant under local chiral symmetry transformations.

Consider the $U(1)_{V}\otimes U(1)_{A}$ gauge theory of massless fermions
with Lagrangian
\begin{eqnarray}
{\cal L}&=& \bar{\psi}i\gamma^{\mu}D_{\mu} \psi+ F^{V}_{\mu\nu}F^{V\mu\nu}+F_{\mu\nu}^{A}F^{A\mu\nu} \label{ZU1},\\
D_{\mu}&=&\partial_{\mu} - (c_{L}+c_{R})V_{\mu}- (c_{R}-c_{L})i\gamma^{5}A_{\mu}, \nonumber\\
F^{X}_{\mu\nu} &= & \partial_{\mu}X_{\nu}-\partial_{\nu}X_{\mu}; \hspace{0.1cm}X=V,A .\nonumber
\end{eqnarray}
It has the classical conserved currents Eqs.(\ref{vec}, \ref{ax}) (with
$\lambda^{0}=1$)
\begin{eqnarray}
\frac{i}{c_{L}+c_{R}}\partial_{\mu}<J^{\mu}>&=&0, \\
\frac{i}{c_{R}-c_{L}}\partial_{\mu}<J_{5}^{\mu}>&=&0; \hspace{0.1cm}c_{R}\neq c_{L}. \label{cc}
\end{eqnarray}
The generating functional of the theory is
\begin{eqnarray*}
{\cal Z} = \int DVDAD\psi D\bar{\psi}\exp(i\int d^{4}x {\cal L}).
\end{eqnarray*}
To rigorously define the path-integral measure, the fermion fields
are expanded in terms of complete eigenfunctions $\phi_{n}$ of the Dirac 
operator $\gamma^{\mu}D_{\mu}$, which are defined by 
\begin{eqnarray}
\psi=\sum_{n} a_{n}\phi_{n}, & & \bar{\psi}=\sum_{n}\phi^{\dagger}_{n}b_{n},
\label{eigtran} \\
\int d^{4}x \phi_{n}^{\dagger}\phi_{m} &=& \delta_{nm},\nonumber
\end{eqnarray}
where $a_{n}$, $b_{n}$ are Grassmann numbers. 
The fermion integral measure is invariant under the transformation Eq.(\ref{eigtran}) and the generating functional $Z$ becomes
\begin{eqnarray*}
Z= \int DZ\prod_{n}Db_{n} \prod_{m}Da_{m} \exp(i\int d^{4}x {\cal L}),
\end{eqnarray*}
where here the Grassmann measures are defined as the left-derivative.

Under the local chiral transformations
\begin{eqnarray}
\psi \to \exp{i\alpha(x)\gamma^{5}}\psi, & \bar{\psi} \to \bar{\psi}\exp{i\alpha(x)\gamma^{5}},
\end{eqnarray}
the Lagrangian and path integral measure transform as
\begin{eqnarray}
{\cal L}& \to &{\cal L}-\bar{\psi} \gamma^{\mu} \partial_{\mu} \alpha(x)\gamma^{5}\psi, \label{lag2}\\
\prod_{m}Da_{m} & \to & (\det B)^{-1} \prod_{m}Da_{m},  \label{mez1} \\
\prod_{n}Db_{n} & \to & \prod_{n}Db_{n} (\det B)^{-1},  \label{mez2}.\\
B_{k\ell}  & = & \int d^{4}x \phi^{\dagger}_{k} \exp(i\alpha(x)\gamma^{5})\phi_{\ell}.
\end{eqnarray}
That is, neither the action nor the path-integral measure are invariant under 
chiral transformations. If the path integral is to remain invariant Eq.(\ref{lag2}) must compensate for Eqs. (\ref{mez1},\ref{mez2}).
For small $\alpha(x)$ the determinant is computed as
\begin{eqnarray*}
(\det B)^{-1} & = & \exp(i\sum_{m,n}\int d^{4}x\alpha(x)\phi^{\dagger}_{n}\gamma^{5}\phi_{m}) +O(\alpha^{2}) \nonumber \\
 &\equiv & \exp( i\int d^{4}x\alpha(x){\cal A}(x)).
\end{eqnarray*}
The compensation condition is then
\begin{eqnarray*}
2 \textrm{Tr}({\cal A})=<-i\int d^{4}x \partial_{\mu}\alpha(x)
\bar{\psi}\gamma^{\mu}\gamma^{5}\psi>,
\end{eqnarray*}
where the trace $\textrm{Tr}$ is understood to contain path integrals.
Performing the $x$ integration by parts on the right-hand side and neglecting
an inessential surface term, we see the classical axial current Eq.(\ref{ax}).
is no longer conserved (cf Eq.(\ref{cc}))
\begin{eqnarray*}
\frac{1}{c_{R}-c_{L}}\partial_{\mu}<J_{5}^{\mu}>=-2i {\cal A}.
\end{eqnarray*}
After regularisation of $\textrm{Tr}(\alpha(x)\gamma^{5})$, for example,
as discussed in \cite{Swanson}, a lengthy expansion in plane waves yields the expression
\cite{Bardeena}
\begin{eqnarray*}
{\cal A}& =& \frac{1}{16 \pi^{2}} \epsilon^{\mu\nu\rho\sigma}
\left( F^{V}_{\mu\nu}F^{V}_{\rho\sigma}+
\frac{1}{3} F^{A}_{\mu\nu}F^{A}_{\rho\sigma}+\frac{32}{3}A_{\mu}A_{\nu}A_{\rho}A_{\sigma} \right. \nonumber \\
& & \left. -\frac{8}{3}(F^{V}_{\mu\nu}A_{\rho}A_{\sigma}+A_{\nu}F^{V}_{\nu\rho}A_{\sigma}+A_{\mu}A_{\nu}F^{V}_{\rho\sigma})\right).
\end{eqnarray*}
While, conventionally, gauge theories (such as the Standard Model)
require anomaly cancellation in order to define renormalisable perturbations,
recent study of anomalous gauge theories in 2 dimensions \cite{Kieu1}, \cite{Kieu2}, \cite{cas1}, \cite{cas2} should be briefly mentioned.
It has been shown for the chiral Schwinger model
\begin{eqnarray*}
{\cal L}_{\chi SM}=-F_{\mu\nu}F^{\mu\nu}+\bar{\psi}\gamma^{\mu}(i\partial_{\mu}+eA_{\mu}\chi_{+})\psi,
\end{eqnarray*}
either the introduction \cite{cas1}, \cite{cas2} of Wess-Zumino  \cite{WZ}, \cite{Witten} fields, or non-gauge-invariant quantisation {\cite{Kieu1}, \cite{cas1},
enables construction of a unitary, perturbatively renormalisable theory.
Little work, aside from \cite{Kieu2}, has been done in higher dimensions.

\section{Effective action}\label{sec:CJT}
The effective action is a standard method by which to study the behaviour
of vacuum degrees of freedom in quantum field theories. Specifically, from
the point of view of dynamical chiral symmetry breaking it enables
calculation of permissible non-vanishing vacuum-expectation values of scalar
condensates. Historically concepts such as the classical order parameter 
were borrowed from
superfluidity and since Coleman and Weinberg \cite{ColeWein72} have been
successfully applied to relativistic theories.
 
Given a scalar theory with fields $\phi_{i}$ and Lagrangian density ${\cal L}(\phi_{1} \ldots \phi_{n})$ (we shall discuss the generalisation to
fermions presently), the generating functional $Z$ is defined as
\begin{eqnarray}
Z[J,I]&=&<0| \int \prod_{i=1}^{n} D\phi_{i} T \exp \{ i \int d^{4}x ({\cal L}(\phi_{1}(x),\ldots \phi_{n}(x))  \nonumber \\
&&\mbox{}+\Phi(x). J(x)) +\int d^{4}x d^{4}y \Phi(x)_{i}\Phi(y)_{i}I^{ii}(x,y)\} |0>,
\label{genf}
\end{eqnarray}
where $\Phi(x)=(\phi_{1}(x),\ldots,\phi_{n}(x))$ and $J(x)=(J_{1}(x),\ldots,
J_{n}(x))$, with $J_{i}(x)$ a classical external source term coupling to field
$\phi_{i}(x)$. $I_{ii}(x,y)$ is a bilocal source.
It is clear that these sources require the same statistics, thus the sources 
coupling to fermion fields are Grassmann numbers.
Eq.(\ref{genf}) has an series expansion in terms of Greens functions
\begin{eqnarray*}
Z[J,I]= 1+ \sum^{\infty}_{n=1} \frac{i^{n}}{n!}\int d^{4}x_{1}\ldots
d^{4}x_{n} G^{(n)}(x_{1},\ldots ,x_{n})J_{1}(x_{1}) \ldots J_{n}(x_{n}), 
\end{eqnarray*}
where the $n$-point Greens function $G^{(n)}$ in the presence of source $J$
is given by  
\begin{equation}
G^{(n)}(x_{1},\ldots ,x_{n};J)=(-i)^{n} \frac{\delta^{n}Z[J,I]}{\delta J(x_{1})\ldots \delta J(x_{n})}  =<0|T \phi_{1}(x_{1})\ldots \phi_{n}(x_{n})|0>_{J},
\label{gfn}
\end{equation}
and care needs to be taken with minus signs arising from the anticommutativity
of Grassmann sources. The notation $\delta$ denotes a functional derivative.
For a discussion of Grassmann differentiation and integration. See, for example, \cite{Swanson}.

The connected generating functional, so named because it is expanded in
terms of Greens functions $G^{(n)}_{c}$ which have connected Feynman diagrams
in the corresponding perturbation theory, is defined by
\begin{equation}
W[J,I]= -i \textrm{Ln} Z[J,I] \label{W},\\ 
\end{equation}
or, in terms of connected Greens functions,
\begin{eqnarray}
W[J,I]&=& \sum^{\infty}_{n=1} \int d^{4} x_{1}\ldots d^{4} x_{n} d^{4}y_{1}\ldots d^{4}y_{n}G^{(n)}_{c}(x_{1},\ldots ,x_{n};J)I_{1}(x_{1},y_{1})\ldots I_{n}(x_{n},y_{n}), \nonumber
\end{eqnarray}
where $G^{(n)}_{c}$ is readily determined by comparison of Eqs.(\ref{genf},\ref{W})

With the restriction to one-particle-irreducible amplitudes, the vacuuum 
expectation value of operator $\phi(x)$ in the presence of current $J$ is
\begin{eqnarray*}
\phi(x;J)&\equiv&\frac{<0|\phi(x)|0>_{J}}{<0|0>_{J,I}}, \nonumber\\
& = & \frac{i}{Z[J,I]}\frac{\delta}{\delta J(x)}Z[J,I], \nonumber \\
& = & \frac{\delta}{J(x)}W[J,I]. \label{phic}
\end{eqnarray*}
If $\phi(x;J)$ has the value $\phi_{c}(x)$ for the particular currents 
$J(x)=J_{c}(x)$, $I(x,y)=I_{c}(x,y)$ and  
\begin{eqnarray*}
\frac{\delta W[J,I]}{\delta I(x,y)}= \phi_{c}(x)\phi_{c}(y)-iG(x,y),
\end{eqnarray*}
where $G$ is the 2-point Greens function in Eq.(\ref{gfn}),
then the effective action is defined as the double Legendre
 transform
\begin{eqnarray}
\Gamma[\phi,G]&=&W[J_{c},I_{c}]-\int d^{4}x \phi_{c}(x)J(x)\nonumber\\
& & \mbox{}+\int d^{4}xd^{4}y(iG(x,y)I_{c}(x,y)-\phi_{c}(x)\phi_{c}(y)I_{c}(x,y)).  \label{effact}
\end{eqnarray}
The physical significance of this is that, analogous to the action of
a classical theory, in the absence of fictitious external currents the 
permissible values of $\phi_{c}(x)$ are given by the stationary points of 
$\Gamma$. Taking a functional derivative of Eq. (\ref{effact}) with respect to $\phi_{c}$
\begin{equation}
\frac{\delta \Gamma[\phi,G]}{\delta \phi_{c}(x)}=-J_{c}(x)-2 \int d^{4}y \phi_{c}(y)I(x,y),
\end{equation}
the right-hand side vanishes readily when $J_{c}(x)=0=I(x,y)$.
Note that the other stationarity equation
\begin{equation}
\frac{\delta \Gamma}{\delta G(x,y)}=0,
\end{equation}
corresponds to the Schwinger-Dyson equation for propagator $G$, discussed in 
section \ref{sec:DSE} below.

\section{Schwinger-Dyson equations}\label{sec:DSE}
The Schwinger-Dyson equations (SDEs) of a quantum field theory determine
relations between its Greens functions. They are derived from the 
fact that the total functional derivative of a path integral, given 
appropriate boundary conditions, vanishes.
For example, consider the generating functional for the 
$U(1)_{V}\otimes U(1)_{A}$ theory with Lagrangian Eq. (\ref{ZU1})
\begin{eqnarray*}
Z[J_{V},J_{A},\eta,\bar{\eta}]&=&
{\cal N} \int DVDAD\psi D\bar{\psi}
\exp\{i S[V,A,\psi,\bar{\psi}]+ \nonumber \\
& & \mbox{}+\int d^{4}x (J_{V}^{\mu}V_{\mu}+J_{A}^{\mu}A_{\mu}+ \bar{\eta}\psi
+\bar{\psi}\eta)\}, \label{zG} \\
S[V,A,\psi,\bar{\psi}]& = & \int d^{4}x \{
 \bar{\psi}(\partial_{\mu}-g_{V}V_{\mu}-ig_{A}\gamma^{5}A_{\mu})\gamma^{\mu} \psi +F^{V}_{\mu\nu}F^{V\mu\nu},\nonumber \\
& & \mbox{} +F_{\mu\nu}^{A}F^{A\mu\nu} -(\frac{1}{2v}\partial^{\mu}V_{\mu})^{2}
-(\frac{1}{2a}\partial^{\mu}A_{\mu})^{2} \}
\label{seff2}
\end{eqnarray*}
where $J_{V}$, $J_{A}$ and the Grassmann-valued functions $\eta$, $\bar{\eta}$ are 
external sources coupling to the gauge bosons and fermions respectively.
The conventional covariant gauge-fixing prescription is used, with 
$v$ and $a$ the gauge parameters for vector and axial fields respectively. 
In the renormalisation group (RG) picture outlined below, $S$ is a regulated,
low energy effective action and $Z$ has a cutoff dependence via the running 
couplings, however this shall be taken as implicit.
While the ABJ anomaly is present the theory is unregulated, and several of
the results below are defined at a formal level, further refinement is
needed for well-defined quantities.

The SDEs for the two-point Greens functions follow from the four formal
equations
\begin{equation}
\left( \frac{\delta}{\delta \phi}S[V,A,\psi,\bar{\psi}]+ J_{\phi}\right)Z[J_{V},J_{A},\eta,\bar{\eta}]=0, \label{zero}
\end{equation}
where $\phi$ with corresponding source $J_{\phi}$ ranges over
the fields $\{V,A,\psi,\bar{\psi}\}$.
Following the previous section, the connected generating functional is 
defined, formally at least, by
\begin{eqnarray*}
W[J_{V},J_{A},\eta,\bar{\eta}]=-i \textrm{Ln}Z[J_{V},J_{A},\eta,\bar{\eta}].
\end{eqnarray*}
It is customary to decompose $W$ into parity-even and -odd parts
\begin{eqnarray*}
W[J_{V},J_{A},\eta,\bar{\eta}]=W_{+}[J_{V},J_{A},\eta,\bar{\eta}]+W_{-}[J_{V},J_{A},\eta,\bar{\eta}],
\end{eqnarray*}
where the even part $W_{+}$ is understood to be regularisable and its 
computation follows the standard procedure used for vector theories. 
Calculation of the $W_{-}$ term, which may also be rendered finite
\cite{salcedo} contains the ABJ anomaly and a $2\pi i $ multivaluation,
will be considered in a later chapter. To illustrate the derivation of
SDEs simply we shall consider only the parity-even part below.

The effective action corresponding to $W_{+}$ is obtained from the
Legendre-transform (c.f. Eq.(\ref{effact}))
\begin{equation}
i \Gamma[V,A,\psi,\bar{\psi}]=W[J_{V},J_{A},\eta,\bar{\eta}]-\int d^{4}x (J_{V}^{\mu}V_{\mu}+J_{A}^{\mu}A_{\mu}+ \bar{\eta}\psi
+\bar{\psi}\eta), \label{eff2}
\end{equation}
where, for $Z=V,A$
\begin{eqnarray*}
Z_{\mu}(x)= \frac{\delta W}{i\delta J_{Z}^{\mu}(x)}, & 
\psi(x)=\frac{\delta W}{i\delta \bar{\eta}(x)},&
\bar{\psi}(x)=-\frac{\delta W}{i\delta \eta(x)}, \\
J_{Z}^{\mu}(x)=-\frac{\delta \Gamma}{\delta Z_{\mu}(x)}, & 
\bar{\eta}(x)=\frac{\delta \Gamma}{\delta \psi(x)}, &
\eta(x)=-\frac{\delta \Gamma}{\delta \bar{\psi}(x)}. 
\end{eqnarray*}
With vanishing external currents Eq.(\ref{zero}) with $\phi=V$ leads to
\begin{equation}
\frac{\delta \Gamma(V,A,\psi,\bar{\psi})}{\delta V_{\mu}(x)}=
(\partial^{2}g_{\mu\nu}-(1-v)\partial_{\mu}\partial_{\nu})V^{\nu}(x)
- i g_{V} \textrm{tr}\left(\gamma^{\mu}\frac{\delta^{2}\Gamma}{\delta\psi
\delta\bar{\psi}}(x,x)\right)^{-1}, \label{veq1}
\end{equation}
where the identity
\begin{eqnarray*}
\int d^{4}x \frac{\delta^{2}G}{\delta\eta(y)\delta\bar{\eta}(x)}
\frac{\delta^{2}\Gamma}{\delta\psi(x)\delta\bar{\psi}(z)}=
i \delta^{4}(y-z); \hspace{0.1cm} \begin{array}{c}
\eta=0=\bar{\eta}, \\
\psi=0=\bar{\psi}, \end{array}
\end{eqnarray*}
has been used. Taking the functional derivative of Eq.(\ref{veq1}) with 
respect to $V$ and setting all fields to zero, the SDE for the inverse two-point vector Greens function is obtained:
\begin{eqnarray}
(D_{V}^{-1})^{\mu\nu}(x,y)&\equiv &\left(\frac{\delta^{2}\Gamma}{\delta V^{\mu}(x)\delta V^{\nu}(y)}\right)_{V=A=0=\psi=\bar{\psi}} \nonumber \\
& = & (D_{V0}^{-1})^{\mu\nu}(x,y)+\Pi_{V}^{\mu\nu}(x,y), \label{pse1}
\end{eqnarray}
where $D_{V0}^{\mu\nu}(x,y)$ denotes the bare vector boson propagator and
$\Pi_{V}^{\mu\nu}(x,y)$ is the vacuum polarisation:
\begin{eqnarray*}
(D^{-1}_{V0})^{\mu\nu}(x,y)&=&(\partial^{2}g^{\mu\nu}-(1-v)\partial^{\mu}\partial^{\nu})\delta^{4}(x-y), \label{bareprop1}\\
\Pi_{V}^{\mu\nu}(x,y)&=&i g_{V} \textrm{Tr}(\gamma^{\mu}S\Gamma^{V\nu}S).
\end{eqnarray*}
Here the trace includes integration over fermion loops ,
while $S$ and $\Gamma^{V\mu}$ are the inverse 2-point fermion 
and irreducible vector-fermion vertex Greens functions respectively. 
The latter is distinguished from the effective action $\Gamma$ by the 
presence of a Lorentz index. 
These functions are defined by
\begin{eqnarray*}
S(x,y) & = & \left(\frac{\delta^{2}\Gamma}{\delta\bar{\eta}(x)\delta\eta(y)}\right)^{-1}_{V=A=0=\psi=\bar{\psi}} , \label{FSE1}\\
\Gamma^{V\mu}(x,y,z)& = &\left(\frac{\delta^{3}\Gamma}{\delta V_{\mu}(x)\delta\bar{\eta}(y)\delta\eta(z)}\right)_{V=A=0=\psi=\bar{\psi}},
\end{eqnarray*}
and may be computed from Eq.(\ref{eff2}) accordingly.
The propagator for the axial boson, given by 
\begin{eqnarray*}
 D^{A}_{\mu\nu}(x,y)\equiv \left(\frac{\delta^{2}\Gamma}{\delta A^{\mu}(x)\delta A^{\nu}(y)}\right)^{-1}_{V=A=0=\psi=\bar{\psi}}, \label{pse2}
\end{eqnarray*}
is computed analogous to the vector propagator above. With the irreducible
axial-fermion vertex defined by
\begin{eqnarray*}
\Gamma^{A}_{\mu}(x,y,z)= \left(\frac{\delta^{3}\Gamma}{\delta A^{\mu}(x)\delta\bar{\eta}(y)\delta\eta(z)}\right)_{V=A=0=\psi=\bar{\psi}},
\end{eqnarray*} 
the full set of (unrenormalised) 2-point SDEs for the theory Eq.(\ref{ZU1}) is now given, in 
momentum space by
\begin{eqnarray*}
D_{V}^{\mu\nu}(k) &= &(k^{2} g^{\mu\nu}+ \frac{v-1}{v}k^{\mu}k^{\nu})-\Pi_{V}^{\mu\nu}(k),\\
\Pi_{V}^{\mu\nu}(k)&=&\int \frac{d^{4}q}{(2 \pi)^{4}}
\textrm{tr}\left((-ig_{V}\gamma^{\mu})(iS(q))(-i \Gamma_{V}^{\nu}(q,k-q))iS(k-q)\right), \\
D_{A}^{\mu\nu}(k) &= &(k^{2} g^{\mu\nu}+ \frac{a-1}{a}k^{\mu}k^{\nu})- \Pi_{A}^{\mu\nu}(k),\\
\Pi_{A}^{\mu\nu}(k)&=&\int \frac{d^{4}q}{(2 \pi)^{4}}
\textrm{tr}\left((-ig_{A}\gamma^{5}\gamma^{\mu})(iS(q))(-i \Gamma_{A}^{\nu}(q,k-q))iS(k-q)\right), \nonumber \\
S^{-1}(p) & = & \gamma^{\mu}p_{\mu} - \Sigma(p), \\
\Sigma(p) & = &\int \frac{d^{4}q}{(2 \pi)^{4}}\left(
(-ig_{V}\gamma^{\mu})(iS(q))(-i \Gamma_{V}^{\nu}(q,q-p)))D^{V}_{\mu\nu}(q-p)\right. \nonumber\\
& & \left. \mbox{}+(-ig_{A}\gamma^{5}\gamma^{\mu})(iS(q))(-i \Gamma_{A}^{\nu}(q,q-p))D^{A}_{\mu\nu}(q-p) \right).
\end{eqnarray*}
Note, that due to the presence of the 3-point functions $\Gamma_{V,A}^{\mu}$ 
the system does not close and constitutes a countably infinite tower of coupled
equations. The standard computational procedure, which truncates the set
(typically at the 3-point level) with a set of approximations, will
be followed in chapters below.

\section{Renormalisability and triviality}\label{sec:RG}
While QED is often cited as the most successful theory in terms of
experimental data, this is entirely within a perturbative framework. 
The ``zero charge problem''  encountered at very short distances
from a point charge was first reported in 1955 \cite{Landau}, and has remained 
mathematically unresolved since.  
The effect is common to all NAF gauge theories, where the bare coupling is 
allowed to run unchecked at short distances
leading to a total screening of the bare charge by the vacuum. Such a 
screening renders the theory ``trivial'' or non-interacting.
In this section we briefly review the salient features of  Wilson's  \cite{Wilson}, \cite{WK}
renormalisation group (RG) as presented in \cite{Muta}.

Consider an action on $D$-dimensional spacetime
\begin{equation}
S=\int d^{4}x\sum_{i}g_{i}O_{i}, \label{act}
\end{equation}
where the $g_{i}$ are coupling constants and the operators $O_{i}$ 
are local monomials of elementary fields, $\phi$, with canonical 
dimension $d_{i}$.
The standard definition of a perturbatively renormalisable theory is
one for which $S$ is composed of operators with $d_{i}\leq D$.

Upon the introduction of a cutoff, the resulting action $S_{\Lambda}$
may be considered as that of a low-energy (with respect to $\Lambda$)
approximation to a ``fundamental'' theory $S_{0}$. This requires the
(non-trivial) assumption that the fields appearing in the path integral 
can be divided into high-and low energy parts
\begin{eqnarray*}
\phi=\phi_{-}+\phi_{+},
\end{eqnarray*}
so that
\begin{eqnarray*}
\int D\phi_{-}D\phi_{+}e^{iS_{0}[\phi_{-},\phi_{+}]}=
\int D\phi_{-}e^{iS_{\Lambda}[\phi_{-}]}.
\end{eqnarray*}
That is, $S_{\Lambda}$ is obtained by integrating out the heavy
degrees of freedom:
\begin{equation}
e^{iS_{\Lambda}[\phi_{-}]}=\int D\phi_{+}e^{iS_{0}[\phi_{-},\phi_{+}]}.
\label{seff}
\end{equation}
Above the cutoff $S_{\Lambda}$ is nonlocal, but in the low energy 
region it may be approximated by a local expansion of the form Eq.(\ref{act})
The operators $O^{i}$ are classified in terms of their dimension;
those with $d<D$ correspond to relevant or superrenormalisable interactions,
while those with $d=D$ are called marginal. A perturbatively renormalisable 
theory, composed of these types of interaction monomial contains only a 
finite number of divergent diagrams and can be formulated independently
of a given cutoff value $\Lambda$.
Non-renormalisable interactions, those for which $d_{i}>4$ lead to new
divergences at each order of perturbation are accommodated by the assumption \cite{Wilson} that their (dimensionless) couplings $\lambda_{i}$
are of the order of unity. Such a term,
\begin{eqnarray*}
\frac{\lambda_{i}}{\Lambda^{d_{i}-D}}O_{i},
\end{eqnarray*}
is then suppressed by positive powers of $\Lambda$. 
Its divergent diagrams are now removed by relevant or marginal 
counterterms, or equivalently absorbed into ``running'' bare quantities.
 
For large, i.e., non-perturbative) couplings this running may change
the dimension $d_{i}$ significantly, so that, e.g., perturbatively 
irrelevant interactions become marginal (such as in the GNJL below)

The transformation of shifting renormalisation scale, readily demonstrated 
to have semigroup properties, hence dubbed the ``renormalisation group''
is the most suitable method for analysing such running behaviour.
The differential equations describing the response of the Greens functions
and couplings of a theory to a small shift in renormalisation scale
are known as renormalisation group equations (RGEs). 
Consider a perturbatively renormalisable theory, such as four dimensional QED with,
 for simplicity, a single fermion flavour:
\begin{eqnarray*}
{\cal L}_{QED}=
\bar{\psi}(i\gamma^{\mu}(\partial_{\mu}+ e A_{\mu})-m_{0})\psi
-F_{\mu\nu}F^{\mu\nu}-\frac{1}{2z}(\partial^{\mu}A_{\mu})^{2}, \label{LQED}
\end{eqnarray*}
where $z$ is the covariant gauge-fixing parameter and $m$, $e$ denote
the bare mass and coupling respectively. The renormalisability
property means that the divergences may be removed order-by order upon 
the field redefinitions 
\begin{eqnarray*}
\psi=\sqrt{Z_{2}}\psi, & & A_{\mu}=\sqrt{Z_{3}}A_{\mu}, 
\end{eqnarray*}
and vertex renormalisation
\begin{eqnarray*}
\gamma^{\mu}=Z_{1}\gamma^{\mu}.
\end{eqnarray*}
Consider now the 2-point fermion Greens function renormalised at an arbitrary
scale $\mu$:
\begin{eqnarray*}
S(p,g_{R},m_{R};\mu)=Z_{2}(\mu)S(p,g_{0},m_{0}). \label{ren}
\end{eqnarray*}
The statement that the unrenormalised function S is independent of the 
renormalisation scale for fixed $g_{0}$, $m_{0}$
\begin{eqnarray*}
\frac{d}{d\mu}S(p,g_{0},m_{0})|_{g_{0},m_{0}}=0,
\end{eqnarray*}
can thus be rearranged, on inverting Eq.(\ref{ren})
as
\begin{equation}
\left(\mu\frac{\partial}{\partial\mu}+ \beta\frac{\partial}{\partial g_{R}} -
\gamma_{m}m_{R}\frac{\partial}{\partial m_{R}}- 2\gamma \right)S|_{g_{0},m_{0}}=0,
\end{equation}
where 
\begin{eqnarray}
\beta&=&\mu\frac{\partial g_{R}}{\partial \mu}|_{g_{0},m_{0}}, \label{bfn}\\
\gamma_{m}&=&-\frac{\mu}{m_{R}}\frac{\partial m_{R}}{\partial \mu}|_{g_{0},m_{0}} ,\\
\gamma&=&\frac{\mu}{2Z_{2}}\frac{\partial Z_{2}}{\partial \mu} |_{g_{0},m_{0}}.\label{gfn2}
\end{eqnarray}
The actual parameter-dependence of these functions varies with the RG
method used.
For example, in the Gell-Mann Low scheme $m_{R}$ is fixed as the physical
fermion mass, $\gamma_{m}=0$, while $g_{R}$, $Z_{2}$ are defined in terms of 
the off-shell subtraction $p^{2}=-\mu^{2}$. In this case
$\beta=\beta(g_{R}, m_{R}/\mu)$, $\gamma=\gamma(g_{R}, m_{R}/\mu)$ and the
two point function can be written in terms of dimensionless variables as
\begin{equation}
S(p,g_{R},m_{R})=\mu^{4-2\times 3/2}\tilde{S}(\frac{p}{\mu}, g_{R}, \frac{m_{R}}{\mu}).
\label{dimless}
\end{equation}
For a shift in scale $p\to \lambda p$ the RG equation Eq.(\ref{dimless}) can
be re-expressed as
\begin{eqnarray*}
(-\lambda \frac{\partial}{\partial \lambda}
+\beta(g_{R}, \frac{m_{R}}{\mu})\frac{\partial}{\partial g}+
4-2(\frac{3}{2}+\gamma))\tilde{S}(\frac{\lambda p}{\mu}, g_{R}, \frac{m_{R}}{\mu}). \label{gs}
\end{eqnarray*} 
In the massless case, on setting $t=\ln \lambda$,
this may be solved (the result coincides for 
a number of RG schemes) to give \cite{Muta}
\begin{eqnarray}
\tilde{S}(\frac{\lambda p}{\mu}, g_{R}, \frac{m_{R}}{\mu})
& = & \tilde{S}(p, \bar{g}(t), \mu) \exp \{t-2 \int_{0}^{t}dt^{'}\gamma(\bar{g}(t^{'})), \label{rco}\\
t &= & \int_{g}^{\bar{g}(t)}\frac{dg^{'}}{\beta(g^{'})}. \label{rcu}
\end{eqnarray}
The second expression, Eq.(\ref{rcu}), gives the behaviour of the running coupling
in the following way: Any root of $\beta$ at the corresponds to a fixed 
point of Eq.(\ref{rcu}). If $\beta(g_{c})=0$ for some coupling $g_{c}$ then
the behaviour at asymptotic scales is given by
\begin{eqnarray*}
\frac{d \beta}{d\bar{g}}|_{g=g_{c}}>0, & & \bar{g}(t)\to g_{c}; \hspace{0.1cm}
t\to \infty, \\
\frac{d \beta}{d\bar{g}}|_{g=g_{c}}<0, & & \bar{g}(t)\to g_{c}; \hspace{0.1cm}
t\to -\infty. 
\end{eqnarray*}
The point $g_{c}$ is then known as a UV or IR (Gaussian) fixed point 
respectively. 
At such a point, scale invariance is recovered and the solution of
Eq.(\ref{rco}) becomes
\begin{eqnarray*}
\tilde{S}(p,g_{c})\lambda^{4-2(3/2 +\gamma)},
\end{eqnarray*}
that is, the fermion fields have dynamically acquired an anomalous
dimension $\gamma$, given by Eq.(\ref{gfn2}), in the neighborhood of the fixed point. 
It is apparent that at this scale operators in Eq.(\ref{act}) which are monomials in these
fields have a different dimension than expected from naive power-counting.
It is in this way that, for example, irrelevant 4-fermi operators become
marginal and could mix with the QED interaction close to such a point
(see section \ref{sec:gnjl} below).

If, however, there is no nontrivial root the coupling grows without limit and 
the theory is described as being non-asymptotically free (NAF). Finite answers are only
obtained for the noninteracting limit, $g(t)=0$, in which case the theory
is said to be trivial.
The running behaviour of the coupling $\beta$ is readily computed via loop
corrections
\begin{eqnarray*}
\beta(g_{R})\simeq \sum \beta_{i}g^{2i+1}_{R},
\end{eqnarray*}
and it is precisely the perturbative approximation which leads to 
triviality of NAF theories. For example, substituting the one-loop QED 
result $\beta=\beta_{1}\bar{g}^{3}$ into Eq.(\ref{rcu}) gives
\begin{eqnarray*}
\bar{g}^{2}(t)=g^{2}_{R}\frac{1}{1-2\beta_{1}g^{2}_{R} t},
\end{eqnarray*}
which has a singularity at the scale
\begin{eqnarray*}
\Lambda_{L}= e^{1/2\beta_{0}g^{2}}.
\end{eqnarray*}
The singularity persists for higher loop calculations, moreover being 
driven to lower scales. For example, in two loop perturbation theory the 
``Landau pole''  occurs \cite{G1} at
\begin{eqnarray*}
\Lambda_{L}=m_{R} \exp\{\frac{e_{R}^{2}}{\beta_{1}}
\left( \frac{\beta_{2} e_{R}^{2}}{\beta_{1}+\beta_{2}e_{R}^{2}}\right)^{\beta_{2}/\beta_{1}^{2}}\},
\end{eqnarray*}
which, with the particle spectrum of the standard model gives 
$\Lambda_{L} \simeq 10^{34}$GeV. For ``realistic'' SUSY theories \cite{G1} the singularity 
may be pushed as low as $10^{17}$GeV, adding to the neccessity for resolution of the Landau ghost
problem in NAF gauge theories.
The problem is of a perturbative nature , therefore, in the absence of higher 
physics, the solution must be non-perturbative. G\"{o}ckeler {\it et al.} in a lattice 
study \cite{G1} suggested that the singularity is avoided due to spontaneous breakdown 
of chiral symmetry. Their results however were consistent with triviality. Moreover 
\cite{G2} it was subsequently suggested that the chiral $U(1)$ suffers a similar fate.

As will be discussed in sec \ref{sec:gnjl} there is growing evidence that the 
best hope for nontrivial, singularity-free NAF theories such as the vector and chiral 
$U(1)$ gauge theories is the existence of a UV fixed point, whereby marginal 4-fermi 
operators mix with the gauge interaction. 
In this picture triviality is argued \cite{Reenders} to be averted due to the strong fermi 
self-interactions, suppressing the charge-screening effect.

\subsection{Wegner-Houghton RGE}
We conclude the discussion of RG methods with a brief summary of the
Wegner-Houghton \cite{WH} scheme, used to determine the variation of the
effective action $S_{\Lambda}$ appearing in Eq.(\ref{seff}). 
Let us define the dimensionless scale parameter as 
\begin{eqnarray*}
t=\ln \Lambda_{0}/\Lambda,
\end{eqnarray*}
where now $\Lambda_{0}$ denotes the cutoff of Eq.(\ref{seff})
and explicitly write the field-dependence $S_{\Lambda}=S_{\Lambda}[\phi_{i},\ldots,\phi_{n}]$.
The scaling response of the effective action to the infinitesimal shift in 
cutoff $\Lambda_{0} \to \Lambda_{0}-\delta \Lambda$ is given by 
\begin{eqnarray}
\frac{\partial S_{\Lambda}}{\partial t}
&=&4 S_{\Lambda}-\int\frac{d^{4}p}{(2\pi)^{4}}\phi_{i}(p)\left(
\frac{2 d_{i}-4-\gamma_{i}}{2}-p^{\mu}\frac{\partial}{\partial p^{\mu}}
\right)\frac{\delta S_{\Lambda}}{\delta \phi_{-}(p)} \nonumber \\
& & -\frac{1}{2}\int\frac{d^{4}p}{(2\pi)^{4}} \delta(|p|-1)
\{\frac{\delta S_{\Lambda}}{\delta \phi_{i}(p)}
\left(\frac{\delta^{2} S_{\Lambda}}{\delta \phi_{i}(p)\delta \phi_{j}(-p)}\right)^{-1}
\frac{\delta S_{\Lambda}}{\delta \phi_{j}(-p)}\nonumber  \\
& & -\textrm{tr}\ln\left(
\left(\frac{\delta^{2} S_{\Lambda}}{\delta \phi_{i}(p)\delta \phi_{j}(-p)}\right)\right)\}. \label{whs}
\end{eqnarray}
Here $d_{i}$ and $\gamma_{i}$ denote the canonical and anomalous dimensions
of field $\phi_{i}$ respectively, while the trace appearing in the second 
line is graded
(for fermion fields the sign changes).
The most common simplification, the local potential approximation (LPA) 
corresponds to keeping only the evolution of the potential part. 
That is, radiative corrections to operators containing derivatives are ignored,
as are field renormalisations and anomalous dimensions.
If the potential term of the action is denoted $V_{\Lambda}$, then in the
LPA approximation the Wegner-Houghton RGE Eq.(\ref{whs}) reduces to
\begin{equation}
\frac{\partial V_{\Lambda}}{\partial t}
=4 V_{\Lambda}-\frac{4-d_{i}}{2}\phi_{i}\frac{\partial V_{\Lambda}}{\partial \phi_{i}}+\frac{1}{4 \pi^{2}}\ln \left(1+\frac{\partial^{2}V_{\Lambda}}{\partial\phi_{i}\partial_{\phi_{j}}}\right). \label{veff}
\end{equation}
This result shall be used to compute the $\beta$ functions of the 4-fermi 
couplings in section \ref{sec:triv}, for direct comparison with the GNJL 
result \cite{Aoki2}.

\chapter{Models} \label{chap:mods}
Before introducing the model in the next chapter it is necessary to describe 
the phenomena we seek to explain. The other purpose of this chapter is to 
provide motivation for the model by discussing analogues from other
physical systems. Section \ref{sec:SM} introduces the electroweak sector of the 
Standard Model and briefly illustrates how $CP$ violation, mass and fermion 
generations are accommodated with a lack of predictive power. The material 
presented is readily found in 
textbooks and lecture notes, here we follow \cite{Pich94} and \cite{Aitchison}.
The second half of the chapter, section \ref{sec:gnjl}, is a review of the progress
made in understanding dynamical chiral symmetry breakdown, specifically within
QED and the GNJL.

\section{Electroweak model}\label{sec:SM}
The Glashow-Weinberg-Salam (GSW) electroweak model 
\cite{Glashow}, \cite{Weinberg}, \cite{Salam}
is a chiral gauge theory which, at low energies, 
has its full symmetry group spontaneously broken to an electromagnetic 
subgroup: 
\begin{eqnarray*}
SU(2)_{L}\otimes U(1)_{Y} \supset U(1)_{QED}. \label{decomp}
\end{eqnarray*}
The gauge sector of the electroweak Lagrangian contains 4 boson fields,
three $SU(2)$ bosons $\vec{W^{\mu}}=(W^{\mu}_{1},W^{\mu}_{2},W^{\mu}_{3})$
plus the Abelian field $B^{\mu}$. The kinetic terms (without gauge-fixing) are
\begin{eqnarray}
{\cal L}_{G}&=& -\frac{1}{4}(\vec{W_{\mu\nu}}\vec{W^{\mu\nu}}+B_{\mu\nu}B^{\mu\nu}),  \label{boskin}\\
\vec{W^{\mu\nu}}&=&\partial^{\mu}\vec{W^{\nu}} -\partial^{\nu}\vec{W^{\mu}} 
+g_{2}\vec{W^{\mu}}\times\vec{W^{\nu}},\nonumber \\
B^{\mu\nu}&=&\partial^{\mu}B^{\nu} -\partial^{\nu}B^{\mu}.\nonumber
\end{eqnarray}
The weak boson fields couple only to left-handed fermions, which
transform as $SU(2)_{L}$ isodoublets, one doublet each for quarks and 
leptons, while right-handed fermions transform as isoscalars.
In terms of generic fermion isodoublets 
\begin{eqnarray*}
\Psi_{i}=\left(\begin{array}{c}
            \psi_{\uparrow i} \\
            \psi_{\downarrow i}
            \end{array}\right),
\end{eqnarray*}
with $U(1)$ hypercharge $y_{i}$, the fermion-gauge interaction term is:
\begin{eqnarray}
{\cal L}_{I}&=& \sum_{j=q,\ell}\bar{\Psi}_{j} i\gamma^{\mu}.D_{j\mu} \Psi_{j}, \label{ferm} \\
D_{j\mu}&=&\partial_{\mu}-i g_{2}\frac{\vec{\tau}}{2}.W_{\mu}\chi_{L} -i g_{1}y_{j}B_{\mu}.\nonumber
\end{eqnarray}

Here $\tau$ are the $SU(2)$ isospin matrices, $g_{2}$ and $g_{1}$
are respectively the $SU(2)$ and $U(1)$ couplings. Here it is understood that isospin indices are suppressed and
that right-handed neutrinos $\nu_{R}$ have $y=0$.
Note the Lagrangian ${\cal L}_{I}$ contains a sum over 
isodoublets and can in principle accommodate any number of fermions.

\subsection{Generations and anomaly cancellation}
The importance of the generation structure becomes apparent when
considering anomalous processes \cite{Pich94}, which arise from
contributions of the triangle loop diagrams $\Delta^{abc}$ of Fig.\ref{figtri}.
The total sum of anomalous contributions $\Delta^{abc}$, ${\cal A}^{abc}$ is 
proportional to \cite{Swanson}
\begin{eqnarray*}
{\cal A}^{abc} \sim \textrm{tr} ( \{ \lambda^{a},\lambda^{b}\} \lambda^{c})_{L} -\textrm{tr} (\{\lambda^{a},\lambda^{b} \} \lambda^{c})_{R},
\end{eqnarray*}
where the group generators are $\lambda=\tau /2$, $Y$.
The possible combinations of $\lambda$'s are 
\begin{eqnarray*}
\textrm{tr} (\{ \tau^{a},\tau^{b} \} \tau^{c}) \sim 0 & &
\textrm{tr} (Y^{2} \tau^{c})\sim 0, \\
\textrm{tr} (\{ \tau^{a},\tau^{b} \} Y) \sim Q & &
\textrm{tr} (Y^{3}) \sim Q,
\end{eqnarray*}
where $Q$ is the sum of electromagnetic charges of the contributing internal 
fermions. If $Q$ is summed over a generation:
\begin{equation}
Q=\sum Q_{i}=Q_{\ell}+Q_{\nu}+N_{C}(Q_{q_{\downarrow}}+Q_{q_{\uparrow}})=\frac{N_{C}}{3}-1,
\end{equation}
and thus the electroweak anomalies cancel precisely when quarks
come in 3 colours. However while the requirement of anomaly cancellation
suggests a classification of fermions into families, the number of such 
families is still a free parameter. 
The remarkable conspiracy of fermions within a generation to have a 
vanishing sum of charges $Q$ is often touted as evidence for higher
physics (for a recent review see \cite{Doff}) although the problem
of anomaly cancellation shall not concern us here.

\subsection{Spontaneous symmetry breaking and mass}
As seen in section \ref{sec:chi}, bare fermion masses mix left- and 
right-handed states, violating the chiral gauge-invariance and hence 
necessarily remain absent from a renormalisable theory. Moreover
the addition of mass terms $m^{2}W^{2}$ to the gauge-boson Lagrangian Eq.(\ref{boskin}) in order to accommodate the massive $W^{\pm}$ and $Z$ bosons
also violates $SU(2)_{L}$ gauge-invariance.
In the pure SM gauge-boson and fermion masses are accommodated via the 
Higgs-Kibble mechanism \cite{Higgs}, \cite{Kibble} of spontaneous symmetry 
breaking (SSB).
That is, an extra $SU(2)_{L}$ isodoublet $\phi$ of scalars fields is 
postulated
\begin{eqnarray*}
\phi=\left(\begin{array}{c}
            \phi^{+} \\
            \phi^{0} \end{array}\right),
\end{eqnarray*}
with the corresponding gauged scalar Lagrangian
 \begin{eqnarray}
{\cal L}_{H}&=& (D_{\mu}\phi_{i})^{\dagger} D^{\mu}\phi_{i} -\mu^{2}\phi^{\dagger}\phi + \lambda (\phi^{\dagger}\phi)^{2}, \\
D_{\mu} & =& \partial_{\mu}-ig_{2}\frac{\vec{\tau}}{2}.\vec{W}_{\mu}-
i\frac{g_{1}}{2}B_{\mu}.\nonumber
\end{eqnarray}
For $\lambda>0$, $\mu^{2}<0$ the scalar potential terms have the famous 
``Mexican hat'' shape with a continuum of vacuum minima satisfying
\begin{eqnarray*}
|<0|\phi|0>|=\sqrt{-\frac{\mu^{2}}{2\lambda}}\equiv\frac{\nu}{\sqrt{2}}.
\end{eqnarray*}
Upon choosing a single ground state, the electroweak symmetry
is broken to its QED subgroup Eq.(\ref{decomp}) and from Goldstone's
theorem (see section \ref{sec:chi}) three massless states $\vec{\theta}$
appear. Rewriting $\phi$ in terms of the real fields $\vec{\theta}$
and the Higgs field $H$
\begin{eqnarray*}
\phi=\exp\left(\frac{i}{2}\vec{\tau}.\vec{\theta}\right)
\frac{1}{\sqrt{2}} \left(\begin{array}{c}
                         0 \\
                        \nu+H(x) \end{array}\right),
\end{eqnarray*}
$SU(2)_{L}$-invariance allows the unphysical (due to gauge invariance)
$\vec{\theta}$ states to be rotated away, in which case ${\cal L}_{H}$ 
becomes
\begin{eqnarray}
{\cal L}_{H}&=&\frac{1}{4}\lambda\nu^{4}+\frac{1}{2}(\partial H)^{2}-\frac{1}{2}M_{H}^{2}(H^{2}-\frac{1}{\nu}H-\frac{1}{4\nu^{2}}H^{4}) \nonumber \\
& & \mbox{} + (1 + \frac{H}{\nu})^{2} (M^{2}_{W} W^{2}+\frac{1}{2} M^{2}_{Z} Z^{2}), \label{lhig}\\
M_{H}&=& \sqrt{2\lambda}\nu,\nonumber \\
M_{W}&=&\frac{\nu g}{2}=M_{Z}\cos \theta_{W}.\nonumber
\end{eqnarray}
where now, due to the symmetry breaking, it is more convenient physically to 
rewrite the Lagrangian in terms of the intermediate gauge fields
$A_{\mu}$, $Z_{\mu}$, $W^{\pm}_{\mu}$ defined by
\begin{eqnarray}
\left(\begin{array}{c}
 W^{3}_{\mu} \\
      B_{\mu}
\end{array}\right)
& = &
\left(\begin{array}{cc}
      \cos\theta_{W} & \sin\theta_{W} \\
      \sin\theta_{W} & \cos\theta_{W}
\end{array}\right)
\left(\begin{array}{c}
      Z_{\mu} \\
      A_{\mu}
\end{array}\right), \label{nuf1}\\
W^{\pm}_{\mu}& =&\frac{1}{\sqrt{2}}(W^{1}_{\mu}\pm W^{2}_{\mu}), \label{nuf2}
\end{eqnarray}
where $\theta_{W}$ is the Weinberg mixing angle.
The SSB has been constructed in such a way that these fields are
identifiable with the photon of QED and the experimentally observed 
massive intermediate vector bosons. That is, the unphysical Goldstone 
particles have been transformed into longitudinal gauge degrees of 
freedom and give rise to mass terms $M^{2}_{W}$ and $M^{2}_{Z}$.
In terms of the intermediate fields defined in Eqs.(\ref{nuf1}, \ref{nuf2}) the fermion
Lagrangian ${\cal L}_{I}$ (given by Eq.(\ref{ferm})) becomes
\begin{eqnarray}
{\cal L}_{I}&=& \bar{\Psi}i\gamma^{\mu}\partial_{\mu}\Psi
-\frac{ig_{2}}{2\sqrt{2}}(\bar{\psi}_{\uparrow}\gamma^{\mu}W^{-}_{\mu}\chi_{L}
\psi_{\downarrow}+\bar{\psi}_{\downarrow}\gamma^{\mu}.W^{+}_{\mu}\chi_{L}\psi_{\uparrow}) \nonumber \\
& & \mbox{}-\bar{\Psi}\gamma.\left( A_{\mu}\{g_{2}\frac{\tau_{3}}{2}\sin\theta_{W}+g_{1}y\cos\theta_{W}\} + \right. \nonumber\\
& & \mbox{}\left.
Z_{\mu}\{g_{2}\frac{\tau_{3}}{2}\cos\theta_{W}-g_{1}y\sin\theta_{W}\}\right)\Psi, \label{lint}
\end{eqnarray}
and all that remains to extract QED from the $A$ terms is to impose
\begin{eqnarray*}
g_{2}\sin\theta_{W}&=&g_{1}\cos\theta_{W}\equiv e, \\
y & =&q -\tau_{3}.
\end{eqnarray*}
By postulating the existence of a complex scalar isodoublet there is the
possibility for introducing a fermion mass via a gauge-invariant
scalar-fermion coupling
\begin{equation}
{\cal L}_{Y}= -\kappa \phi \bar{\Psi}_{L}\Psi_{R}-(\kappa \phi)^\dagger \bar{\Psi}_{R}\Psi_{L}. \label{lyuk}
\end{equation}
which for a generic fermion isodoublet $\Psi$, after SSB becomes
\begin{eqnarray*}
{\cal L}_{Y}= -\frac{\nu}{\sqrt{2}}(\nu+H)
(\kappa_{1}\bar{\psi}_{\uparrow}\psi_{\uparrow}+\kappa_{2}\bar{\psi}_{\downarrow}\psi_{\downarrow}), \label{lyuk2}
\end{eqnarray*}
for arbitrary complex parameters $\kappa$. In particular if there are $n$
identical generations $\kappa$ is an $n\times n$ matrix.

Thus, the electroweak model is seen to accommodate fermion and
boson masses in a gauge-invariant way, however the price to be paid 
for multiple generations is the introduction of a proliferation of Yukawa 
couplings $\kappa_{ij}$. 
The main drawback of the Higgs-Kibble mechanism is, however, the lack of 
experimental evidence for a physical Higgs particle
(for a recent review see \cite{higgs}).
A number of alternative mechanisms for generating gauge-invariant
boson masses have been proposed, for example, \cite{Nicholson}, \cite{Leblanc},
which do not suffer the shortcomings of elementary scalar fields.
Dynamical symmetry breaking mechanisms are also used to generate fermion
masses, for example, the technicolour \cite{Weinberg2},\cite{Susskind}
or top-condensation \cite{topcon},\cite{topcon2} scenarios), and in addition
can generate gauge boson masses \cite{Gribovgol}. 

\subsection{Flavour mixing and CP violation}
A priori, the coupling matrix $\kappa$ in Eq.(\ref{lyuk}) may be
non-diagonal and complex, constrained by the requirement that it commute with 
the charge operator $q$ of QED. 
To obtain the mass eigenstates, i.e., the physical fermion basis, one must
diagonalise $\kappa$ by unitary phase-redefinitions of the fermions
\begin{eqnarray*}
\psi_{L}\to \psi^{'}_{L}=U_{L}\psi_{L}, \\
\psi_{R}\to \psi^{'}_{R}=U_{R}\psi_{R}, 
\end{eqnarray*}
such that 
\begin{eqnarray*}
\bar{\psi}_{L}U^{\dagger}_{L}\kappa U_{R}\psi_{R} \to \bar{\psi}_{L}M\psi_{R},
\end{eqnarray*}
however this phase-redefinition transforms fermion interaction terms
as
\begin{eqnarray*}
\bar{\psi}_{\uparrow}\gamma^{\mu}\chi_{L}\psi_{\downarrow} \to 
\bar{\psi_{\uparrow}}\gamma^{\mu}\chi_{L}V \psi_{\downarrow},
\end{eqnarray*}
where $V$ is an $n\times n$ unitary matrix. Essentially the mass and
weak eigenstates are incompatible, and thus flavour-changing charged 
currents are predicted.

As pointed out by Kobayashi and Maskawa \cite{Kobayashi}, for $n$ 
generations the number of complex elements $\kappa_{ij}$ which cannot be 
removed by unitary redefinitions of fermion fields $n_{\epsilon}$ is
\begin{equation}
n_{\epsilon}=(n-1)(n-2)/2 \label{genconst},
\end{equation}
and hence for $n>2$, complex couplings must remain, which lead to 
Charge-conjugation- and Parity- (CP-)violating interactions.
In fact the existence of CP violating electroweak processes had 
been previously confirmed in the decays of neutral kaons \cite{kaons}.
Hence the SM with CP-violation provides a lower bound on the number
of quark generations, $n>2$. 
\noindent
From Eq.(\ref{genconst}), for $n=3$ there exists a single complex phase in 
the quark sector.
With $\psi_{\uparrow}=(u,c,t)$and $\psi_{\downarrow}=(d,s,b)$ the 
isospin-changing, i.e., $W^{\pm}$-fermion) interactions in Eq.(\ref{lint})
become 
\begin{eqnarray*}
\bar{\psi}_{\uparrow}\gamma^{\mu}W^{-}_{\mu}\chi_{L}V_{CKM}\psi_{\downarrow}
+\bar{\psi}_{\downarrow}\gamma^{\mu}W^{+}_{\mu}\chi_{L}V^{\dagger}_{CKM}\psi_{\downarrow},
\end{eqnarray*}
where the Cabbibo-Kobayashi-Maskawa (CKM) matrix is commonly
parametrised in terms of the angles $\theta_{ij}$, the mixing angle
between the $i^{th}$ and $j^{th}$ generations \cite{bklt}:
\begin{eqnarray*}
V_{CKM}&\equiv &\left(\begin{array}{ccc}
                     V_{ud} & V_{us} & V_{ub} \\
                     V_{cd} & V_{cs} & V_{cb} \\
                     V_{td} & V_{ts} & V_{tb} \\
                    \end{array}\right) \nonumber \\
& = &           \left(\begin{array}{ccc}
c_{12}c_{13} & s_{12}c_{13} & s_{13}e^{-i\delta_{13}} \\
-s_{12}c_{23} -c_{12}s_{23}s_{13}e^{-i\delta_{13}} &
c_{12}c_{23} -s_{12}s_{23}s_{13}e^{-i\delta_{13}} & s_{23}c_{13} \\
s_{12}s_{23}-  c_{12}c_{23}s_{13}e^{-i\delta_{13}} &
-c_{12}s_{23}-  s_{12}c_{23}s_{13}e^{-i\delta_{13}} & c_{23}c_{13} 
                      \end{array}\right),
                      \end{eqnarray*}
where $c_{ij}$ and $s_{ij}$ denote $\cos\theta_{ij}$ and $\sin\theta_{ij}$
respectively.
Note that with recent experimental evidence for neutrino masses 
\cite{superK},\cite{superK2} a CKM matrix and CP-violation potentially 
exist in the leptonic sector also.
As only one CP-violating parameter $\delta_{13}$ is predicted
considerable experimental focus is currently being placed on tests of 
the CKM model (for a recent review see \cite{Rosner}, for a survey
of generalised CP violation in the SM see \cite{Espriu}).
Recently it has been argued \cite{Einhorn} that only three types
of interactions in gauge theories lead to CP violation, fermion-scalar
interactions (as above), scalar-scalar interactions, for example,
\cite{Hall},\cite{Zarikas}) and gauge terms $F\tilde{F}$ as seen in e.g.,the strong CP problem.
However it has also been pointed out \cite{transmut} that 
non-perturbative sources of flavour mixing also exist as the SM fermion 
mass matrices also undergo rotation as a result of scale changes.

\subsection{Non-perturbative aspects}
With historical emphasis in theoretical physics on grand unification 
the non-perturbative region of the SM has been largely ignored. As pointed
out in \cite{BassThomas1996}, at large momenta the hypercharge sector 
dominates, due to the asymptotic freedom of the colour and flavour 
interactions. Not only is the $U(1)$ hypercharge NAF, but numerical
simulations \cite{Beg1989a} suggest that for elementary Higgs the renormalised
quartic scalar coupling also vanishes. 
Indeed the fact that triviality can apparently be averted by embedding the 
SM group in a higher semisimple group adds to the appeal of higher physics.
However, as demonstrated in the last chapter the GNJL offers an 
alternative, in the case of NAF, or ``pure'', QED.

In another paper Kiselev \cite{K2} postulated a strongly 
self-interacting (SSIR) (composite) Higgs region intermediate to $\Lambda$ 
and a higher, e.g., unification scale $M$. 
Qualitatively, condensates of three bilocal fermion composites
(corresponding to $\bar{\tau}\tau$, $\bar{t}t$ and $\bar{b}b$)
were found to acquire non zero VEV's below the scale $\Lambda$, breaking
EW symmetry. The masses of these three fermions may be predicted
accurately, e.g. $m_{t}=165\pm 1$ GeV \cite{K2}, from the SSIR IR fixed point 
conditions. The model is also consistent with \cite{Kiselev}, where
three Higgs sharing a specially-constrained potential led to the assertion of 
an extended vacuum with $Z_{3}$-symmetry. Here the existence of multiple 
local vacuum minima was suggested to be an origin for the fermion 
generations.

A realistic description of the fermion generations via this method
requires new symmetry breaking for the lighter fermion generations
to acquire masses, however the successes of \cite{K2} provide further
motivation for a study of the qualitative features of
a hypercharge equivalent of the GNJL.

In conclusion, this section has illustrated how the electroweak model
accommodates as arbitrary parameters the number of fermions
(with some constraint given by anomaly cancellation), the fermion masses,
mixings and CP violation. For three fermion generations, one Higgs isodoublet
and no lepton CKM matrix the poor understanding of the scalar sector
implies 14 {\it ad hoc} parameters (9 Yukawa couplings, 3 mixing angles, 1 
phase and the Higgs VEV) are required.

\section{The NJL model}\label{sec:NJL}

Historically the Nambu-Jona-Lasinio model was the first relativistic quantum
field theory in which dynamical chiral symmetry breaking was studied
\cite{NJL}. It is an important sector of contemporary theories including 
top-condensation models. See, for example, \cite{topcon},\cite{topcon2}) and 
the GNJL \cite{Bardeen1989}. In this section we illustrate a number of its qualitative 
features following Miransky \cite{Miransky}.
The NJL Lagrangian is obtained from the GNJL Lagrangian Eq.(\ref{gnjleq})
upon setting the gauge coupling $e=0$:
\begin{eqnarray}
{\cal L}_{NJL}= i\bar{\psi} \gamma^{\mu}\partial_{\mu} \psi + G\sum_{a=0}^{N^{2}-1}((\bar{\psi}\frac{\lambda^{a}}{2}\psi)^{2}+
(\bar{\psi}\frac{\lambda^{a}}{2}\gamma^{5}\psi)^{2}) \\
=i\bar{\psi} \gamma^{\mu}\partial_{\mu} \psi - \bar{\psi}_{L} M \psi_{R}
- \bar{\psi}_{R} M^{\dagger} \psi_{L}-\frac{1}{2G} \textrm{tr} M^{\dagger}M,
\nonumber
\end{eqnarray}
where the auxilary field $M$ satisfies the Euler-Lagrange
equation
\begin{eqnarray*}
M^{ab}=-2 G \bar{\psi}^{b}_{L} \psi^{a}_{R}, 
\end{eqnarray*}
i.e., the coupling $G$ has canonical dimension $-2$
and in $L_{NJL}$ we have suppressed all colour indices. 
The corresponding generating 
functional is
\begin{eqnarray}
Z_{NJL}(J)&=&{\cal N} \int DM DM^{\dagger}D\psi D\bar{\psi}\exp\{i \int d^{4}x(
L_{NJL}+\sigma)\},\label{njlgen} \\ 
\sigma&=&\bar{\eta}\psi +\bar{\psi}\eta + J_{M} M +J_{M^{\dagger}}M^{\dagger},\nonumber 
\end{eqnarray}
where ${\cal N}$ is a normalisation factor.
Now the fermion fields simply appear in Eq.(\ref{njlgen}) as a Gaussian 
integral and may be formally integrated out:
\begin{eqnarray*}
\int D\bar{\psi} D\psi \exp\{i \int d^{4}x \bar{\psi}\Delta\psi \} \sim
(\textrm{Det}i\Delta)^{N_{c}},\\
\Delta=(i \gamma^{\mu}.\partial_{\mu}- M\chi_{L}- M^{\dagger}\chi_{R}).
\end{eqnarray*}
Upon substituting in Eq.(\ref{njlgen}) the effective NJL action may then be 
read off as
\begin{eqnarray*}
S_{eff}=-N_{c}(i\ln \textrm{Det}i(i \gamma^{\mu}\partial_{\mu}
- M\chi_{L}- M^{\dagger}\chi_{R}) + \frac{1}{2 GN_{c}}\int d^{4}x \textrm{tr}( M^{\dagger}M)).
\end{eqnarray*}
For $N_{c} \to \infty$ , the path integral 
\begin{eqnarray*}
Z_{NJL} \sim \int DM DM^{\dagger}\exp\{-iS_{eff}(M, M^{\dagger})+ \int d^{4}x
\sigma(x) \},
\end{eqnarray*}
is dominated by the stationary points of $S_{eff}$ and the dynamics
is classical. Thus ignoring quantum fluctuations of $M$, 
$M^{\dagger}$, we approximate $M$ by its mean-field value $\nu$.
Using chiral symmetry any vacuum solution may be transformed via chiral 
symmetry into
\begin{eqnarray*}
M=\nu= M^{\dagger},
\end{eqnarray*}
so in momentum space we consider
\begin{eqnarray*}
0=\frac{\delta S_{eff}}{\delta M}=\frac{1}{G N_{c}} \nu - i\textrm{tr}\frac{\nu}{\gamma^{\mu}.p_{\mu} - \nu+ i\epsilon}.
\end{eqnarray*}
Evaluating the trace
\begin{eqnarray*}
\nu=4 iG N_{c} \int \frac{d^{4}p}{(2 \pi)^{4}} \frac{\nu}{p^{2}- \nu^{2}+ i\epsilon}.
\end{eqnarray*}
it is immediately clear that the solution to this equation is constant
and the trivial solution $\nu=0$ exists.
Rotating into Euclidean space and computing the momentum integral yields
the algebraic equation
\begin{equation}
\nu(\frac{G N_{c}\Lambda^2}{4\pi^2} -1) - \frac{G N_{c}}{4\pi^2}\nu^{3}
\ln \frac{\Lambda^{2}+ \nu^{2}}{\nu^{2}}=0. \label{njleqn1} 
\end{equation}
For $0<G< 4 \pi^2 /N_{c}\Lambda^{2}$ both terms on the left-hand side of 
Eq.(\ref{njleqn1}) are negative if $\nu>0$ and in this case only the trivial 
solution $\nu=0$ exists. It follows then that to obtain a non-trivial real value of $\nu$, $G$ must exceed the critical value $4 \pi^{2}/N_{c}\Lambda^{2}$. Moreover this solution is unique. It is clear that in the subcritical case the single zero corresponds to a stable extremum and chiral symmetry is preserved.
However for the supercritical case this zero corresponds to a maximum, while
the nontrivial value is the stable vacuum solution. In this supercritical
phase then a mass term is dynamically generated, spontaneously breaking
the chiral symmetry of the bare Lagrangian.

\section{$QED_{4}$: dynamical chiral symmetry breaking}\label{sec:QED}
A number of early studies \cite{MaskawaNakajima},\cite{FominMiransky1976},\cite{FGM1978},\cite{FukudaKugo} in massless, quenched, rainbow QED 
indicated a 2-phase structure, with a critical coupling value 
$\alpha=\frac{\pi}{3}$ above which chiral symmetry was broken. 
Maskawa and Nakajima\cite{MaskawaNakajima} 
found this to be possible only for a finite cutoff. In general 
\cite{FominMiransky1976} the supercritical solution is found to be 
proportional to the cutoff, requiring judicious fine-tuning of the coupling 
constant to render the solution finite.

The general strategy for studying dynamical fermion mass generation
in the literature has been to truncate the system of Schwinger-Dyson 
equations at the stage of 2-point Greens functions. Typically
only the fermion propagator is retained and a test form is used for
the photon propagator. 
In QED$_{4}$ the 2-point equations to be solved are
\begin{eqnarray}
\Pi(k)&=&-\frac{1}{3}g_{\mu\nu} N \int \frac{d^{4}r}{(2\pi)^{4}}
 \textrm{tr} \{\Gamma^{\mu}_{0}S(r)\Gamma^{\nu}(r,r+k)S(r+k)\} \\
\Sigma (p)&=& \int \frac{d^{4}q}{(2\pi)^{4}},
D_{\mu\nu}(p-q) \Gamma^{\nu}(p,q) S(q)\Gamma^{\mu}_{0},
\label{fse}
\end{eqnarray}
where $D_{\mu\nu}$ is the gauge-boson propagator, while $\Gamma^{\mu}_{0}$
and $\Gamma^{\nu}(q,p)$ denote the bare and dressed vertex functions 
respectively.

From Lorentz invariance and parity conservation the most general 
decomposition of the inverse fermion propagator is, in Euclidean space
\begin{equation}
S^{-1}(p)=iA(p)\gamma.p + B(p)\equiv i\gamma.p +m_{0} + 
\Sigma(p),\label{qedinvprop}
\end{equation}
where $m_{0}$ is the (unrenormalised) bare mass, $A(p)$, $B(p)$
are scalar form factors and $\Sigma(p)$ is the fermion self-energy.
At this point a number of simplifying assumptions must be made for
the photon propagator $D_{\mu\nu}$ and the full vertex $\Gamma^{\nu}(q,p)$
in order to obtain approximate solutions.

\subsection{Quenched rainbow approximation}
In this, and the following subsection the running of the coupling is ignored 
(quenched approximation) and the photon propagator is approximated by its 
bare counterpart:
\begin{equation}
D_{\mu\nu}(k)=(\delta_{\mu\nu}+(z-1)\frac{k_{\mu}k_{\nu}}{k^{2}})\frac{1}{k^{2}}, \label{unq}
\end{equation}
where $z$ is the gauge parameter.
Further, approximating the full vertex 
$\Gamma^{\mu}(q,p)$ by the bare $\gamma^{\mu}$ (rainbow approximation)
and performing the trace over Eq.(\ref{qedinvprop}) yields:
\begin{eqnarray}
A(x)&=&1 +\frac{z \alpha}{4\pi}\int^{\Lambda^{2}}_{0} dy
\frac{A(y)}{y A^{2}(y)+B^{2}(y)}(\frac{y^{2}}{x^{2}}\theta_{+}+\theta_{-}) \label{rainA}, \\
B(x)&=&m_{0}+\frac{(3+z)\alpha}{4\pi}\int^{\Lambda^{2}}_{0} dy
\frac{B(y)}{y A^{2}(y)+B^{2}(y)}(\frac{y}{x}\theta_{+}+\theta_{-}).  \label{rainB}
\end{eqnarray}
where $x$ and $y$ are momentum-squared while $\theta_{\pm}=\theta(\pm (x-y))$ 
is the Heaviside step function. In the Landau gauge, $z=0$, this system 
reduces to a single integral
\begin{eqnarray*}
B(x)=m_{0}+\frac{3\alpha}{4\pi}\int^{\Lambda^{2}}_{0} dy
\frac{B(y)}{y+B^{2}(y)}\{\frac{y}{x}\theta_{+}+\theta_{-}\},  \label{lg0}
\end{eqnarray*}
which was extensively studied by Fukuda and Kugo \cite{FukudaKugo}
in its differential form:
\begin{equation}
y \frac{d^{2}B}{dy^{2}}+ 2 \frac{dB}{dy}+\frac{3\alpha}{4\pi} \frac{B}{y+B^{2}}
=0.
\label{dif}
\end{equation}
Here the integration limits become the boundary conditions
\begin{eqnarray}
\lim_{y\to 0} \frac{d}{dy}(y^2 B)&=&0, \\
\lim_{y\to \Lambda^{2}} \frac{d}{dy}(y B)&=&m_{0}. \label{uvb}
\end{eqnarray}
The expression Eq.(\ref{dif}) has no analytic solution, 
however analyses via bifurcation techniques \cite{bif}, 
\cite{Kondobif} have been performed,
while linearisation $B^{2}(y)\sim B^{2}(0)$ has been shown to be a 
good approximation in both infra-red and ultra-violet regions
\cite{FGM1981}.
Adopting the latter approach, the solution in the UV regime is 
\cite{Miransky}
\begin{eqnarray*}
B(y)&\simeq &
B(0) \{\frac{\Gamma(\omega)}{\Gamma(\frac{1+\omega}{2})
\Gamma(\frac{3+\omega}{2})}(\frac{y}{B^{2}(0)})^{(\omega-1)/2}\nonumber \\
& &+ \frac{\Gamma(-\omega)}{\Gamma(\frac{1-\omega}{2})
\Gamma(\frac{3-\omega}{2})}(\frac{y}{B^{2}(0)})^{-(\omega+1)/2}\} \hspace{0.1cm} ;\alpha <\frac{\pi}{3},   \\
B(y)&\simeq &
\frac{2}{\pi}\frac{B^{2}(0)}{\sqrt{y}}(\ln \frac{y}{B^{2}(0)}+
\ln 16 -2)) \hspace{0.1cm};\alpha =\frac{\pi}{3}, \\
B(y)&\simeq & B^{2}(0) \sqrt{\frac{8 \textrm{coth} \frac{\pi \tilde{\omega}}{2}}{y\pi \tilde{\omega}(\tilde{\omega}^{2}+1)}} \sin (\frac{\tilde{\omega}}{2} \ln \frac{y}{B^{2}(0)}+ \tilde{\omega}(\ln 4 -1)) \hspace{0.1cm} ;\alpha >\frac{\pi}{3}, \nonumber\\
\end{eqnarray*}
where $\omega=\sqrt{\frac{3\alpha}{\pi}-1}$ and $\tilde{\omega}=\sqrt{1-\frac{3\alpha}{\pi}}$.
Substituting these into the UV boundary condition Eq.(\ref{uvb}) yields
\begin{equation}
m_{0}\simeq \left\{ \begin{array}{ll}
         B(0)\frac{\Gamma(\omega)}{\Gamma(\frac{1+\omega}{2})
          \Gamma(\frac{3+\omega}{2})}(\frac{\Lambda^{2}}{B(0)})^{\omega-1}     
                                                 & ;\alpha <\frac{\pi}{3}, \\
  \frac{4 B(0)^{2}}{\pi\Lambda^{2}}(\ln \frac{\Lambda^{2}}{B(0)} + \ln 4 -1))  
                                                 & ;\alpha =\frac{\pi}{3}, \\
          \frac{B(0)^{2}}{\Lambda}\sqrt{\frac{ 2 \textrm{coth}\frac{\pi \tilde{\omega}}{2}}{\pi \tilde{\omega}}}\sin(\tilde{\omega} \ln\frac{\Lambda^{2}}{B(0)}+\textrm{arg}\frac{\Gamma(1+i\tilde{\omega})}{\Gamma^{2}(\frac{1+i\tilde{\omega}}{2})})                                        & ;\alpha >\frac{\pi}{3}.
         \end{array} \right.
\label{bbehav}
\end{equation}
Taking the chiral limit $m_{0}\to 0$ when $\alpha \leq \pi/3$ in 
Eq.(\ref{bbehav}) for finite $\Lambda^{2}$ it is clear that $B(0)$ is also trivial. For $\Lambda^{2} \to \infty$, $m$ cannot naively be set to zero in a 
consistent way. 
The argument \cite{Miransky} follows from the conservation of the 
(renormalised) axial current $j^{\mu}_{5}$, which is required for chiral 
symmetry-breaking 
\begin{equation}
\lim_{\Lambda^{2}\to\infty}\partial_{\mu} j^{\mu}_{5}=
\lim_{\Lambda^{2}\to\infty} iZ_{m} m_{0}(\Lambda^{2})\bar{\psi}\gamma^{5}\psi=0.
\label{subtle}
\end{equation}
Upon substituting Eq.(\ref{subtle}) into the subcritical solution of Eq.(\ref{bbehav}) it follows that only the trivial solution  $B(0)=0$ satisfies the ultra-violet boundary condition Eq.(\ref{uvb}).

However when $\alpha >\pi/3$, infinitely many solutions of Eq.(\ref{bbehav}) are possible. For $\alpha-\pi/3 <<1$ 
\begin{eqnarray*}
\sin (\tilde{\omega} \ln \frac{\Lambda^{2}}{B(0)})=0,
\end{eqnarray*}
or re-arranging for $B(0)$
\begin{eqnarray}
B(0)\sim \Lambda^{2} \exp (-\frac{\pi n}{\tilde{\omega}} + \ln 4) ; n=1,2,\ldots
\label{B0q}
\end{eqnarray}
Miransky \cite{Miransky1985} 
showed that only the case $n=1$ corresponds to a stable vacuum; solutions
with higher values are interpreted as radial Goldstone excitations.

\subsection{Improved quenched approximations}
There are a number of general criteria that an {\it ansatz} for the vector
vertex function $\Gamma^{\mu}(q,p)$ should address as outlined by Burden and 
Roberts \cite{BurdenRoberts1993}. Despite being a useful first study for
dynamical chiral symmetry breaking, quenched rainbow QED$_{4}$ fails to
satisfy several of these.

$\bullet$ $\Gamma^{\mu}(q,p)$  must transform in the same way as the bare 
vertex 
$\gamma^{\mu}$ under Lorentz transformations and satisfy charge conservation.

$\bullet$
It must satisfy the vector Ward-Takahashi Identity
\begin{eqnarray*}
k.\Gamma_{V}(q,p)=S^{-1}(q)-S^{-1}(p), \label{qedwti}
\end{eqnarray*}
In the rainbow approximation, substituting Eq.(\ref{qedinvprop}) and $\Gamma_{V}^{\mu}(q,p)=\gamma^{\mu}$ this gives
\begin{eqnarray*}
(q-p).\gamma = A(q)q.\gamma+B(q) - (A(p)p.\gamma+B(p)),
\end{eqnarray*}
which is satisfied only when $A(q)=1$, that is, in Landau gauge.
However the Ward-Takahashi Identity Eq.(\ref{qedwti}) only defines a 
(``longitudinal'') portion of the vertex $\Gamma^{\mu}$; 
a ``transverse'' piece remains unconstrained. 
The full gauge covariance of QED is guaranteed by the Landau-Khalatnikov
transformations \cite{lk}.
The strategy of making vertex {\it ansatze} satisfying individual 
Ward-Takahashi identities provides only approximate gauge-covariance
However the LK transforms are difficult to implement for many choices of 
$\Gamma^{\mu}(q,p)$ and consequently much of the literature for QED 
\cite{CurtisPennington1993},\cite{Ayse1995},\cite{Ayse1997} is aimed at 
pinning down the behaviour of the transverse piece by the other vertex 
requirements. 
An alternative approach is to also consider the so-called
``Transverse Ward-Takahashi identity'' \cite{Takahashi},\cite{Kondo1996}.
Here, in addition to the divergence 
$\partial_{\mu}\Gamma^{\mu}$, one considers the curl of the vertex 
$\partial_{\mu}\Gamma^{\nu}-\partial_{\nu}\Gamma^{\mu}$. In 2-dimensional QED 
the latter simply reduces to the axial Ward-Takahashi 
Identity \cite{DelbourgoT} and moreover the model is exactly solvable \cite{Kondo1996}.

$\bullet$
The vertex must reduce to $\gamma^{\mu}$ when dressed fermion and boson 
propagators are replaced by bare ones so that perturbation theory
is recovered in the free-field limit. The absence of kinematic  
singularities guarantees a unique $q \to p$ limit for the vertex. 
Such a vertex is given in Feynman gauge to two loops \cite{BallChiu1980}
perturbation theory while a 1 loop, singularity-free form in arbitrary
covariant gauge was given in \cite{Ayse1995}.

$\bullet$
It must allow multiplicative renormalisation of the SDE it appears in.
The restrictions that this places on the transverse vertex piece
have been considered in quenched QED$_{4}$ \cite{CurtisPennington1993}. 
The most general, non-perturbative, multiplicatively-renormalisable form for 
the vertex has also been considered \cite{Ayse1997}.

The methods for applying these constraints to the case of QED  will be 
discussed in what follows.
The most general Lorentz-invariant, spin-$\frac{1}{2}$ vertex may be written 
as \cite{Bernstein}
\begin{eqnarray*}
\Gamma^{\mu}(p,q)=\sum_{i=1}^{12} (f_{i}(p^{2},q^{2})+g_{i}(p^{2},q^{2})\gamma^{5})
v^{\mu}_{i},
\end{eqnarray*}
where $q_{\pm}=(q\pm p)$ and $v_{i}$ are given by
\begin{eqnarray*}
\begin{array}{lll}
v^{\mu}_{1}=q_{+}^{\mu}, & v^{\mu}_{2}=q_{-}^{\mu}, & v^{\mu}_{3}=\gamma^{\mu},\\
v^{\mu}_{4}=\sigma^{\mu\nu}q_{+\nu}, & v^{\mu}_{5}=\sigma^{\mu\nu}q_{-\nu}, & v^{\mu}_{6}=\gamma.q_{+}q_{+}^{\mu}, \\
v^{\mu}_{7}=\gamma.q_{+}q_{-}^{\mu}, & v^{\mu}_{8}=\gamma.q_{-}q_{+}^{\mu}, &
v^{\mu}_{9}=\gamma.q_{-}q_{-}^{\mu}, \\
v^{\mu}_{10}=\gamma^{\mu}\sigma^{\nu\lambda}q_{+\nu}q_{-\lambda}, &
v^{\mu}_{11}=\gamma.q_{+}\gamma.q_{-}q_{+}^{\mu}, & 
v^{\mu}_{12}=\gamma.q_{+}\gamma.q_{-}q_{-}^{\mu}.
        \protect
\end{array}
\end{eqnarray*}
For a parity-conserving theory the form factors $g_{i}$ vanish, and the 
further constraint of charge-conjugation invariance 
\begin{eqnarray*}
C \Gamma_{V}^{\mu}(q,p) C^{-1}= -(\Gamma^{\mu})^{T}(-p,-q),
\end{eqnarray*}
demands that each of the 12 terms be symmetric under $q\iff p$.

The requirement that the vertex satisfy the vector-current Ward identity 
Eq.(\ref{qedwti}), and in particular as $p\to q$ to avoid the appearance of 
kinematic singularities 
\begin{eqnarray}
\frac{\partial S^{-1}(p)}{\partial p^{\mu}}=\Gamma^{\mu}(p,p), \label{wgt}
\end{eqnarray}
was considered in \cite{BallChiu1980}. Indeed a criterion for
choosing the basis vectors is that each of the 12 terms be independently 
singularity-free. Substituting
\begin{eqnarray*}
S^{-1}(p)=A(p^{2})\gamma.p + B(p^{2})
\end{eqnarray*}
into Eq.(\ref{wgt}) yields
\begin{eqnarray}
\Gamma^{\mu}(p,p)=\gamma^{\mu}A(p^{2})+ 2p^{\mu}(\gamma.p \frac{\partial A}{\partial p^{2}}+\frac{\partial B}{\partial p^{2}}),
\end{eqnarray} 
which may be rewritten in a $p$,$q$-symmetric way as 
\begin{eqnarray*}
\Gamma_{BC}^{\mu}(q,p)\equiv\frac{A(q^{2})+A(q^{2})}{2}v_{3}^{\mu} + 
\frac{A(q^{2})-A(p^{2})}{2(q^{2}-p^{2})}v_{6}^{\mu}+
\frac{B(q^{2})-B(p^{2})}{q^{2}-p^{2}}v_{1}^{\mu}.
\end{eqnarray*}
In addition to restricting three of the twelve form factors, the absence of 
$q^{\mu}p^{\nu}\sigma_{\mu\nu}$ terms from Eq. (\ref{qedwti}) causes
another to vanish. 
The resulting (Ball-Chiu) vertex may now be written as
\begin{equation}
\Gamma^{\mu}(q,p)=\Gamma_{BC}^{\mu}+\sum_{i=1}^{8}A_{i}T^{\mu}_{i}, 
\end{equation}
where the Ball-Chiu basis vectors for the transverse part of the vector 
vertex are
\begin{eqnarray*}
T_{1}^{\mu}&=&p.(q-p)q^{\mu}-q.(q-p)p^{\mu},\\
T_{2}^{\mu}&=&\gamma .(q+p)T_{1}^{\mu},\\
T_{3}^{\mu}&=&(q-p)^{2}\gamma^{\mu} - (q-p)^{\mu}\gamma .(q-p),\\
T_{4}^{\mu}&=&p^{\nu}q^{\rho}\sigma_{\nu\rho} T_{1}^{\mu},\\
T_{5}^{\mu}&=&(q-p)_{\nu}\sigma^{\nu\mu},\\
T_{6}^{\mu}&=&\gamma^{\mu}(q^{2}-p^{2})- (q+p)^{\mu}\gamma .(q-p) \label{t6},\\
T_{7}^{\mu}&=&\frac{1}{2}(q^{2}-p^{2})(\gamma^{\mu}\gamma .(q+p)-(q+p)^{\mu}) + (q+p)^{\mu}p^{\nu}q^{\rho}\sigma_{\nu\rho},\\
T_{8}^{\mu}&=&-\gamma^{\mu}p^{\nu}q^{\rho}\sigma_{\nu\rho}+p^{\mu}\gamma .q
- q^{\mu}\gamma .p.
\end{eqnarray*}
To two loops in Feynman \cite{BallChiu1980} gauge $A_{4}=0=A_{5}=A_{7}$ while
if $m_{0}$, $A_{1}=0$ also. The analysis has been repeated to 1 loop in 
arbitrary covariant gauge \cite{Ayse1995} and misprints in the coefficients 
$A_{2}$,$A_{3}$ of \cite{BallChiu1980} corrected

Constraining the transverse vertex by requiring the self-energy Eq.(\ref{fse})
be multiplicatively-renormalisable led to the well-known Curtis-Pennington
vertex {\it ansatz} \cite{CurtisPennington1993}: 
\begin{equation} 
\Gamma_{CP}^{\mu}(q,p)=\Gamma^{\mu}_{BC}(q,p)+\frac{A(q)+A(p)}{2}\frac{T_{6}^{\mu}}{d(q,p)}, \label{curpen}
\end{equation}
where 
\begin{eqnarray*}
d(q,p)=\frac{(q^{2}-p^{2})^{2} + ((B(q)/A(q))^{2}+(B(q)/A(q))^{2})^{2}}{q^{2}+p^{2}}.
\end{eqnarray*}
Substituting this into Eq.(\ref{fse}) and tracing out the vector and scalar form-factors as above one arrives at
\begin{eqnarray}
A(x)&=&1-\frac{\alpha}{4 \pi}\int_{0}^{\Lambda^{2}}\frac{dy}{yA^{2}(y)+B^{2}(y)}
\{\frac{y^{2}}{x^{2}}\theta_{+}( -z A^{2}(y)-\frac{z B(y)}{y}(B(y)-B(x))) \nonumber \\
& & \mbox{}-\theta_{-} z A(x)A(y) 
+ \left(\frac{3B(y)}{2(y-x)}(B(y)-B(x)) \right. \nonumber \\ 
& & \left. \mbox{}+ \frac{3(y+x)}{4(y-x)}(1-\frac{(y-x)^{2}}{d(y,x)})A(y)(A(y)-A(x)) \right)(\frac{y^{2}}{x^{2}}\theta_{+}+\theta_{-}) \} ,\label{ac0}\\
B(x)&=&m_{0}+\frac{\alpha}{4 \pi}\int_{0}^{\Lambda^{2}}\frac{dy}{yA^{2}(y)+B^{2}(y)}\{z(\frac{y}{x}\theta_{+} A(x)B(y)+\theta_{-}A(y)B(x)) \nonumber \\
& & \mbox{}-\frac{3 x}{2(y-x)}(A(x)B(y)-A(y)B(x))(\frac{y^{2}}{x^{2}}\theta_{+}+\theta_{-}) \nonumber \\ 
& & \mbox{}+\frac{3B(y)}{2}(A(y)+A(x)+(A(y)-A(x))\frac{y-x}{d(y,x)})(\frac{y}{x}\theta_{+}+\theta_{-})\}. \label{bcp}
\end{eqnarray}
Naturally the corresponding equations for the Ball-Chiu vertex are obtained
by dropping the terms containing $d(q,p)$. Numerical solutions of this 
system of equations have yielded estimates of the critical coupling in the 
range of $0.91< \alpha_{C} <0.98 $ in Landau gauge \cite{Atkinson1993} and 
$\alpha_{C}\sim 0.92$ for Landau, Feynman and Yennie gauges \cite{CurtisPennington1993}.
The gauge-dependence of the critical coupling has been investigated both
numerically \cite{CurtisPennington1993} and analytically \cite{Atkinson1994}. 
Numerically the ``Euclidean mass'' \cite{CurtisPennington1993}
is associated with the appearance of a fixed point of the equation
\begin{eqnarray*}
M(p^{2}=-m^{2}_{E})=m_{E}; \hspace{0.1cm} M(x)=B(x)/A(x).
\end{eqnarray*}
In the latter study a bifurcation technique \cite{bif}, namely functional
differentiation of this self-mapping, was employed to locate
the change between the oscillatory and non-oscillatory behaviours of the
mass function.
Differentiating Eqs.(\ref{ac0},\ref{bcp}) with respect to $B$ and evaluating
the system at the trivial point $B(x)=0$, the momentum integrals may
be performed \cite{Atkinson1994}. Scale invariance is recovered for the
large cutoff limit, $\Lambda\to \infty$, with the solution
\begin{eqnarray}
A(x)&=&(1+\frac{\alpha z}{8 \pi})(\frac{x}{\Lambda^{2}})^{\nu}, \\
B(x)&=& A(x) x^{-s} ,
\end{eqnarray}
where
\begin{eqnarray*}
\nu=\frac{2\alpha z}{8 \pi + \alpha z}<1,
\end{eqnarray*}
and $0<s<2$ satisfies
\begin{eqnarray}
z&=&\frac{3\nu(\nu-s+1)}{2(1-s)}( 3 \pi \cot\pi (\nu-s)+ 2\pi \cot \pi s
- \pi \cot \pi\nu  \nonumber \\
& &  \hspace{2cm}\mbox{} + \frac{1}{\nu}+\frac{1}{\nu+1}+\frac{2}{1-s}+\frac{3}{s-\nu} +\frac{1}{s-v-1} ).
\label{sconstr}
\end{eqnarray}
A bifurcation point occurs when 2 solutions of Eq.(\ref{sconstr}) become equal,
and, in particular, criticality is reached when a nontrivial solution of 
Eq.(\ref{bcp}) bifurcates away from the trivial one, $B(x)=0$.
In Landau gauge Eq.(\ref{sconstr}) has two solutions, and the critical coupling
was found to be $\alpha_{C}=0.933667$. More than 2 solutions exist in an
arbitrary gauge and only the solutions continuously connected to the
Landau solutions are of interest. 
Differentiating the right-hand side of Eq.(\ref{sconstr}) w.r.t. $s$ gives
\begin{equation}
2 \pi^{2} \csc^{2} \pi s- 3 \pi^{2} \csc^{2} \pi (\nu-s) +\frac{3}{(v-s)^{2}}
-\frac{2}{(1-s)^{2}}+ \frac{1-\frac{2z}{3}}{(1+\nu-s)^{2}}=0.
\label{rootequal}
\end{equation}
Solving Eqs.(\ref{sconstr},(\ref{rootequal}) simultaneously, Atkinson
et al. \cite{Atkinson1994} found only 11$\%$ deviation from the Landau gauge 
result for gauge parameter $-2<z<20$, while for $z<-3$ the solution $B$ 
becomes 
infra-red-divergent and the gauge-dependance of $\alpha_{C}$ increases 
markedly.

Further improvements upon the Curtis-Pennington vertex must ultimately
improve the gauge-dependance of the critical coupling, and 
include the effects of unquenching the photon propagator.

\subsection{Unquenched approximations}
The unquenched behaviour of the SDE equations in QED is of vital 
importance: the running behaviour of the coupling leads problems such as 
the Landau pole in perturbation theory, while removing the cutoff ($\Lambda^{2}
\to \infty$) leads to the requirement that QED (and non-asymptotically free
theories in general) be trivial \cite{Beg1989a}. 
Including fermion loops introduces an extra dimensionless parameter into
the theory, N, the number of flavours of contributing fermions, and is 
responsible for any possible dynamically-generated photon mass.
The effect that quenching the photon propagator has upon the phase
transition in the unquenched theory has been investigated in massless, rainbow 
approximation in a number of studies \cite{OliensisJohnson1990}, \cite{Gusynin1990},\cite{Kondo1992},
\cite{Rakow}. Qualitatively the phase transition has been found to be pushed
to larger coupling scales, for $N=1$ estimates (to be compared with
$\alpha_{C}=\frac{\pi}{3}$ in quenched, rainbow case) include $\alpha_{C}\sim 2.00$\cite{OliensisJohnson1990} and $\alpha_{C}\sim 2.25$ \cite{Rakow}.

The simplest way to improve upon the quenched approximation of section 2.1
is to replace the photon propagator Eq.(\ref{unq}) by 
\begin{eqnarray*}
D_{\mu\nu}(k)=(\delta_{\mu\nu}-\frac{k_{\mu}k_{\nu}}{k^{2}(1+\Pi(k^{2}))})+G\frac{k_{\mu}k_{\nu}}{k^{2}},
\end{eqnarray*}
and re-derive the coupled equations for the fermion form factors $A$ and $B$:
\begin{eqnarray}
A(p)&=& 1 +\alpha\int\frac{d^{4}q}{(2 \pi)^{4}}
\frac{A(q)}{q^{2} A^{2}(q)+B^{2}(q)} \left(\frac{z}{(p-q)^{2}}+\frac{1}{(p-q)^{2}(1+\Pi(q))} \right. \nonumber \\
& & \left. \mbox{} + \frac{2 p.(p-q)}{(p-q)^{2}}(1-\frac{p.q}{q^{2}})(\frac{z}{(p-q)^{2}}-\frac{1}{(p-q)^{2}(1+\Pi(q))}) \right), \label{rainA2} \\
B(p)&=&m_{0}+\alpha\int\frac{d^{4}q}{(2 \pi)^{4}}(\frac{3}{1+\Pi(q)}+z)
\frac{B(q)}{q^{2} A^{2}(q)+B^{2}(q)}. \label{unqwenB}
\end{eqnarray}
The simplest non-trivial {\it ansatz} for $\Pi(k^{2})$ is to take the large momentum
limit of the one-loop contribution of $N$ massless fermion species:
\begin{eqnarray*}
\Pi(k^{2})=\frac{N \alpha}{3\pi} \ln \frac{\Lambda^{2}}{k^{2}}.
\end{eqnarray*}
However the ``Landau ghost problem'' arises at large momenta:
setting $k^{2}=\epsilon \Lambda^{2}$  an unremoveable singularity appears when 
\begin{eqnarray*}
1+\Pi(k^{2})=0 \iff \alpha= -\frac{3 \pi}{N \ln \epsilon^{-1}},
\end{eqnarray*}
i.e., when $ \epsilon \geq 1$. In particular at the cutoff when $\epsilon$
can equal 4, this constrains N to a few permissible values
$N < \frac{3\pi}{\alpha \ln 4}$. Setting $\alpha_{C}=2.00$, $2.25$
it is clear that to avoid the singularity $N \leq 3$. Indeed one study \cite{
Kondo1992} found nontrivial solutions for only $N=1$, $2$. The solution to 
Eqs.(\ref{unqwenB}) was found \cite{Gusynin1990},\cite{Kondo1992} to possess mean field behaviour
\begin{eqnarray*}
B(0)\sim \Lambda^{2} (\alpha-\alpha_{C})^{\frac{1}{2}},
\end{eqnarray*}
compared with the exponential behaviour Eq.(\ref{B0q}) in the unquenched 
case.

\section{Gauged model}\label{sec:gnjl}
The suggestion that 4-fermi operators of the type contained in the NJL 
could mix with the QED gauge interaction was first made by Bardeen {\it et al.} 
\cite{Bardeen1989}. In the quenched rainbow QED approximation in the neighbourhood
of the UV fixed point they showed that the naively irrelevant (6-dimensional)
operators
\begin{eqnarray*}
(\bar{\psi}\psi)^{2} + (\bar{\psi}i \gamma^{5}\psi)^{2},
\end{eqnarray*}
become marginal, i.e., acquire an anomalous dimension $\geq -2$.
In terms of the Wilson RG, the GNJL can be thought of as a low-energy effective action
arising in the following way. Consider the action
\begin{eqnarray*}
S^{\Lambda}_{eff}=\int d^{4}x {\cal L}_{QED}+V(\bar{\psi},\psi),
\end{eqnarray*}
where $V$ is a chiral- and parity-invariant potential. In the local potential 
approximation (LPA),i.e., in the absence of derivative terms, the lowest order
terms are given by 
\begin{eqnarray*}
\frac{g_{S}}{\Lambda^{2}}((\bar{\psi}\lambda^{\alpha}\psi)^{2} + (\bar{\psi}\lambda^{\alpha}i \gamma^{5}\psi)^{2})+\frac{g_{V}}{\Lambda^{2}}((\bar{\psi}\lambda^{\alpha}\gamma^{\mu}\psi)^{2} + (\bar{\psi}\lambda^{\alpha} \gamma^{\mu}\gamma^{5}\psi)^{2}),
\end{eqnarray*}
and chiral symmetry guarantees that the next terms in the expansion of $V$
are of order eight in the fermion fields.
Several analyses \cite{Bardeen1989}, \cite{Aoki1} have shown that for quenched QED, the 
scalar term $\sim g_{S}$ becomes marginal while the vector couplings \cite{Aoki2} remain 
irrelevant.

The GNJL (in covariant gauge) for N fermion flavours is therefore given by:
\begin{equation}
{\cal L}_{GNJL}= {\cal L}_{QED} +G\sum_{a=0}^{N^{2}-1}((\bar{\psi}\frac{\lambda^{a}}{2}\psi)^{2}+
(\bar{\psi}\frac{\lambda^{a}}{2}\gamma^{5}\psi)^{2}),
\label{gnjleq}
\end{equation}
where $G$ is a dimensionful four-fermion coupling constant, 
identified with $g_{S}/\Lambda^{2}$, while
$\lambda^{a}$ are global $U(N_{f})$ flavour generators normalised by
\begin{eqnarray*}
\textrm{tr} \lambda^{a} \lambda^{b} = 2 \delta_{ab},
\end{eqnarray*}
and the QED Lagrangian, Eq.\ref{LQED} is
\begin{equation}
{\cal L}_{QED}=
\bar{\psi}(i\gamma^{\mu}(\partial_{\mu}+ e A_{\mu})-m_{0})\psi
-F_{\mu\nu}F^{\mu\nu}-\frac{1}{2z}(\partial^{\mu}A_{\mu})^{2}.
\end{equation}
The extra 4-fermi term contributing to the fermion self-energy is given by
Figure \ref{fer4}.
\begin{figure}[tbp]
\centering{
\rotatebox{270}{\resizebox{4cm}{5cm}{\includegraphics{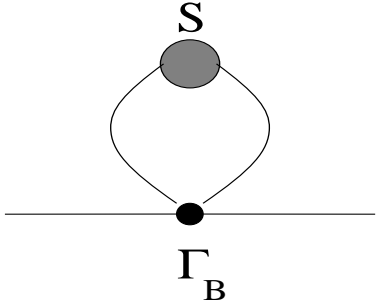}}}
}
\caption{Contribution of 4-fermi interactions to fermion self energy.}
\protect \label{fer4}
\end{figure}
In the quenched rainbow approximation this leads to the modification (cf. Eq.(\ref{rainB}))
\begin{eqnarray}
B(x)&=&m_{0}+G\int^{\Lambda^{2}}_{0} dy
\frac{B(y)}{y A^{2}(y)+B^{2}(y)} \nonumber \\
& & + \frac{(3+z)\alpha}{4\pi}\int^{\Lambda^{2}}_{0} dy
\frac{B(y)}{y A^{2}(y)+B^{2}(y)}\{\frac{y}{x}\theta_{+}+\theta_{-}\}.  \label{rainBG}
\end{eqnarray}
In the Landau gauge, $z=0$, this system is equivalent to the the differential equation
Eq.(\ref{dif})
\begin{equation}
y \frac{d^{2}B}{dy^{2}}+ 2 \frac{dB}{dy}+\frac{3\alpha}{4\pi} \frac{B}{y+B^{2}}
=0,
\label{dif2}
\end{equation}
where the 4-fermi coupling enters through the UV boundary condition Eq.(\ref{uv2})
\begin{eqnarray}
\lim_{y\to 0} \frac{d}{dy}(y^2 B)&=&0, \\
\lim_{y\to \Lambda^{2}} (1+\frac{4\pi G}{3\alpha}\frac{dB}{dy}+B)&=&m_{0}. \label{uv2}
\end{eqnarray}
In this case the large $y$, scale-invariant solution reveals a critical curve in 
coupling constant space given by \cite{bif}
\begin{equation}
G_{c}=\frac{1}{4}(1+\sqrt{\frac{3 \alpha}{4\pi}}).
\end{equation} 
Beyond this approximation, progress is made upon introducing auxilary $\sigma$ and $\pi$
fields of the type used in the NJL. Recently it has been argued in detail \cite{Reenders}
that interactions of these scalar composites with fermions suppress the charge screening
fermion loops.

\chapter{Four-fermion model of generations} \label{chap:4fer}
{\it The crucial point is to make an appropriate choice of variables, able 
to capture the physics which is most important for the problem at hand.}

\hspace{7cm}Antonio Pich \\

\noindent {\it You can use any variables at all to analyze a problem, but if you
 use the wrong variables you'll be sorry.} 

\hspace{7cm}Steven Weinberg\\

In this chapter we construct a toy four-fermion model in order to
develop the qualitative discussion of \cite{BassThomas1996}). 
The key feature of the proposal is for the interactions between
fermion chiralities to occur at separate momentum scales, specifically
\begin{eqnarray*}
\lambda^{2}_{R}< \lambda^{2}_{X}< \lambda^{2}_{L}, \label{hier}
\end{eqnarray*}
where $R$, $X$ and $L$ are the scales for right-right, left-right and
left-left couplings respectively. This is anticipated to lead to a rich set 
of phase transitions and fermion-antifermion condensates. 
Under the proposal of \cite{BassThomas1996} a ``fundamental'' fermion 
consisting of three flavours corresponding to the observed generations may
then be self-consistently introduced

We begin with the large momentum-limit of the standard model
where, due to asymptotic freedom, the $SU(3)$ colour and $SU(2)_{L}$ isospin
interactions have become negligible. The dynamics is governed by the
non-asymptotically free hypercharge $U(1)$ interaction. 
To this end we consider the generating functional ${\cal Z}$ for a chiral 
Abelian theory containing, for simplicity, a single charged, massless fermion:
\begin{eqnarray*}
{\cal Z}&=&{\cal N}\int D\bar{\psi}D\psi DZ e^{i S(\bar{\psi},\psi,Z)} \nonumber \\
S(\bar{\psi},\psi,Z)&=& \int d^{4}x
\bar{\psi}\gamma^{\mu}(i\partial_{\mu}+(c_{L}\chi_{-}+c_{R}\chi_{+})Z_{\mu})\psi-Z^{\mu}\Delta_{\mu\nu}Z^{\nu}
\end{eqnarray*}
Here ${\cal N}$ is a normalisation factor and $\Delta_{\mu\nu}$ is the 
inverse boson propagator. In the absence of any anomaly-matching, strictly 
speaking the parity-odd part of the action $S$ contains not only a 
$2 \pi i $-multivalued chiral invariant remainder, but also a Wess-Zumino-Witten functional which saturates the anomaly
\begin{eqnarray*}
S_{-}(\bar{\psi},\psi,Z)\equiv \Gamma_{WZW}+2 \pi i \eta(\bar{\psi},\psi,Z).
\end{eqnarray*}
 We shall postpone treatment of the anomaly until the next chapter, with the 
understanding that the following discussion holds for the parity-even part of 
the action at a formal level.

Completing the square in $Z$ and formally integrating out the resulting Gaussian leads to the appearance of three effective interactions, proportional
to $c^{2}_{L}$,  $c^{2}_{R}$,and $c_{L}c_{R}$ respectively:
\begin{eqnarray}
{\cal Z}&=&{\cal N}^{'}(\det \Delta^{-1})^{-1/2}\int D\bar{\psi}D\psi 
\exp(\int d^{4}x(\bar{\psi}i \gamma.\partial \psi +{\cal L}_{eff}))
\nonumber \\
{\cal L}_{eff}&\sim& (c_{L}\psi_{L}\gamma^{\mu}\psi_{L}+c_{R}\psi_{R}\gamma^{\mu}\psi_{R})
\Delta_{\mu\nu}(c_{L}\psi_{L}\gamma^{\nu}\psi_{L}+c_{R}\psi_{R}\gamma^{\nu}\psi_{R})
\label{3int}
\end{eqnarray}
In order to proceed one must make an approximation to the quartic
fermion terms. In the conventional GNJL, the bosons acquire an effective
mass below the chiral symmetry breaking scale $\Lambda_{\chi}$:
\begin{eqnarray*}
\Delta_{\mu\nu}(k)\sim g_{\mu\nu}/\Lambda_{\chi}.
\end{eqnarray*}
Here the various terms could be critical at {\it separate} scales, however
the simplest form for the massive boson propagator is
\begin{equation}
\tilde{g}^{2} \Delta_{\mu\nu}(k) \sim -\left( \tilde{c}^{2}_{R}+\tilde{c}^{2}_{L}+\tilde{c}^{2}_{X}\right)g_{\mu\nu}. \label{simp}
\end{equation}
where it is understood that the couplings denoted with a ``$\sim$''
have absorbed the scale factor $\Lambda_{\chi}$ and hence
have canonical dimension $-2$.

Some {\it ansatz} must, in general be made for the separately-running SM couplings $c_{i}(k)$, however we shall initially consider three independent coupling constants. The effective Lagrangian density is now:
\begin{eqnarray}
{\cal L} & =& \bar{\psi}i\gamma^{\mu}\partial_{\mu} \psi-
\tilde{c}^{2}_{R}\bar{\psi}_{R}\gamma^{\mu}\psi_{R}\bar{\psi}_{R}\gamma_{\mu}\psi_{R}
-\tilde{c}^{2}_{L}\bar{\psi}_{L}\gamma^{\mu}\psi_{L}\bar{\psi}_{L}\gamma_{\mu}\psi_{L} \nonumber \\
& & \mbox{} -\tilde{c}^{2}_{X}(\bar{\psi}_{R}\gamma^{\mu}\psi_{R}\bar{\psi}_{L}\gamma_{\mu}\psi_{L}+
\psi_{L}\gamma^{\mu}\bar{\psi}_{L}\bar{\psi}_{R}\gamma_{\mu}\psi_{R}). \label{imedt1}
\end{eqnarray}
Using Fierz identities (see Appendix \ref{chap:fierz}) the general effective
quartic interaction Eq.(\ref{3int}) is arrived at:
\begin{eqnarray}
{\cal L}_{eff}  &=& -
\tilde{g}_{1}\{ (\bar{\psi}\gamma^{\nu}\psi)^{2}+(\bar{\psi}\gamma^{5}\gamma^{\nu}\psi)^{2}\}
+ \tilde{g}_{2}\{(\bar{\psi}\psi)^{2}+(\bar{\psi}i\gamma^{5}\psi)^{2}\} \nonumber\\
& & +\tilde{g}_{3}( \bar{\psi}\gamma^{\mu}\psi \bar{\psi}\gamma_{\mu}\gamma^{5}\psi
+\bar{\psi}\gamma^{\mu}\gamma^{5}\psi \bar{\psi}\gamma_{\mu}\psi) \label{genform}
\end{eqnarray}
The first two terms are just the extended NJL if we identify
\begin{eqnarray*}
\tilde{g}_{1}&=&\tilde{c}^{2}_{R}+\tilde{c}^{2}_{L}, \\
\tilde{g}_{2}&=&2 \tilde{c}^{2}_{X}. \label{cupdef}
\end{eqnarray*}
The third, anomalous term $\sim \tilde{g}_{3}$ violates parity, chiral symmetry and contributes to the odd part of the effective
action. 
It is therefore clear that an analysis of D$\chi$SB of the even part would proceed along the 
same lines as the extended NJL.

\section{Digression}
Due to the ABJ anomaly, chiral symmetry in the single-fermion hypercharge theory would be broken by gauge transformations. In a first attempt therefore to distinguish the current model from the extended NJL we could 
add chiral-breaking 4-fermi interactions. 
These interactions would be constrained by the requirement
that the resulting Dirac operator $D$ has non-negative eigenvalues in order 
for the theory to be regularisable in Euclidean space. 
In computing the quantity $\ln \det D$, and in the absence of external vector 
and axial fields, the most general expression for the mass matrix is (see, e.g,. appendix A of \cite{Gasser},)
\begin{eqnarray*}
(s^{2}+(ip)^{2}) + 4 p^{2} - i \gamma^{5}(sp+ps)
\end{eqnarray*}
where $s$ and $p$ are, respectively, external scalar and pseudoscalar flavour 
matrices. Here the first term is recognisable as the chiral-invariant NJL 
terms, while the second and third are new. The above expression is Hermitian 
in the limit of vanishing $p$, however recently it has been demonstrated 
\cite{bender} that Hermiticity, while a sufficient condition for 
non-negativity is not necessary. Invariance under PT transformations was 
conjectured \cite{bender} to be the fundamental requirement.

Assuming {\it a priori} that there are no difficulties in regulating $S$,e.g.,
 via the $\zeta$-function approach \cite{Gasser}), our toy model will 
have 4-fermi interactions 
\begin{eqnarray}
{\cal L}_{4F} & =& 
- \tilde{g}_{2}\{(\bar{\psi}\psi)^{2}+(\bar{\psi}i\gamma^{5}\psi)^{2}\}
+i \tilde{g}_{3}\{\bar{\psi}\psi\bar{\psi}i\gamma^{5}\psi + \bar{\psi}i\gamma^{5}\psi\bar{\psi}\psi \} 
\nonumber \\
& \equiv &
- 2 \tilde{g}_{2} \bar{\psi}_{R}\psi_{L}\bar{\psi}_{L}\psi_{R}
-  \tilde{g}_{3}((\bar{\psi}_{L}\psi_{R})^{2}- (\bar{\psi}_{R}\psi_{L})^{2}) \label{chiform1}
\end{eqnarray}
where couplings $\tilde{g}_{2}\equiv \tilde{x}$ and $\tilde{g}_{3}$ are taken to be real.

Since chiral symmetry is explicitly broken a chiral rotation 
\begin{eqnarray*}
\psi \to \exp (i\gamma^{5}\omega)\psi,  & & \bar{\psi} \to \bar{\psi} \exp (i\gamma^{5}\omega),
\end{eqnarray*}
has the effect of transforming the anomalous coupling into a complex 
quantity 
\begin{eqnarray}
{\cal L}_{4F}^{'}&= &- 2 \tilde{x} \bar{\psi}_{R}\psi_{L}\bar{\psi}_{L}\psi_{R}
- \tilde{r}(\omega) (\bar{\psi}_{L}\psi_{R})^{2}- \tilde{r}^{*}(\omega) (\bar{\psi}_{R}\psi_{L})^{2} \label{chiform2} \\
&\equiv & 2 \frac{x}{\Lambda_{\chi}} \bar{\psi}_{R}\psi_{L}\bar{\psi}_{L}\psi_{R}
- \frac{r(\omega)}{\Lambda_{\chi}} (\bar{\psi}_{L}\psi_{R})^{2}- \frac{r^{*}(\omega)}{\Lambda_{\chi}}(\bar{\psi}_{R}\psi_{L})^{2} \nonumber
\end{eqnarray}
where $\Lambda_{\chi}$ is some renormalisation scale, which we set to the
scale of chiral symmetry breaking.
As in the case without explicit chiral symmetry breaking,
progress is made via  the mean field approximation, however here
the choice of auxilary field varies differs for both
couplings $\tilde{x}$, $\tilde{r}$ nonzero.

When $\tilde{r}=0$, $\tilde{x}\neq 0$ we encounter the familiar NJL auxilary field
\begin{eqnarray}
M&=&-2 \tilde{x} \bar{\psi}_{L}\psi_{R}, \label{aux1} 
\end{eqnarray}
For $\tilde{x}=0$, $\tilde{r}\neq 0$ the choice will be
\begin{eqnarray}
\mu &=&-\tilde{r} \bar{\psi}_{R}\psi_{L}, \label{aux2}
\end{eqnarray}
while for $\tilde{x},\tilde{r}\neq 0$ it is the complex, chiral matrix
\begin{eqnarray}
\nu &=&-2\tilde{x} \bar{\psi}_{L}\psi_{R}-\frac{\tilde{r}}{2\tilde{x}} \bar{\psi}_{R}\psi_{L}. \label{aux3} 
\end{eqnarray}
For convenience of analysis, let us consider the general case by introducing 
fermions 
$\psi_{1}$, $\psi_{2}$, $\psi_{3}$, with respective coupling values
\begin{eqnarray}
(\tilde{x},\tilde{r})=\{(\tilde{x},0),(0,\tilde{r}),(\tilde{x},\tilde{r})\}. \label{art}
\end{eqnarray}
The general 4-fermi terms in  Eq.(\ref{chiform2}) may then be 
rewritten as a (Hermitean) mass matrix as
\begin{eqnarray}
{\cal L}_{M}&= & \bar{\psi}_{1L}M\psi_{1R}+\bar{\psi}_{2L}\mu\psi_{2R}
+\bar{\psi}_{3L}\nu\psi_{3R} +\hspace{0.1cm} h.c. \nonumber\\
& &- \frac{1}{\tilde{x}}MM^{\dagger}
-\frac{1}{2}\left( \frac{\mu^{2}}{\tilde{r}}+\frac{(\mu^{\dagger})^{2}}{\tilde{r}^{*}}\right)\nonumber \\ 
& & \mbox{}- \frac{\epsilon}{2\tilde{x}}\left(\frac{1}{\epsilon}\nu\nu^{\dagger}-\frac{\tilde{r}}{2\tilde{x}}\nu^{2}-\frac{\tilde{r}^{*}}{2\tilde{x}}(\nu^{\dagger})^{2}\right)
\label{mt}
\end{eqnarray}
where 
\begin{equation}
\epsilon=1+ \frac{\tilde{r}\tilde{r}^{*}}{\tilde{x}^{2}} \label{ee}
\end{equation}

\section{Mass generation}
\label{sec:MG}
Although chiral symmetry is explicitly broken in the model 
Eq.(\ref{chiform1}), we shall now investigate the case where the magnitude of 
this effect is small. 
An analysis of dynamically-generated mass associated with the breaking of
the chiral-symmetric component can still be legitimately compared to that in 
other quantum field theories. 
Meanwhile the other two features of the toy model, fermionic generations 
and CP violation are entirely dependent on the possiblity of mass generation.

We consider the theory with three introduced fermion species with the
coupling constants specified in (\ref{art}). With the mass terms of Eq.(\ref{mt})
the Lagrangian density is
\begin{eqnarray}
{\cal L}&=&\sum_{j=1}^{3}\bar{\psi}_{j}(i\gamma.\partial-m_{j}\chi_{R}-m^{\dagger}_{j}\chi_{L})\psi_{j}+ {\cal L}_{M} \\ \label{theor}
{\cal L}_{M}&= & \sum_{j=1}^{3}(\bar{\psi}_{jL}m_{i}\psi_{jR}+ \hspace{0.1cm}h.c.)- \frac{1}{\tilde{x}}m_{1}m_{1}^{\dagger}
-\frac{1}{2}\left( \frac{m^{2}_{2}}{\tilde{r}} +\frac{(m_{2}^{\dagger})^{2}}{\tilde{r}^{*}}\right)
\nonumber\\
& & -\frac{\epsilon}{2\tilde{x}}\left(\frac{1}{\epsilon}m_{3}m_{3}^{\dagger}
-\frac{\tilde{r}}{2\tilde{x}}m_{3}^{2}-\frac{\tilde{r}^{*}}{2\tilde{x}}(m_{3}^{\dagger})^{2}\right) \nonumber
\end{eqnarray}
with $\epsilon$ given by Eq.(\ref{ee}). From the corresponding generating functional 
\begin{eqnarray*}
Z={\cal N}^{'}\int \left[D\bar{\psi}\right]\left[D\psi\right]
\left[D\mu\right]\left[D\mu^{\dagger}\right] \exp(i\int d^{4}x {\cal L}(x)).
\end{eqnarray*}
one formally integrates out the fermions to obtain the effective action 
\begin{eqnarray*}
S^{eff}_{4P}= -i \sum_{j=1}^{3} \textrm{Tr}\textrm{Ln} i G_{j}
- \int d^{4}x {\cal L}_{m}
\end{eqnarray*}
where the Dirac operators $G_{j}$ are, as discussed previously, assumed 
non-negative in Euclidean space and therefore regularisable.
If this action is dominated by the stationary points
\begin{eqnarray*}
\frac{\delta S^{eff}_{4P}}{\delta m_{j}}=0=\frac{\delta S^{eff}_{4P}}{\delta m^{\dagger}_{j}}
\end{eqnarray*} 
there are three pairs of coupled NJL-like equations, which in momentum space
are found to be
\begin{eqnarray}
\textrm{tr}[G_{1}^{-1}\chi_{R}]&=&\frac{1}{\tilde{x}}m_{1}^{\dagger},\label{emr1} \\
\textrm{tr}[G_{1}^{-1}\chi_{L}]&=&\frac{1}{\tilde{x}}m_{1},\label{eml1}\\
\textrm{tr}[G_{2}^{-1}\chi_{R}]&=&\frac{1}{\tilde{r}}m_{2},\label{emr2} \\
\textrm{tr}[G_{2}^{-1}\chi_{L}]&=&\frac{1}{\tilde{r}^{*}}m_{2}^{\dagger},\label{eml2}\\
\textrm{tr}[G_{3}^{-1}\chi_{R}]&=&\frac{\epsilon^{2}}{2\tilde{x}}(\frac{2}{\epsilon} m_{3}^{\dagger}-\frac{\tilde{r}}{\tilde{x}}m_{3}), \\
\label{emr3}
\textrm{tr}[G_{3}^{-1}\chi_{L}]&=&\frac{\epsilon^{2}}{2\tilde{x}}(\frac{2}{\epsilon} m_{3}-\frac{\tilde{r}^{*}}{\tilde{x}}m^{\dagger}_{3}), 
\label{eml3}\end{eqnarray}
Neglecting the momentum dependence of the couplings, in the mean field 
approximation this system has a constant solution.
The momentum integrals are evaluated in polar coordinates and it is 
convenient to define the integrand function
\begin{eqnarray*}
{\cal F}(m)=\int^{\Lambda}_{0}dz \frac{mz}{z+m^{2}}= m\Lambda-m^{3}\ln \frac{\Lambda+m^{2}}{m^{2}}
\end{eqnarray*} 
where $z$ is the Euclidean squared momentum. Given that $m_{j}$ is in general a
complex quantity, $m_{j}=a_{j}+ib_{j}$, we shall frequently express this function as
\begin{eqnarray*}
{\cal F}(m)&\equiv& {\cal F}_{1}+i{\cal F}_{2} \\
{\cal F}_{1}& = & a\Lambda-(a^{3}-3ab^{2})L+(3a^{2}b-b^{3})A \nonumber\\
{\cal F}_{2}& = & b\Lambda-(a^{3}-3ab^{2})A-(3a^{2}b-b^{3})L \nonumber\\
L&\equiv& \frac{1}{2}\log \left(1+\frac{\Lambda^{2}+2\Lambda(a^{2}-b^{2})}{a^{2}+b^{2}}\right)\nonumber \\
A& \equiv& \arctan \left(\frac{2ab}{\Lambda+a^{2}-b^{2}}\right)-
\arctan \left(\frac{2ab}{a^{2}-b^{2}}\right)\nonumber
\end{eqnarray*}
where for the complex logarithm the usual branch cut along the negative
imaginary axis has been made and it has been assumed that $a>b$.

The mean-field expectation values thus obtained are complex, however
in the results obtained below it is understood that a suitable unitary
redefinition of the fermion field in question enables the phase to be 
eliminated. 
For the question of mass generation it therefore suffices to seek non-trivial 
real values $a$, which minimise the effective potential.

\subsection{Conventional chiral symmetric phase}
Let us first consider the chiral-invariant limit of the model, $r=0$, where the 
behaviour should emulate that of the conventional NJL. That is, above a 
critical coupling value the chiral-symmetric vacuum becomes unstable
and undergoes a phase transition to  a configuration where chiral symmetry
is dynamically broken. 
Integrating Eqs.(\ref{emr1},\ref{eml1}) up to the UV cutoff $\Lambda$ yields
\begin{eqnarray*}
\frac{1}{\tilde{x}} m_{1}& = &\frac{1}{4\pi^{2}}{\cal F}(\Lambda,0)m^{\dagger}_{1}, \\
\frac{1}{\tilde{x}} m_{1}^{\dagger}& = &\frac{1}{4\pi^{2}}{\cal F}(\Lambda,0)m_{1}.
\end{eqnarray*}
In terms of real and imaginary parts $m_{1}=a_{1}+ib_{1}$ 
there are two ``gap'' equations
\begin{eqnarray}
a_{1}&=&\frac{\tilde{x}}{4\pi^{2}}{\cal F}_{1} \label{njleqn1a}\\
b_{1}&=&-\frac{\tilde{x}}{4\pi^{2}}{\cal F}_{2}\label{njleqn1b} 
\end{eqnarray}
In the case of real fermion mass, $b_{1}=0$, ${\cal F}_{2}=0$ and 
Eq.(\ref{njleqn1a}) is just the mean-field NJL equation (\ref{njleqn1}).
Hence nontrivial solutions exist for the former equation when
both sides have the same sign, i.e.,
\begin{equation}
\tilde{x}_{c}\geq 4\pi^{2}/ \Lambda,
\end{equation}
or, in terms of the dimensionless couplings defined in Eq.(\ref{chiform2}),
\begin{eqnarray*}
x_{c}\geq 4\pi^{2} \frac{\Lambda_{\chi}}{\Lambda},
\end{eqnarray*}
The trivial solution $a_{1}=0$ also exists for both phases, but it 
corresponds to a vacuum maximum when $\tilde{x} > 4\pi^{2}/ \Lambda$ (see section \ref{sec:NJL}). 

\subsection{Explicitly broken phase}
Consider now the purely explicit chiral-breaking interaction. Integrating
over the fermion loop, Eqs.(\ref{emr2}, \ref{eml2}) give
\begin{eqnarray*}
\frac{1}{\tilde{r}} m_{2}& = &\frac{1}{4\pi^{2}}{\cal F}(m^{\dagger}_{2}), \label{g2}\\
\frac{1}{\tilde{r}^{*}} m_{2}^{\dagger}& = &\frac{1}{4\pi^{2}}{\cal F}(m_{2}).
\end{eqnarray*}
Writing the coupling as $\tilde{r}=\tilde{\rho}+i\tilde{\sigma}$, the real and imaginary gap equations then read
\begin{eqnarray}
a_{2}&=&\tilde{\rho}{\cal F}_{1}-\tilde{\sigma}{\cal F}_{2} \\
b_{2}&=&\tilde{\rho}{\cal F}_{2}+\tilde{\sigma}{\cal F}_{1}
\end{eqnarray}
Real, nontrivial solutions are therefore obtained when the imaginary
component of the coupling $\tilde{\sigma}$ vanishes. The NJL equation (\ref{njleqn1})
is obtained again, with the explicit chiral-breaking coupling $\tilde{r}$ 
having the critical value
\begin{equation}
\tilde{r}_{c}=\rho_{c}=4\pi^{2}/ \Lambda \label{ncr}
\end{equation}

\subsection{Anomalous phase}
Finally we integrate the equations for $G_{3}$ up to scale 
$\Lambda$ yielding
\begin{eqnarray*}
\frac{\epsilon^{2}}{2\tilde{x}}(\frac{2}{\epsilon}m_{3}^{\dagger}-\frac{\tilde{r}}{\tilde{x}}m_{3}) & = & \frac{1}{4\pi^{2}}{\cal F}(m_{3}) \\
\frac{\epsilon^{2}}{2\tilde{x}}(\frac{2}{\epsilon}m_{3}- \frac{\tilde{r}^{*}}{\tilde{x}}m^{\dagger}_{3}) & = & \frac{1}{4\pi^{2}}{\cal F}(m^{\dagger}_{3}) 
\end{eqnarray*}
In this case when $b_{3}$=0, $\tilde{\sigma}=0$ the NJL gap equation is found 
where the critical coupling curve is given by
\begin{equation}
\tilde{\ell}=\frac{\tilde{x}}{\epsilon}\frac{1}{1-\tilde{\rho} \epsilon/2\tilde{x}}=\frac{4\pi^{2}}{\Lambda}
\end{equation}
and $\epsilon$ is given by Eq.(\ref{ee}).

\section{Fermion generations}
\label{sec:FG}
It was seen in the previous section that the three fermion species introduced 
in Eq.(\ref{mt}) each, living in their own 4-fermi potential, experience
a two-phase vacuum structure analogous to the NJL. 
Moreover there are three phase transitions in the theory, corresponding
to the scales where  $\tilde{x}$, $\tilde{r}$ and $\tilde{\ell}$ each attain 
the critical value
$4 \pi^{2}/\Lambda$. This is the behaviour envisaged in 
\cite{BassThomas1996} and fermion generations can be introduced
as the components of a self-consistently defined fundamental fermion
$\Psi=(\psi_{1},\psi_{2},\psi_{3})$ via the mechanism of Kiselev \cite{Kiselev},
which we now briefly outline.

In the original study \cite{Kiselev} 3 Higgs scalars were introduced into
a specially-constrained potential leading to the possibility of an extended 
$Z_{3}$-degenerate vacuum. Fermionic fields with $Z_{3}$ components 
corresponding to fermion generations are then self-consistently introduced into 
the theory.

For the model under consideration we can rewrite the three ``masses'' as 
rewritten in the polar form
\begin{equation}
m_{j}=\tilde{g}_{j}\nu_{j}e^{i(\phi_{j}+\omega_{j}\gamma^{5})}; \hspace{0.1cm} j=r,x,\ell \label{scal}
\end{equation}
where $\tilde{g}_{j}$ are understood to be (real) Yukawa constants.
These quantities could replace the three complex VEVs in \cite{Kiselev} if we
 anticipate the full gauged version of the toy model in the following chapter.
Without loss of generality the quantities $\phi_{1}$, $\omega_{1}$, $\omega_{2}$ 
could be eliminated in the unitary gauge with the remaining phases 
$\phi_{2,3}$, $\omega_{3}$ potentially leading to observable mass effects.
The fermion ``mass'' terms could be recast in terms of auxilary fermions, 
for example, $\bar{\Psi}_{L}M_{LR}\Psi_{R}$ (c.f. \cite{Kiselev},
where $\omega_{3}=0$) given by
\begin{eqnarray}
\bar{\Psi}_{L} & =& \frac{1}{\sqrt{3}}\left( \begin{array}{c}
\bar{\psi}_{L} \\ e^{2 i \phi_{2}} \bar{\psi}_{L} \\ e^{2 i (\phi_{3}+\omega_{3}\gamma^{5})} \bar{\psi}_{L}\end{array}\right)\nonumber \\
\Psi_{R}& = &\frac{1}{\sqrt{3}}\left(\begin{array}{ccc}
\psi_{R},& e^{-i \phi_{2}} \psi_{R},&e^{-i (\phi_{3}+\omega_{3}\gamma^{5})} \psi_{R}
\end{array}\right), \nonumber\\
M_{LR}& =& 
\left(\begin{array}{ccc}
\nu_{1} & \nu_{2}e^{3i \phi_{r}} & \nu_{3}e^{3i(\phi_{3}+\omega_{3}\gamma^{5})}\\
\nu_{2} & \nu_{3}e^{i(\phi_{2}+\phi_{3}+\omega_{3})\gamma^{5})} & \nu_{1}e^{i(2\phi_{3}-\phi_{2}+2\omega_{3}\gamma^{5})} \\
\nu_{3} & \nu_{1}e^{i(2\phi_{3}-\phi_{2}+2\omega_{3}\gamma^{5})} & \nu_{2}e^{i(\phi_{2}+\phi_{3}+\omega_{3}\gamma^{5})} 
\end{array}\right)
\label{auxfer}
\end{eqnarray}
and similarly for the conjugate term $\sim M_{RL}$.
As Kiselev pointed out, all $\phi$-dependence is removed from the mass 
matrices if $\phi_{r}=-\phi_{\ell}= 2k \pi/3$. More generally it is clear
that if in addition $\omega_{3}= k \pi$, the above mass 
matrix is also scalar.

In \cite{Kiselev} the discrete $\phi$ phase values result from a special choice 
of Higgs potential, whereby the vacuum configuration has a $Z_{3}$ 
degeneracy. With the extended vacuum state 
\begin{eqnarray*}
|0,0,0> = |0_{1}>\otimes |0_{2}>\otimes |0_{3}>
\end{eqnarray*} 
a $Z_{3}$ symmetric model of fermions $\Psi$, $\bar{\Psi}$ can be 
self-consistently defined. See \cite{Kiselev} for full details of the scalar
potential. 
Once the chiral phase is also allowed, the required symmetry here is
$Z_{3}\otimes Z_{2}$. 
Although we have not yet considered internal fermionic degrees of freedom
the $Z_{2}$ factor could be relevant to the isospin structure of leptons or 
quarks, particulary in the context of left-right symmetric models 
(see section \ref{sec:ol} for further discussion).

We have outlined in this section how the existence of condensates associated
with the three critical scales provides a mechanism for fermion generations
to be self-consistently introduced.
We  note however that this auxilary field technique is quite general 
and may in principle be extended beyond the current 2- and 3-generation models 
\cite{Kiselev}, \cite{K2}.
Here the motivation for considering three generations is given by the
observation that, once the requirement of chiral symmetry is dropped, 
there are three possible types of 4-fermi potential. 

\section{CP-violation}
\label{sec:CP}
The model contains several potential sources of CP-violation. Firstly, as 
noted, the masses in Eq.(\ref{scal}) are complex, chiral objects. 
As outlined above, upon introducing three generations
the phases may be eliminated if $\phi_{j}=2 k \pi/3$, $\omega_{j}=2k \pi$. 
That is, if one appeals to the existence of a chiral $Z_{3}\otimes Z_{2}$ symmetry.

In this case flavour mixing also arises from the fact that the mass matrix in 
Eq.(\ref{auxfer}) is only determined up to an ambiguity; an internal rotation 
which cancels at the level of the $Z_{3}$ constituent fields, leaving the 
eigenvalues of $M_{LR}$ invariant.
These CKM/MNS mixing angles are naturally expressed in terms of the
ratios of quark/lepton generation masses and Kiselev has shown 
\cite{Kiselev} that when the augmented mass 
matrix is expressed in the form 
\begin{eqnarray*}
M=\left(\begin{array}{ccc}
\eta_{1} & \zeta+\theta  & \zeta \\
\zeta+\theta & \eta_{2} &\zeta-\theta \\
\zeta & \zeta-\theta & \eta_{3} \end{array}\right),
\end{eqnarray*}
the resulting estimates for the CP-violating phase and certain CKM matrix 
elements are statistically equivalent to best current experimental limits
(see \cite{Kiselev}).

In the absence of the discrete chiral symmetry, the complex chiral phases
will lead to further sources of CP violation. In the general 
parametrisation there are such three CP parameters: $\phi_{2}$, $\phi_{3}$, 
$\omega_{3}$.

\subsection{Triviality} \label{sec:triv}
In summarising this section to date, we have introduced a toy fermionic
theory which appears to contain mass generation and, upon self-consistent
introduction of three fermion generations, CP-violating behaviour. 
It was constructed in such a way that perturbative renormalisability
was implicit. However the magnitude of the  running of the coupling was 
ignored until now.
If the model is to represent a high-energy
matter-only sector, rescuing the $U(1)$ hypercharge from triviality
analogous to the GNJL, the model must itself be well-behaved in the UV
and IR limits.

The running behaviour of the dimensionless couplings $x$, $r$, $r^{*}$ in the 
LPA approximation of the Wegner-Houghton RG scheme are readily obtained
from using Eq.(\ref{veff}) with the Wilson potential given by 
${\cal L}_{4F}^{'} $ in Eq.(\ref{chiform2}).
Equivalently the form of the one-loop $\beta$ functions can be deduced
from the diagrams given in Appendix \ref{sec:NPRG}. 
For vanishing bare fermion mass the running couplings are described by 
(c.f. \cite{Aoki2})
\begin{eqnarray}
\frac{d x}{d t} & = & -2x+ (x^{2}+2rr^{*}) f, \label{cup1}\\
\frac{d r}{d t} & = & -2r+ (2 rx+r^{2}) f,\nonumber\\
\frac{d r^{*}}{d t} & = & -2r^{*}+ (2 r^{*}x +r^{*2})f,\nonumber 
\end{eqnarray}
where $f=1/4\pi^{2}$ is a constant associated with the evaluation of the
fermion loop in figure \ref{4fer}, and the dimensionless scale parameter
$t$ is defined in terms of the fixed scale $\Lambda_{X}$ as
\begin{equation}
t=\ln \Lambda_{X}/\Lambda.
\end{equation}
Equivalently the equations for $r(t)$, $r^{*}(t)$ may be split into real and 
imaginary parts, as above:
\begin{eqnarray}
\frac{d\rho}{dt}&=&-2 \rho (xf+1)+ f(\rho^{2}-\sigma^{2}),\label{cup2} \\
\frac{d\sigma}{dt}&=&-2 \sigma (fx+1) + 2f\sigma\rho. \label{cup3}
\end{eqnarray} 
The system of equations (\ref{cup1},\ref{cup2},\ref{cup3}) is found, for 
$\rho(t)=0=\sigma(t)$, to have the same qualitative properties as existing 
analyses of the conventional 4-Fermi theory \cite{Aoki1}, \cite{Branchina}. 
In this case Eq.(\ref{cup1}) has the analytic solution
\begin{equation}
x(t)=\frac{2}{f}\frac{1}{1+ce^{2t}}, \label{brach}
\end{equation}
where $c$ is the constant of integration.
This solution evidently flows to a vanishing IR point ($t\to \infty$) 
and the non-trivial UV fixed point ($t \to -\infty$) at $x=2/f$.

There is also a well-behaved analytic solution for Eqs.(\ref{cup2}, \ref{cup3}) when
$x=0$
\begin{eqnarray}
r(t)&=&\frac{2}{f}\frac{1}{1+de^{2t}}, \label{brach2}\\
r^{*}(t)&=&\frac{2}{f}\frac{1}{1+d^{*}e^{2t}}, \label{brach3}
\end{eqnarray}
i.e., it has vanishing IR and finite UV behaviour, similar to Eq.(\ref{brach}) 
above.

Once the symmetry-breaking couplings are switched on, the full system 
of equations can only be solved numerically. Again the model flows to a 
UV fixed point, however in the general complex case the IR behaviour has been altered.
Given the IR boundary conditions
\begin{eqnarray*}
x(t_{0})&=&x_{0}, \\
r(t_{0})&=&r_{0}, \\
\sigma(t_{0})&=&\sigma_{0},
\end{eqnarray*}
it is found that for $\sigma_{0}=0$, the values $x_{0}$, $r_{0}$
are scale-independent, however $x(t)$ and $r(t)$ rapidly become singular below $t<t_{0}$.

The reason for this behaviour becomes clear when these equations are combined:
\begin{equation}
\frac{d}{dt}(x+r+r^{*})=(x+r+r^{*})(-2+f(x+r+r^{*})).
\end{equation}
That is, when $x(t)=-2 \rho(t)$, the flow through coupling constant space
remains null, defining another renormalised trajectory.
Along this path the RG equations simplify to
\begin{eqnarray}
\frac{dx}{dt}&=&-2x(t)+f(\frac{3}{2}x^{2}(t)+2\sigma^{2}(t)),\label{crt1}\\
\rho(t)&=& -\frac{1}{2}x(t), \label{crt2}\\
\sigma(t) &\sim & \exp (-\int dt (2-fx(t))). \label{crt3}
\end{eqnarray}
From Eq.(\ref{crt3}) it is clear that a possibility for finite nontrivial
solutions exists if the exponent is positive, i.e., for 
``small'' enough $x(t)$.
\begin{figure}[htb]
\centering{
\rotatebox{0}{\resizebox{7.5cm}{8cm}{\includegraphics{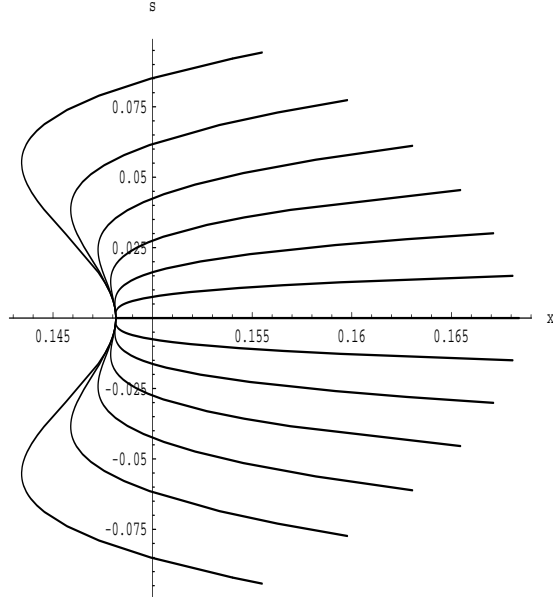}}}
}
 \caption{Behaviour of the solution of Eqs. \ref{crt1}-\ref{crt3}.
Here $s$ denotes the imaginary coupling $\sigma$.}
 \protect \label{traj1}
 \end{figure}
With vanishing imaginary coupling $\sigma$ the running couplings are, 
from Eqs.(\ref{crt1},\ref{crt2})
\begin{equation}
x(t)  = \frac{3}{2f}\frac{1}{c e^{2t}+1} = -\frac{1}{2}r(t).  
\end{equation}
This phase of the theory could, therefore, with suitable matching conditions upon $x_{0}$, $r_{0}$ at some scale $\lambda$
be interpreted as a high ($p^{2}>\lambda$) energy limit of some
lower effective theory.

At this stage it is helpful to classify the three interesting coupling-constant trajectories.
We shall refer to the subspace defined by 
\begin{eqnarray*}
x(t)&=&0, \\
r(t) & =& \frac{2}{f}\frac{1}{1+de^{2t}} =r^{*}(t),\label{ty1}
\end{eqnarray*}
as region I. It is characterised by a real, positive coupling and
the relevant low-energy 4-fermi operator is
\begin{equation}
O_{1}=\tilde{r}(\bar{\psi}_{1L}\psi_{1R})^{2}+\tilde{r}^{*}(\bar{\psi}_{1R}\psi_{1L})^{2}. \label{o1}
\end{equation}
Region II is defined to be the trajectory
\begin{eqnarray*}
x(t) & =& \frac{2}{f}\frac{1}{1+be^{2t}}, \label{ty2}\\
r(t) & = & 0 = r^{*}(t), 
\end{eqnarray*}
which is just the conventional attractive scalar interaction
of the GNJL
\begin{equation}
O_{2}=\tilde{x}\bar{\psi}_{2L}\psi_{2R}\bar{\psi}_{2R}\psi_{2L}.\label{o2}
\end{equation}
Finally there is region III, where both couplings are
non-zero
\begin{eqnarray}
x(t) & =& \frac{2}{f}\frac{1}{1+ce^{2t}}, \label{ty3}\\
\textrm{Re}(r(t)) & = &  -x(t)/2, 
\end{eqnarray}
where if $x$ is now an attractive scalar interaction, $r$
is necessarily repulsive, a similar situation to the 
1+1-dimensional Luttinger liquid model of interactions between
left- and right-moving charge densities.
The relevant operator in this case is 
\begin{equation}
O_{3}=\tilde{x}(\bar{\psi}_{3L}\psi_{3R}-\bar{\psi}_{3R}\psi_{3L})^{2}.\label{o3}
\end{equation}
Up until now our analysis has only considered one scale, however
with the hierarchy $\lambda_{1}<\lambda_{2}<\lambda_{3}$ the set of
phase transitions suggested in \cite{BassThomas1996} could naturally be described by the following scenario. 

Consider a chiral gauge theory defined at momentum scales $p^{2}<<\lambda_{1}$. Upon evolution to higher scales, an attractive,
irrelevant scalar  operator $O_{1}$ [Eq.(\ref{o1})] acquires a large 
anomalous dimension, and begins to mix with the gauge interaction, suppressing
the non-asymptotically free coupling before becoming marginal
in the neighborhood of $\lambda_{1}$. This chiral symmetry- and parity-breaking
interaction is analogous to the chiral condensates generated at the
right-right critical scale in \cite{BassThomas1996}.
At progressively larger scales fluctuations of operator $O_{2}$ [Eq.(\ref{o2})]
become apparent, before, at the scale $\lambda_{2}$, also becoming
relevant. As noted previously it is the left-right hypercharge interaction 
which gives rise to such a term, it is natural therefore to associate 
$\lambda_{2}$ with this critical scale.
Finally, at $\lambda_{3}$ both couplings are switched on,
however in this phase $r$ has been driven negative, the marginal operator
is now $O_{3}$ [Eq.(\ref{o3})]. 
This phenomenon of competing attractive and repulsive scalar interactions
also occurs in the Luttinger liquid models of strongly correlated
electrons in 1+1 dimensions (see section \ref{sec:ol} for further discussion).

\subsection{Summary}
In this section we proposed a toy 4-fermion model which, upon analysis
was demonstrated to have behaviour matching that of the hypothesis
\cite{BassThomas1996} of dynamical mass, generations and CP violation.

To the best of our knowledge it is the first toy model exhibiting such 
dynamical features and can be considered essentially intermediate to
two differing approaches to the generation problem, \cite{transmut}, 
\cite{Kiselev}, both of which are independent of the nature of 
the origin of mass and make good tree-level predictions for CKM matrix
elements, heavy/light mass ratios etc. 

The study of Kiselev \cite{Kiselev} illustrates how, with the self-consistent
introduction of $Z_{3}$ symmetric fermions and a special 3-Higgs potential,
the generation structure emerges. Under our proposal the three Higgs scalars 
are composites, the underlying dynamics being the four-fermion operators
needed to rescue the chiral $U(1)$ model from triviality in the Wilsonian
RG picture. This is in direct analogy to the GNJL as the extended version
of QED.

The latter \cite{transmut} notes that RG evolution of the SM 
fermion mass matrices offers a mechanism for flavour mixing and the fermion 
mass hierarchy. Given that these ``masses'' are coupling-dependent
linear combinations of the conventional NJL auxilary fields, the running of 
the couplings would generate flavour mixing in a similar manner.

The main drawback with this model is the origin and nature of the explicit
chiral symmetry breaking.
In the Fierz reordering of the (chiral-symmetric) hypercharge effective
action above, such terms cancelled for the separate right-right and 
left-left interactions, also for the combination of left-right and right-left 
terms. The most probable cause is an anomalous mixing of pseudoscalar
and vector/ axial fermion couplings, in which case the effect is small
at low scales.

Secondly there is the question of how the operators $O_{1}$, $O_{3}$ 
of Eqs.(\ref{o1},\ref{o3}) which 
seemingly break the chiral symmetry but, in the mean-field approximation 
at least, impart no fermion mass at subcritical scales.
With two of the three fermion species ($\psi_{1}$, $\psi_{3}$) behaving 
in a different manner to $\psi_{2}$, it moreover appears logical to
associate the former with the heavier, unstable generations.

A resolution to this may lie in the ``pseudogap'' phenomenon of 
strongly-correlated electrons \cite{Babaev}:
It is well-known that mean-field, Bardeen-Cooper-Schriffer (BCS) type theories 
of superconductivity are inadequate outside the regimes of weak coupling and
high carrier density. A separation of the temperatures of pair formation
and pair condensation occurs and, in the intermediate temperature range
Cooper pairs exist, however due to large fluctuations no condensate is 
formed. 

Recently it was shown \cite{Babaev} that the phenomenon exists in the
Gross-Neveu model and it was suggested to apply to other relativistic 
matter-only models. 
In the theory under consideration the implication is that the 
Dirac vacuum is, in fact such a ``pseudogap''-like phase.
Although the pairing operators $O_{1}$, $O_{3}$ should be written
down, in the absence of a condensate the chiral symmetry is not broken.

While the origin of such an effect is speculative, the model 
contains all the desirable features needed for the hypothesis
\cite{BassThomas1996}. Thus, instead of rejecting the model in Eq.(\ref{chiform1}) 
outright, we shall now use the insights gained to attempt to construct a 
well-defined theory from an alternative approach.

\section{Four-fermion model}\label{sec:alter}
We now proceed to investigate the effective 4-fermi theory 
containing chirally-invariant, parity-violating terms Eq. (\ref{genform}).
In terms of left and right auxilary fields we may write it as
\begin{eqnarray}
{\cal L}&=&\bar{\psi}(i \gamma^{\mu}\partial_{\mu} + M\chi_{R}+M^{\dagger}\chi_{L}
+\gamma^{\mu}\chi_{R}R_{\mu}+\gamma^{\mu}\chi_{L}L_{\mu})\psi \nonumber \\
& & -\textrm{tr}(\frac{1}{x} MM^{\dagger}+\frac{1}{r}R^{2}+\frac{1}{\ell}L^{2}),
 \label{alt1}\\
M & =& -2 \tilde{x} \bar{\psi}_{R}\psi_{L}, \nonumber\\
M^{\dagger} & =& -2 \tilde{x} \bar{\psi}_{L}\psi_{R}, \nonumber\\
R_{\mu} & =& -\tilde{r} \bar{\psi}_{R}\gamma_{\mu} \psi_{R}, \nonumber\\
L_{\mu} & =& -\tilde{\ell} \bar{\psi}_{L}\gamma_{\mu} \psi_{L}. \nonumber
\end{eqnarray}
In order to study chiral symmetry breakdown we shall improve on the mean-field
approximation and consider dynamical fluctuations in the composite
chiral bosons. The fermion self-energy SDE is readily obtained from the
generating functional of Eq.(\ref{alt1}) via a similar procedure to that in 
section \ref{sec:DSE}. 
Alternatively it may be written diagrammatically as shown
in Figure \ref{fsefig}:
\begin{figure}[htb]
\centering{
\rotatebox{270}{\resizebox{3cm}{12cm}{\includegraphics{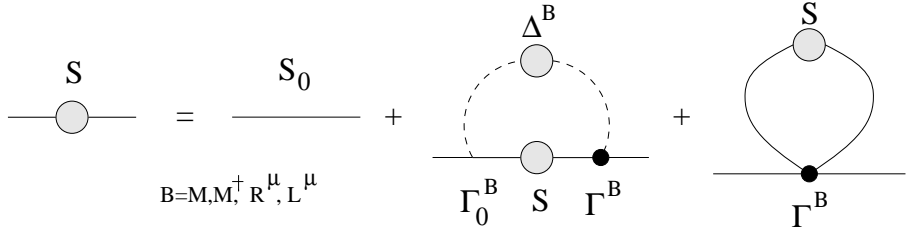}}}
}
 \caption{SDE for the fermion self-energy from the model Eq.(\ref{alt1}).}
 \protect \label{fsefig}
 \end{figure}
The most general form of the inverse fermion propagator consistent with
Lorentz invariance we consider is, in Euclidean space,
\begin{equation}
S(p)=\{(A(p) +a(p)\gamma^{5}) i \gamma .p+ B(p)+i b(p)\gamma^{5} \}^{-1},
\end{equation}
which may be rewritten in a number of ways, however
we shall project out the chiral 
components of the propagator on the left
simplifies to
\begin{eqnarray}
 S^{-1}(p)&=&\chi_{+}( i A_{+}(p)\gamma .p + B_{+}(p))^{-1}
     +\chi_{-}(i A_{-}(p)\gamma .p+B_{-}(p))^{-1},\nonumber \\
& &  \label{invprop} \\
S(p)& = &
\frac{(A_{+}(p)i\gamma.p-B_{+}(p))}{D_{+}(p)}\chi_{+}+
\frac{(A_{-}(p)i\gamma.p-B_{-}(p))}{D_{-}(p)}\chi_{-},\label{prop}\\
D_{\pm}(p)&=&A_{\pm}(p)^{2}p^{2} +B_{\pm}(p)^{2},
 \end{eqnarray}
where the chiral form factors are defined by
\begin{eqnarray*}
B(p)&=& (B_{+}(p)+B_{-}(p))/2, \\
b(p)&=& (B_{+}(p)-B_{-}(p))/2, \\
A(p)&=& (A_{+}(p)+A_{-}(p))/2, \\
a(p)&=& (A_{+}(p)-A_{-}(p))/2,
\label{ff}
\end{eqnarray*}
and the Euclidean mass ``poles'' and wavefunction normalisations
are read off as
\begin{eqnarray}
M_{\pm}(p) & =&  \frac{|B_{\pm}(p)|}{|A_{\pm}(p)|}, 
\label{mod}\\
Z_{\pm}(p) & =&\frac{1}{|A_{\pm}(p)|} \label{wav}. 
\end{eqnarray}
The presence of chirality-dependent form factors
in the numerator of Eq.(\ref{prop}) can be interpreted as
the fermion interacting with a non-trivial background.
For example, in the presence of an external, spatially
varying pseudoscalar field it has been shown \cite{joyce} that a bias exists between particles and antiparticle distributions of the same chirality. We shall encounter this  plus the case
when Eq.(\ref{prop}) contains two distinct terms, 
$|A_{+}|\neq |A_{-}|$, $|B_{+}|\neq |B_{-}|$, in section\ref{sec:rain}.

The fermion self energy is now written down from Fig \ref{fse} as
\begin{eqnarray}
\Sigma(p)&=&
\int \frac{d^{4}q}{(2\pi)^{4}}\textrm{tr}\left(
\Gamma^{M}_{0}S(q)+ \Gamma^{M^{\dagger}}_{0}S(q)\right) \nonumber \\
& & +\sum_{B=R^{\mu},L^{\mu}}
\int \frac{d^{4}q}{(2\pi)^{4}}\Gamma^{B}_{0}S(q)\Gamma^{B}(q,p)\Delta^{B}(q-p),
\label{fse2}
\end{eqnarray}
where the quantities $\Gamma^{B}_{0}$, $\Gamma^{B}(q,p)$ denote
the bare and full boson vertices and $\Delta^{B}$ the inverse boson 
propagators. The bare vertices appear in Eq.(\ref{alt1}) while the boson 
propagators satisfy the SDEs
\begin{eqnarray*}
\Delta^{-1}_{B}(k) & = & \Delta^{-1}_{0B}(k)+ \Pi_{M}(k), \\
\Pi_{B}(k) & = & \int \frac{d^{4}q}{(2\pi^{4})}
\textrm{tr}\left( (-i\Gamma^{0}_{B})(iS(q))(-i\Gamma_{B}(q,k-q))
(iS(k-q))\right).
\end{eqnarray*}
Here the bare boson propagators are 
\begin{eqnarray*}
\Delta^{-1}_{0M}(k)& = & \frac{1}{\tilde{x}^{2}}=\Delta^{-1}_{0M^{\dagger}}(k), \\
(\Delta^{\mu\nu}_{0R})^{-1}(k)& = &\frac{\delta^{\mu\nu}}{\tilde{r}^{2}}, \\
(\Delta^{\mu\nu}_{0L})^{-1}(k)& =&\frac{\delta^{\mu\nu}}{\tilde{\ell}^{2}}. 
\end{eqnarray*}
The full vertices $\Gamma_{B}$ also satisfy SDEs, however in addition to the 
two-point functions defined here they depend upon 4-point Greens functions. 
As mentioned in section \ref{sec:DSE} the full set of SDEs for a theory is a 
countably infinite set of coupled equations. In order to obtain an 
approximate solution it is necessary to truncate the system. Here we shall
adopt the standard approach and truncate at the two-point level, making
approximations to the vertex functions.

\subsection{Mass generation}
The simplest approximate solution of the fermion self-energy Eq.(\ref{fse2})
is obtained by replacing all full vertices and boson propagators by 
their bare counterparts, for the latter this is achieved by neglecting
the vacuum polarisation terms $\Pi_{B}$.

The (Euclidean space) self energy is also defined through the inverse fermion 
propagator Eq.(\ref{invprop})
\begin{eqnarray} 
\Sigma(p)&=&S^{-1}(p)-i\gamma.p-m \nonumber\\
& = &\chi_{+} \{ i\gamma.p(A_{+}(p)-1)+ (B_{+}(p) - m_{+})\}\nonumber \\
& & \mbox{}+ \chi_{-}\{ i\gamma.p(A_{-}(p)-1)+ (B_{-}(p) - m_{-})\}. \label{lor}
\end{eqnarray}
Equating Eq.(\ref{lor}) and Eq.(\ref{fse2}) projected into chiral components
\begin{eqnarray*}
\Sigma(p)=\chi_{+}\Sigma_{+}(p)+\chi_{-}\Sigma_{-}(p),
\end{eqnarray*} 
expressions for the chiral form factors $A_{\pm}$, $B_{\pm}$ may then be 
traced out. Computing the momentum integral in polar coordinates, these 
expressions are found to be: 
\begin{eqnarray}
\textrm{tr}(\gamma \Sigma_{\pm})/4=A_{\pm}(y)&=& 1, \label{mfa} \\
\textrm{tr}( \Sigma_{\pm})/4=B_{\pm}(y)=B_{\pm} &=& \frac{\tilde{x}}{16\pi^{2}}\int^{\Lambda}_{0} dz 
\frac{B_{\mp}}{z+B_{\mp}^{2}}, \label{mfb}
\end{eqnarray}
which have an analytic solution given by [c.f. the NJL gap
equation Eq.(\ref{njleqn1})]:
\begin{eqnarray*}
B_{\pm}=\frac{\tilde{x}^{2}}{16 \pi^{2}}B_{\mp}{\cal F}(B_{\mp}).
\end{eqnarray*}
As for the NJL model, $B_{+}=B_{-}$ are real solutions 
and one finds the critical coupling 
\begin{equation}
\tilde{x}^{2}_{c}=16 \pi^{2}/\Lambda^{2}. \label{xC}
\end{equation}

To move beyond this approximation it is necessary to promote the bosons to dynamical particles. This is achieved by dropping the auxilary ``mass'' terms and adding the kinetic piece
\begin{eqnarray*}
{\cal L}_{kin}= \frac{1}{x^{2}}\partial_{\mu}M\partial^{\mu}M^{\dagger}+
\frac{1}{r^{2}}{\cal F}_{R}^{\mu\nu}{\cal F}_{R\mu\nu}+
\frac{1}{\ell^{2}}{\cal F}_{L}^{\mu\nu}{\cal F}_{L\mu\nu}, 
\end{eqnarray*}
with 
\begin{eqnarray*}
{\cal F}_{R}^{\mu\nu}= \partial^{\mu}X^{\nu}-\partial^{\nu}X^{\mu},
\end{eqnarray*}
to the Lagrangian Eq.(\ref{alt1}). The  bare boson propagators $\Delta_{B}$ are 
thus modified:
\begin{eqnarray*}
\Delta_{M}=\Delta_{M^{\dagger}}& = & \tilde{x}^{2} \frac{1}{k^{2}}, \\
\Delta_{R}^{\mu\nu}(k) & = & 
\tilde{r}^{2}\frac{1}{k^{2}}(\delta^{\mu\nu}-\frac{k^{\mu}k^{\nu}}{k^{2}}),\\
\Delta_{L}^{\mu\nu}(k) & = & 
\tilde{\ell}^{2}\frac{1}{k^{2}}(\delta^{\mu\nu}-\frac{k^{\mu}k^{\nu}}{k^{2}}).
\end{eqnarray*}
With these substitutions in Eq.(\ref{fse2}), Eqs. (\ref{mfa}, \ref{mfb}) become
\begin{eqnarray}
A_{\pm}(y) & =  & 1, \label{mf2a}\\
B_{\pm}(y) & = & \frac{\tilde{x}^2}{16\pi^{2}}\int^{\Lambda^{2}}_{0} dz
\frac{x^{2}B_{\mp}(z)}{D_{\mp}(z)}\{\frac{z}{y}\theta_{+}+\theta_{-}\},  \label{mf2b}
\end{eqnarray}
where, as before, $\theta_{\pm}=\theta(\pm(y-z))$
is the Heaviside step function and the squared momenta are $z=q^{2}$, $y=p^{2}$.

The solutions of the system Eqs.(\ref{mf2b}) are again real, $B_{+}=B_{-}\equiv B$ and given by the equation of  quenched rainbow QED in the Landau gauge with the 4-fermion interaction strength  $\tilde{x}^{2}$ playing the role of the 
fine-structure constant $\alpha$.
Analysis of the critical coupling may proceed in an almost identical 
fashion as in \cite{RobertsWilliams1994}, for example, converting Eq.(\ref{mf2b}) into 
a differential form
\begin{eqnarray*}
y\frac{d^{2}B}{dy^{2}}+2\frac{dB}{dy}+ \frac{\alpha}{16\pi^{2}}\frac{B}{y+|B|^{2}}=0, \\
\lim_{y\to 0}\frac{d}{dy}(y^{2}B)=0, \\
\lim_{y\to \Lambda^{2}} \frac{d}{dy}(yB)=0. 
\end{eqnarray*}
These equations are scale invariant for $y>>B^{2}$ and have the 
solution
\begin{eqnarray*}
B_{<}(y) & = & y^{(-1+\sqrt{1-\frac{\alpha}{16 \pi^{3}}})2}; \hspace{0.1cm} \alpha \leq 16 \pi^{2}/3, \\
B_{>}(y)& = & y^{-0.5} e^{i(\sqrt{\frac{\alpha}{16 \pi^{3}}-1}\ln y)/2}; \hspace{0.1cm} \alpha\geq 16 \pi^{2}/3 .
\end{eqnarray*}
Only the latter of these satisfies the ultraviolet 
boundary condition nontrivially:
\begin{eqnarray*}
B(0)=\Lambda e^{-\frac{\pi}{\alpha/16\pi^{2}-1}}.
\end{eqnarray*}
Therefore real positive solutions 
\begin{eqnarray*}
B(y)=\textrm{Re}(B_{>\pm}(y)),
\end{eqnarray*}
are obtained above the dimensionless critical coupling defined by 
(c.f. Eq.(\ref{xC}))
\begin{eqnarray*}
\alpha_{c}= x^{2} =\frac{16 \pi^{2}}{3}.
\end{eqnarray*}

In the 4-fermi model therefore, one sees that in this
approximation the couplings $r$, $\ell$ enter the expressions for the scalar form factors $B_{\pm}$
on an equal footing and that the momentum integral
of the kernel in Eq.(\ref{mf2a}) vanishes.
This vanishing of the chiral dependency of form 
factors is a natural artifact of the choice of bare propagators and vertices. Thus while such an 
approximation is useful for study of mass generation
there is no opportunity for CP-violation. 

An improved vertex {\it ansatz}, leading to non-vanishing $A_{\pm}$
contributions could be expected to remedy this problem. However study of the gauged 4-fermion model in the next chapter is also seen to lead to non-vanishing contributions of the required type.

\chapter{Quenched hypercharge}\label{chap:fin}
In this chapter we wish to investigate the question of criticality in the hypercharge theory. 
In contrast to QED there are three couplings and the behaviour of the two models could therefore reasonably 
be expected to differ. 

In \ref{sec:rain} for the (non-anomalous) quenched, rainbow approximation we find an indication of different dynamics for the left- and right- fermion chiralities 
in the DSE for the self-energy.
Specifically the departure from the equalities $A_{+}=A_{-}$, $B_{+}=B_{-}$ in non-Landau gauges 
($G\neq 0$) leads to the appearance of two poles
in the fermion propagator, associated with
two types of scalar fermion pairing and the dynamical breaking
of parity.

In addition to being badly gauge-dependent, these 
poles are degenerate in the Landau gauge. 
In section \ref{sec:nunon} therefore, following the successful Ball-Chiu \cite{BallChiu1980} and Curtis-Pennington \cite{CurtisPennington1993} approaches for QED we
propose a (naively) multiplicatively-renormalisable hypercharge vertex. A bifurcation analysis of these equations
suggests the separation of the chiral poles, while a small effect, persists 
for all choices of gauge.

The second part of the chapter is concerned with 
incorporating composite scalars into the theory, 
preliminary to considering the full gauged 4-fermi model.
In \cite{BassThomas1996} it was suggested that at the lowest (right) critical 
scale the lightest generation ``froze out'' of the theory; fermions with right chirality formed a condensate $\bar{\psi}_{L}\psi_{R}$, 
consistent with the presence of two poles in the propagator.
The left-handed vacuum was highly excited relative to its right counterpart and via gauge transformations the ABJ anomaly caused a flux of (dynamical) left fermions over into the condensed phase. It is for this reason 
that both chiralities are expected to feel the same scalar potential. 
Moreover in the presence of global chiral symmetry 
breaking, inclusion of Goldstone bosons is necessary
for construction of a unitary, low-energy effective
theory. 
To this end we consider the contribution of the dynamically-generated composite Goldstones to the fermion self-energy and their mixing with the gauge boson via the
anomalous correction to the vertex WTI are thus investigated in \ref{sec:av}.

\section{D$\chi$SB in hypercharge $U(1)$} 
\label{sec: U1}
In the same way that approximate solutions to the QED SDEs led to 
early suggestions that non-perturbative QED was capable of dynamically
generating a fermion mass, we commence with the analogous equations
for the $U(1)$ hypercharge model. The relevant SDEs were derived in chapter
\ref{chap:pre}. With a quenched boson propagator the main  
qualitative difference to QED is that the fermion gauge-boson coupling is a
linear combination of vector and axial vector contributions; 
The bare (respectively right-and left-handed) fermion-boson vertices have 
the form 
\begin{equation}
\Gamma_{0}^{\mu}=\gamma^{\mu}(c_{+}\chi_{+}+c_{-}\chi_{-}).
\label{bare}
\end{equation}
We shall adopt the same form of the inverse fermion propagator in Euclidean space, 
Eq.(\ref{invprop}), as used in section \ref{sec:alter}
\begin{equation}
 S^{-1}(p)=\chi_{+}( i A_{+}(p)\gamma .p + B_{+}(p))^{-1}+\chi_{-}  (i A_{-}(p)\gamma .p+B_{-}(p))^{-1}.
 \end{equation}

\subsection{Rainbow approximation}
\label{sec:rain}
From Eq.(\ref{invprop}) we can write the fermionic self- energy as
\begin{eqnarray}
\Sigma(p) & = &\chi_{+}\{ i\gamma.p(A_{+}(p)-1)+ (B_{+}(p) - m_{+})\}\nonumber \\
& & \mbox{}+ \chi_{-}\{ i\gamma.p(A_{-}(p)-1)+ (B_{-}(p) - m_{-})\} \label{lor2} \nonumber \\
 & = & \int \frac{d^{4}q}{(2\pi)^{4}} D_{\mu\nu}(p-q) \Gamma^{\nu}(p,q) S(q)\Gamma^{\mu}_{0}\label{int2}.
\label{fsech}
\end{eqnarray}
Approximating the full vertex $\Gamma^{\nu}(p,q)$ by Eq. (\ref{bare}) 
and the gauge boson propagator $D_{\mu\nu}$ by its quenched form
\begin{equation}
D_{\mu\nu}(k)=(\delta_{\mu\nu}+(G-1)\frac{k_{\mu}k_{\nu}}{k^{2}})\frac{1}{k^{2}}, \label{unq2}
\end{equation}
one arrives at the analogue of the quenched rainbow approximation for the 
fermion self-energy. 
Tracing out the various projected components yields: 
\begin{eqnarray}
A_{\pm}(y) & =  & 1 +\frac{G c^{2}_{\mp}}{16\pi^{2}}\int^{\Lambda^{2}}_{0} dx
\frac{A_{\pm}(x)}{D_{\pm}(x)} \{\frac{x^{2}}{y^{2}}\theta_{+}+\theta_{-}\}, \label{impA} \\
B_{\pm}(y) & = & m_{\pm}  + \frac{(3+G)c_{+}c_{-}}{16\pi^{2}}\int^{\Lambda^{2}}_{0} dx
\frac{B_{\mp}(x)}{D_{\mp}(x)} \{\frac{x}{y}\theta_{+}+\theta_{-}\},  \label{impB} 
\end{eqnarray}
where $x=q^{2}$, $y=p^{2}$ and $m_{\pm}=m_{e} \pm m_{o}$

In the Landau gauge $A_{\pm} \to 1$
and mass generation is obtained from the
behaviour of the two coupled integral equations in $B_{\pm}$.
In the limit $c_{+}=c_{-}$ the system reduces to the integral
equations of the bare-vertex 4 fermi approximation  Eqs.(\ref{mf2a},\ref{mf2b}).
That is, the solutions are of the form $B_{+}(p)=B_{-}(p)$
and can be considered as equivalent to the QED scalar form factor.
Converting Eq.(\ref{impB}) to differential form and repeating the latter analysis one then obtains the critical coupling
\begin{equation}
\frac{c_{+}c_{-}}{16 \pi^{2}}=\frac{\pi}{3}. \label{lcu}
\end{equation}
That is, like quenched rainbow QED the hypercharge
theory appears to have two phases separated by a transition associated with breakdown of chiral symmetry.

There are two important distinctions however:
Eq.(\ref{lcu}) now represents a curve (hyperbola) in 2-dimensional coupling constant space rather than a point on a line. Secondly the degeneracy $A_{+}=A_{-}$ 
in Eq.(\ref{impA}) is lifted and consequently in Eq.(\ref{impB}), $B_{+}\neq B_{-}$. 
Indeed solving Eqs.(\ref{impA}, \ref{impB}) numerically by an iterative process 
confirms this. See, for example, Fig. \ref{nn}.
\begin{figure}[htb]
\centering{
\rotatebox{0}{\resizebox{9cm}{6cm}{\includegraphics{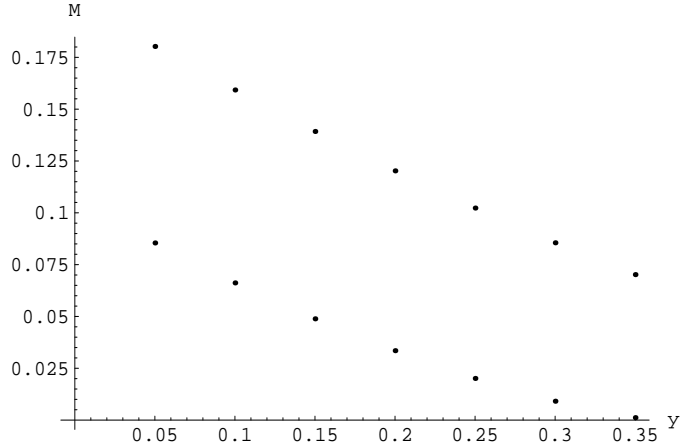}}}
}
 \caption{Mass functions $M_{+}$ (upper) and $M_{-}$ (lower) as a function
of momentum-squared in the Feynman gauge. Coupling values used are
$c_{+}=7$, $c_{-}=5$.}
 \protect \label{nn}
\end{figure}
The appearance of two poles $M_{\pm}$ in the fermion 
propagator Eq.(\ref{prop}) has a natural interpretation in terms of an extra mode 
of fermion pairing \cite{pisarski}. Moreover the pairings were found to
become marginal at separate critical scales. Figure \ref{run2} shows the coupling-dependence 
of the fixed point equations, 
\begin{eqnarray*}
M_{\pm}(p^{2}=-M_{\pm}^{2})=M_{\pm},
\end{eqnarray*}
analogous to the determination of the QED critical coupling 
\cite{CurtisPennington1993}. In this example, for fixed $c_{+}$, because 
$c_{-}>c_{+}$ the left-handed fermion acquires a gap first at 
$(c_{-},c_{+})\sim (4.44,4)$ while the right chirality remains massless until the point $(c_{-},c_{+})\sim (6.605,4)$. The behaviour of the chiralities is switched
upon changing the values $c_{+}\iff c_{-}$.

\begin{figure}[htb]
\centering{
\rotatebox{0}{\resizebox{9cm}{6cm}{\includegraphics{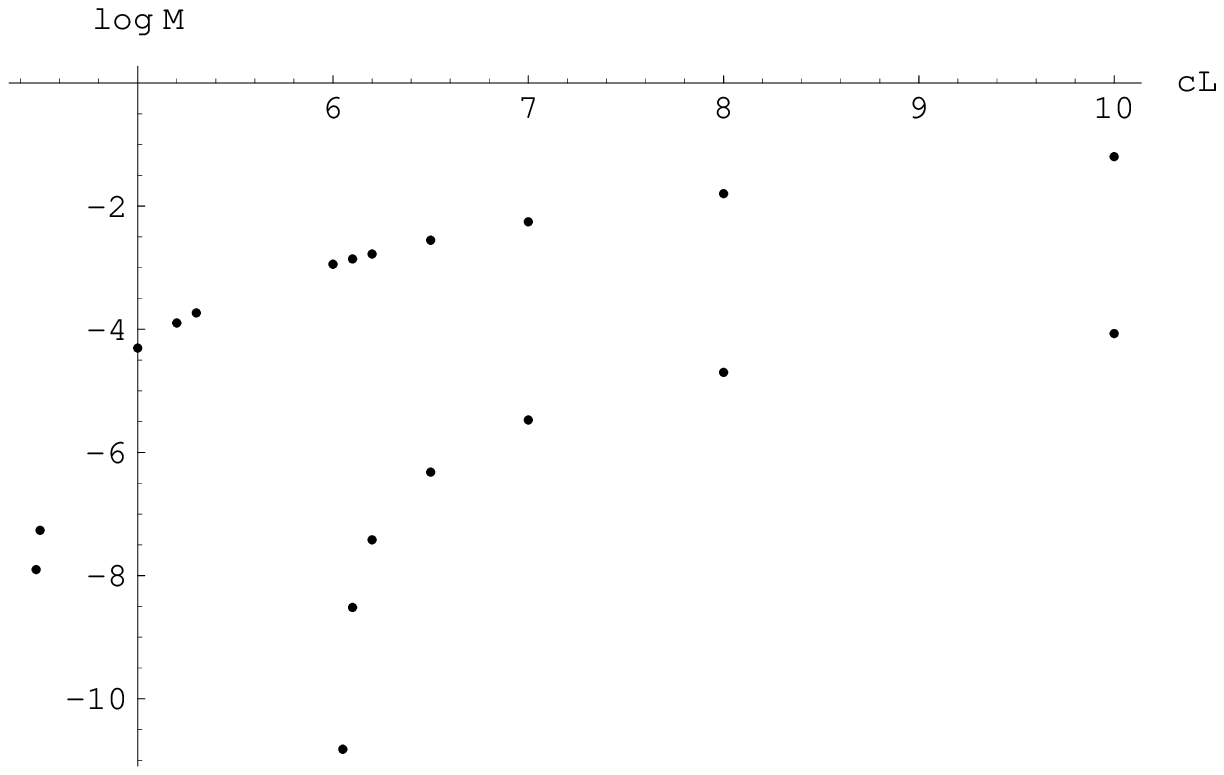}}}
}
 \caption{Fixed point of the mass functions $M_{-}$ (upper) and $M_{+}$ (lower) as 
a function of the coupling $c_{-}$ in the Feynman gauge. The right coupling
is fixed to the value $c_{+}=4$.}
 \protect \label{run2}
\end{figure}
This is precisely the behaviour anticipated \cite{BassThomas1996} at the 
right-right and left-left critical coupling points, where right and 
left-handed condensates were expected to form.

It is clear that a physical, i.e., Hermitean, Lorentz-invariant fermion
``mass'' term requires
\begin{equation}
M^{\dagger}_{+}=M_{-}, \label{stab}
\end{equation}
and moreover, that unless $c_{+}=c^{*}_{-}$, due to Eq. (\ref{impA}) 
this relation will not be satisfied. 
In \cite{BassThomas1996} the mass term becomes possible due to
the ABJ anomaly-driven collapse of the (dynamical) chiral sector
with largest mass. In the case shown in Figure \ref{run2}, 
above the left critical scale, there is a nett loss of gapped (left) fermions 
and gain of gapless right fermions, while the former remains as a resonant 
state. In this way both left and right fermion pairings are on an equal 
footing, allowing condensation, hence physical mass generation, to occur.   

Unlike QED, therefore, dynamical chiral symmetry breaking in hypercharge
is a necessary but not sufficient condition for dynamical mass generation. 
The extra condition, namely the (anomalous) effective coupling of
composite scalars with fermions will be considered in \ref{sec:gol}.

We conclude this subsection with the observation that
that, as is the case for QED, only the choice of Landau gauge in the 
quenched, rainbow approximation satisfies the vertex WTI, required for 
gauge-invariant renormalisability. Clearly any physically realistic 
truncation of the 3-point vertex Greens function DSE must also have this 
property. However as evidenced in Eqs.(\ref{impB},\ref{lcu})
the choice $G=0$ is precisely where the degeneracy of the chiral condensates 
occurs, leading to qualitatively QED-like behaviour. It is vital 
therefore to move beyond rainbow approximation. 
 
\subsection{Renormalisable vertex}
\label{sec:nunon}
We now attempt to reduce the gauge-dependence of the dynamical
mass by building a vertex respecting the appropriate WTI. 
Given the success of the method of Ball and Chiu \cite{BallChiu1980} and 
improvement by Curtis and 
Pennington \cite{CurtisPennington1993} for the dressed, multiplicatively-renormalisable QED vertex {\it ansatz}, it 
is logical to emulate this approach here. 

We begin by considering the analogue of the Ball-Chiu {\it ansatz} 
\cite{BallChiu1980},
i.e., a vertex form which satisfies the 3-point Ward-Takahashi identity
and reduces to the bare vertex in the perturbative limit.
The WTI follows from the general vector and axial current identities 
\begin{eqnarray*}
(q-p).\Gamma^{V}(q,p)& = & S^{-1}(q)-S^{-1}(p), \\
(q-p).\Gamma^{A}q,p)& = & S^{-1}(q)\gamma^{5}+\gamma^{5}S^{-1}(p)-
\bar{F}(q,p),
\end{eqnarray*}
by forming the relevant linear combinations:
\begin{eqnarray}
ik.\Gamma^{\pm}(q,p) &= &
c_{\pm}((S(q)\chi_{\pm}-\chi_{\mp}S(p)) \mp \bar{F}(q,p)) \nonumber \\
& = & \mbox{} c_{\pm}(S(q)-S(p))\chi_{\pm}\pm  c_{\pm}(\eta\chi_{\pm}
-\bar{F}(q,p)),
\label{wti}
\end{eqnarray}
where $k=q-p$. There are a number of immediate qualitative differences to 
QED. The first two terms of the right-hand side can be made to be 
singularity-free in the limit $k\to 0$, analogous to the QED identity 
Eq.(\ref{qedwti}). However, while the currents satisfying Eq.(\ref{wti}) are
gauge invariant, they do not represent a true functional variation.
As a result the two extra terms, the dynamical
Goldstone term (c.f..\cite{Matsuda})
\begin{equation}
\eta(p,q)= (B_{+}(q)-B_{-}(q)), \label{gb}
\end{equation}
and a hard component $\bar{F}(q,p)$, related to the ABJ anomaly
can give rise to kinematic singularities in the vertex.
Initially we shall proceed by naively neglecting these corrections.

The remainder of the vertex function $\hat{\Gamma}^{\mu}(q,p)$ itself may be 
expanded in the same manner as that of QED:
\begin{eqnarray*}
\hat{\Gamma}^{\mu}(q,p)= \Gamma^{\mu}_{0}(q,p)+\sum^{12}_{r=5}\zeta_{r}(q,p)T_{r}^{\mu},
\end{eqnarray*}
where $T_{r}^{\mu}$ are the Ball-Chiu \cite{BallChiu1980} basis vectors.
The longitudinal part of the vertex constrained by eq. Eq.(\ref{wti})
which is free of kinematic singularities is given by 
\begin{equation}
\Gamma_{L}^{\mu}(q,p) = \left(\zeta_{1}\gamma.(q+p)(q+p)^{\mu}
+ \zeta_{2}\gamma^{\mu} +i\zeta_{3}(q+p)^{\mu}\right)(c_{+}\chi_{+}+c_{-}\chi_{-}),\end{equation}
with
\begin{eqnarray}
\zeta_{1}(q,p)& = &\frac{1}{2}\frac{A_{-}(q)-A_{-}(p)}{q^{2}-p^{2}}\chi_{+}+
\frac{1}{2}\frac{A_{+}(q)-A_{+}(p)}{q^{2}-p^{2}}\chi_{-},\nonumber\\
\zeta_{2}(q,p) & =&\frac{1}{2}(A^{-}(q)+A^{-}(p))\chi_{+}+
\frac{1}{2}(A^{+}(q)+A^{+}(p))\chi_{-},\nonumber\\
\zeta_{3}(q,p)& =&\frac{B^{+}(q)-B^{+}(p)}{q^{2}-p^{2}}\chi_{+}+
\frac{B^{-}(q)-B^{-}(p)}{q^{2}-p^{2}}\chi_{-}. \label{z3}
\end{eqnarray}
Setting $\Gamma^{\mu}(q,p)=\Gamma^{\mu}_{L}(q,p)$ in 
Eq.(\ref{fse2})
gives the longitudinal contribution to the self energy $\Sigma_{L}(p)$.
Tracing out the various components yields 
\begin{eqnarray}
A^{L}_{\pm}(y) & = & 1-\frac{c^{2}_{\mp}}{16 \pi^2}\int_{0}^{\Lambda^{2}}dx
\{ (\frac{x^2}{y^2}\theta_{+}+\theta_{-}) \nonumber \\ 
& & \frac{3}{2(x-y)}
\left(\frac{B_{\mp}(x)}{D_{\mp}(x)}(B_{\mp}(x)-B_{\mp}(y))
+ \frac{x+y}{2}\frac{A_{\pm}}{D_{\pm}(x)}(x)(A_{\pm}(x)-A_{\pm}(y))\right) \nonumber\\
 & & \mbox{}-G\frac{x^2}{y^2}\theta_{+}\left(\frac{A_{\pm}(x)A_{\pm}(x)}{D_{\pm}(x)}
+ \frac{1}{x}\frac{B_\mp(x)}{D_{\mp}(x)}(B_{\mp}(x)-B_{\mp}(y)) \right. \nonumber \\
& & \left. \mbox{}- G\theta_{-} \frac{A_{\pm}(x)A_{\pm}(y)}{D_{\pm}(x)}\right)
 \}, \label{Alon}\\
B^{L}_{\pm}(y) & =& m_{\pm}+\frac{c_{+}c_{-}}{16\pi^2}\int_{0}^{\Lambda^{2}}\frac{dx}{D_{\mp(x)}}
\{(\frac{x^{2}}{y^{2}}\theta_{+}+\theta_{-})  \nonumber \\
& & \frac{3x}{2(x-y)}\left( A_{\mp}(x)B_{\mp}(y)-B_{\mp}(y)A_{\mp}(x)\right)  \nonumber \\
& &\mbox{}+\frac{3}{2}B_{\mp}(x)\left( A_{\mp}(x)+A_{\mp}(y)\right)(\frac{x}{y}\theta_{+}+\theta_{-}) \nonumber \\ 
& & \mbox{}+\frac{x}{y}\theta_{+} G B_{\mp}(x) A_{\mp}(y) \nonumber \\ 
& & \mbox{}+\theta_{-} G B_{\mp}(x)A_{\mp}(y) \},\label{Blon} 
\end{eqnarray}
where $x=q^{2}$ and $y=p^{2}$.

The form of the transverse part of the vertex, that is,
the term associated with the eight Ball-Chiu
basis vectors, is constrained by a number of considerations. It is possible to derive
a transverse Ward-Takahashi \cite{Kondo1996} identity
of the form $\partial_{\mu}\Gamma_{\nu}-\partial_{\nu}\Gamma_{\mu}$
or to enforce physical requirements such as that of multiplicative renormalisability (MR) \cite{CurtisPennington1993}.
Finding closed forms for the transverse WTI is difficult
for more than two dimensions, where it may be written exactly, while the MR requirement has been highly successful in quenched QED. We therefore choose to
consider restrictions of the latter kind.

For the (massless) hypercharge theory, the renormalisation
process involves rescaling the fields by
\begin{eqnarray*}
\psi^{R}_{\pm}=1/\sqrt{Z^{\pm}_{2}} \psi_{\pm},& &
Z^{R}_{\mu}=1/\sqrt{Z_{3}}Z_{\mu}, 
\end{eqnarray*}
and the couplings by 
\begin{eqnarray*}
c_{\pm}=Z^{\pm}_{2}\sqrt{Z_{3}}/Z^{\pm}_{1} c^{R}_{\pm}.
\end{eqnarray*}
where the superscript $R$ denotes renormalised quantities.
For the quenched approximation $Z_{3}=1$ and the 
(non-anomalous) WTI imposes $Z^{\pm}_{1}=Z_{2}^{\pm}$.
The renormalised Lagrangian then reads
\begin{eqnarray*}
{\cal L}&=&Z^{+}_{2}\bar{\psi}_{+}\gamma.(i\partial+
c_{+}Z)\psi_{+}+
Z^{-}_{2}\bar{\psi}_{-}\gamma.(i\partial+c_{-}Z)\psi_{-} \nonumber\\
& & -F^{R}.F^{R}-\frac{1}{2G}(\partial.Z^{R})^{2}, 
\end{eqnarray*}
from which the renormalised DSEs may be obtained, 
as in section \ref{sec:DSE} above.
The renormalised chiral vector form functions must satisfy 
\cite{CurtisPennington1993}
\begin{equation}
A^{R}_{\pm}(y; \mu_{\pm})=Z^{\pm}_{2}(\Lambda,\mu_{\pm})A_{\pm}(y; \Lambda),
\label{anr}
\end{equation}
where $\mu_{\pm}$ are the arbitrary renormalisation scales for the left- 
and right-sectors and for simplicity we assume they have the same UV cutoff.

Following \cite{CurtisPennington1993}, Eq.(\ref{anr}) must have the perturbative, 
leading logarithm expansion
\begin{eqnarray*}
A_{\pm}(y; \Lambda)=1+\sum^{\infty}_{i=1}\frac{1}{i!}
\left(\frac{c^{2}_{\mp}}{4 \pi} f \ln(\frac{y}{\Lambda^{2}})\right)^{i},
\end{eqnarray*}
where $f$ is the coefficient associated with the 1-loop correction.
The 1-loop vertex correction for $p>>q$ is readily obtained as
\begin{eqnarray*}
\Gamma^{\mu}_{1\pm}(q+p,p)&=&
\frac{Gc^{2}_{\pm}}{16\pi}(-\gamma^{\mu}-2 p^{\mu}\gamma.q /p^{2}
+\frac{(q+p)^{\mu}}{q^{2}-p^{2}}\gamma.(q+p)))\ln q^{2}/p^{2} \nonumber \\
&\simeq & -\frac{Gc^{2}_{\pm}}{16\pi^{2}} (\gamma^{\mu}(q^{2}-p^{2})-(q+p)^{\mu}\gamma.(q-p)).
\end{eqnarray*} 
Substituting this vertex form into Eq.(\ref{fse2}) yields the coefficient 
\begin{eqnarray*}
f=\frac{G}{8\pi},
\end{eqnarray*}
and thus the transverse vertex choice (c.f. \cite{CurtisPennington1993})
\begin{eqnarray*}
\Gamma^{\mu}_{T\pm}(q,p)=\frac{A_{\pm}(q)-A_{\pm}(p)}{2}T^{\mu}_{6}\frac{1}{p^{2}}, \label{tc6}
\end{eqnarray*}
guarantees renormalisability to 1 loop, where $T^{\mu}_{6}$ is the
Ball-Chiu basis vector
\begin{eqnarray*}
T^{\mu}_{6} = \gamma^{\mu}(q^{2}-p^{2})-(q+p)^{\mu}\gamma.(q-p).
\end{eqnarray*}
This expression is obtained under the assumption $p>>q$, thus removal of
the apparent singularity in Eq.(\ref{tc6}) can be justified by assuming
the denominator is valid only to $O(q^{2}/p^{2})$. 

A form for this modified factor may be obtained upon consideration of the
massive 1-loop fermion correction.
In this case there are extra renormalised parameters 
\begin{eqnarray*}
m^{\pm}_{R}=m_{\pm}/Z^{\pm}_{m}, 
\end{eqnarray*}
appearing in the Lagrangian. Of course for the case of physical fermions,
Eq.(\ref{stab}) must hold, in which case it is well known that the WTI 
Eq.(\ref{wti}) fails unless $Z^{+}_{2}=Z^{-}_{2}$ and $c_{+}=c_{-}$.
However, as pointed out above, it is the anomaly which catalyses
mass generation, in which case gauge-invariant renormalisation is spoiled
anyway.  We shall thus focus here on reducing the gauge-dependance
of the calculated quantities, with the understanding that the
``mass'' terms $M_{\pm}$ obtained need not correspond to physical fermion 
masses.

The term associated with the ``mass'', computed from
the vertex diagram is for $q^{2}>>p^{2}>>m^{2}_{\pm}$
readily verified to be
\begin{eqnarray}
\Gamma^{\mu}_{T2\pm}(q,p)= (3+G)\frac{c_{+}c_{-}}{4\pi}\frac{q^{\mu}}{q^{2}}m_{\mp}
\ln\frac{q^{2}}{p^{2}}, \label{nub}
\end{eqnarray}
while perturbatively, the fermion self-energy to first order is
\begin{eqnarray*}
\Sigma_{\pm}(p)=m_{\pm}-\frac{3c_{+}c_{-}}{16\pi^{2}}m_{\mp} \ln\frac{p^{2}}{\Lambda^{2}}.
\end{eqnarray*}
It is apparent that substitution of Eq.(\ref{nub}) in Eq.(\ref{fse2}) leads to the correct 1-loop self-energy result.
The sum of vertex terms Eqs.(\ref{z3}, \ref{tc6}) are thus 
seen to reproduce the (naively) renormalised 1-loop
self-energy in the leading log approximation.
While, due to decoupling of the chiral sectors, the vertex for massless
fermion corrections could be anticipated to be MR to all orders of the 
leading logarithm expansion as for QED \cite{CurtisPennington1993}
we have only verified it explicitly here at the one-loop level.
This is for simplicity and because the interesting, i.e., massive fermion) 
case and the ABJ anomaly (see section \ref{sec:anom}) in general spoil 
chiral-invariant renormalisability, therfore negating the motivation to 
proceed to higher orders analogous to \cite{CurtisPennington1993}.

All that remains to complete the vertex is to find an $O(q^{2}/p^{2})$ 
modification for the denominator of Eq.(\ref{tc6}) with 
the correct charge- conjugation property, i.e., symmetric
under $q \iff p$). To this end we propose a generalised version of the function used in \cite{CurtisPennington1993}
\begin{equation}
d_{\pm}(x,y)=\frac{(x-y)^{2}}{x+y}+\frac{((B_{\mp}(x)/A_{\mp}(x))^{2}+(B_{\mp}(y)/A_{\mp}(y))^{2})^{2}}{x+y}.
\end{equation}
The kernel of the 1-loop renormalisable transverse vertex 
contribution is thus given by the contribution
\begin{eqnarray*}
\Gamma_{0}^{\mu}S(q)\left(\frac{A_{+}(q)-A_{+}(p)}{2d_{+}(q,p)}c_{+}\chi_{+}+\frac{A_{-}(q)-A_{-}(p)}{2d_{-}(q,p)}c_{-}\chi_{-}\right)T^{\nu}_{6}D_{\mu\nu}(q-p).
\end{eqnarray*}
Computation of the angular integrals and tracing over the vector and scalar parts
yields
\begin{eqnarray}
A^{T}_{\pm}(y)& =&-\frac{3c^{2}_{\mp}}{16\pi^{2}}\int_{0}^{\Lambda}dx\frac{A_{\pm}(x)}{D_{\pm}(x)}(A_{\pm}(x)-A_{\pm}(y))
\frac{x^{2}-y^{2}}{d_{\pm}(y,x)}(\frac{x^{2}}{y^{2}}\theta_{-}+\theta_{+}),\nonumber\\
& & \label{at6} \\
B^{T}_{\pm}(y)& =&-\frac{3c_{+}c_{-}}{16\pi^{2}}\int_{0}^{\Lambda}dx
\frac{B_{\mp}(x)}{D_{\mp}(x)}(A_{\mp}(x)-A_{\mp}(y))
\frac{x-y}{d_{\mp}(y,x)}(\frac{x}{y}\theta_{-}+\theta_{+}). \nonumber \\
& & \label{bt6} 
\end{eqnarray}
The full, 1-loop DSE equations are thus given by 
the combination of Eqs.(\ref{Alon}, \ref{at6}) and Eq.(\ref{Blon}) with Eq.(\ref{bt6}):
\begin{eqnarray}
A_{\pm}(y)&= & A^{L}_{\pm}(y)+A^{T}_{\pm}(y), \label{acp}\\
B_{\pm}(y)&= & B^{L}_{\pm}(y)+B^{T}_{\pm}(y). \label{b2cp}
\end{eqnarray}
Following \cite{bif} we can undertake a bifurcation analysis of
Eqs.(\ref{acp}, \ref{b2cp}), the idea being that at the critical coupling their solution
bifurcates into nontrivial terms. This is achieved by taking the
functional derivative of the equations and enumerating the result at
the trivial value $B_{\pm}(x)=0$. It is clear that only the terms linear in 
$B_{\pm}$  will survive this procedure.
The behaviour of the pure gauge-interaction  part of the theory is 
then obtained by proceeding along similar lines to bifurcation analyses of quenched QED \cite{bif}, \cite{Kondobif}, \cite{Atkinson1994}.
Performing the functional derivative of Eqs.(\ref{acp}, \ref{b2cp}) and evaluating
them at the points $B_{\pm}(y)=0$, the simplified equations then read,
to $O(B^{2}_{\pm})$,
\begin{eqnarray}
A_{\pm}(y)&=&1-\frac{Gc^{2}_{\mp}}{16\pi^{2}}\int_{0}^{\Lambda^{2}}
dx \left(\frac{x}{y^{2}}\theta_{+}+
A_{\pm}(y)\frac{1}{A_{\pm}(x)x}\theta_{-} \right), \label{zba}\\
B_{\pm}(y)&=& \frac{c_{+}c_{-}}{16\pi^{2}} \int_{0}^{\Lambda^{2}}\frac{dx}{x} \left(
\frac{3B_{\mp}(x)}{2A_{\mp}(x)}(1+\frac{A_{\mp}(y)}{A_{\mp}(x)}+\frac{1+x}{1-x}(1-\frac{A_{\mp}(y)}{A_{\mp}(x)})(\frac{x}{y}\theta_{+}+\theta_{-}) \right. \nonumber \\
& & \mbox{} \left. -\frac{3y}{2(x-y)}
(\frac{A_{\mp}(y)}{A_{\mp}(x)}(\frac{B_{\mp}(x)}{A_{\mp}(x)}
-\frac{B_{\mp}(x)}{A_{\mp}(x)})
(\frac{x^{2}}{y^{2}}\theta_{+}+\theta_{-})\right. \nonumber \\
& & \left. +G\frac{B_{\mp}(x)}{A_{\mp}(x)} \frac{A_{\mp}(y)}{A_{\mp}(x)} \theta_{+} +G\frac{B_{\mp}(y)}{A_{\mp}(y)} \frac{A_{\mp}(y)}{A_{\mp}(x)} \theta_{-}\right). \label{zbb}
\end{eqnarray}
The first integral in Eq.(\ref{zba}) is readily computed leaving
\begin{eqnarray*}
A_{\pm}(y)= 1-\frac{Gc^{2}_{\mp}}{16\pi^2}(\frac{1}{2}+
A_{\pm}(y)\int^{\Lambda^{2}}_{0} dx \frac{1}{xA_{\pm}(x)}).
\end{eqnarray*}
These are of the same form as the corresponding equation for QED (see \cite{Atkinson1994}), and upon converting
to differential form
\begin{eqnarray*}
(1-\frac{Gc^{2}_{\mp}}{32\pi^2})\frac{d}{dy}\left(\frac{1}{A_{\pm}(y)}\right)
-\frac{1}{yA_{\pm}(y)}=0,
\end{eqnarray*}
are readily seen to have the unique solutions
\begin{equation}
A_{\pm}(y)= (1+\frac{c^{2}_{\mp}G}{32 \pi^{2}})\left( 
\frac{y}{\Lambda^{2}}\right)^{-\mu_{\pm}}, \label{aexp}
\end{equation}
with the exponent
\begin{eqnarray*}
\mu_{\pm}=2G/(\frac{32 \pi^{2}}{c^{2}_{\mp}}+G).
\end{eqnarray*}
Substituting Eq.(\ref{aexp}) now into Eq.(\ref{zbb}) one obtains the equation
(c.f. \cite{Atkinson1994})
\begin{eqnarray}
\frac{B_{\pm}(y)}{A_{\pm}(y)}&=&
\frac{c_{+}c_{-}}{\mu_{\pm}c^{2}_{\mp}}
\left(\frac{y}{\Lambda^{2}}\right)^{\mu_{\mp}}\int_{0}^{\Lambda^{2}}dx \left(
\frac{3B_{\mp}(x)}{2A_{\mp}(x)}
(1+(\frac{x}{y})^{\mu_{\mp}}) \right. \nonumber\\
& &\left. +\frac{y+x}{y-x}(1-(\frac{x}{y})^{\mu_{\mp}})
(\frac{x}{y}\theta_{+}+\theta_{-}) \right. \nonumber \\
& & \left. -\frac{3y}{2(x-y)}
(\frac{x}{y})^{\mu_{\mp}}(\frac{B_{\mp}(x)}{A_{\mp}(x)}-
\frac{B_{\mp}(y)}{A_{\mp}(y)})(\frac{x^{2}}{y^{2}}\theta_{+}+\theta_{-}) \right. \nonumber \\
& & \left.+(\frac{x}{y})^{\mu_{\mp}+1}\frac{B_{\mp}(x)}{A_{\mp}(x)}( \frac{x}{y} \theta_{+}+\theta_{-})\right). \label{sgh}
\end{eqnarray}
Upon taking the limit $\Lambda^{2} \to \infty$ we look for
scale-invariant solutions to the equations (\ref{sgh}), signalling
the presence of a UV fixed point. The obvious candidate is
\begin{eqnarray*}
M_{\pm}(x)\equiv \frac{B_{\pm}(x)}{A_{\pm}(x)}=x^{s_{\pm}},
\end{eqnarray*}
where, due to the coupling of chiral form factors in Eqs.(\ref{b2cp})
\begin{eqnarray*}
x^{s_{\pm}}=\kappa_{\pm}x^{s_{\mp}}=\kappa_{+}\kappa_{-}x^{s_{\pm}}.
\end{eqnarray*}
That is, the exponents $s_{\pm}$ must now be roots of (c.f. the QED case
Eq.(\ref{sconstr}))
\begin{eqnarray}
1-\kappa_{+}\kappa_{-}=0, \label{2sol}
\end{eqnarray}
where
\begin{eqnarray*}
\kappa_{\pm}&=&\frac{3c_{\pm}}{2c_{\mp}G}
\frac{\mu_{\mp}(\mu_{\mp}-s_{\mp}+1)}{1-s_{\mp}}(3 \pi \cot\pi (\mu_{\mp}-s_{\mp})- \pi \cot \pi\mu_{\mp}+2\pi \cot \pi s_{\mp}\nonumber \\
& &  \hspace{2cm}\mbox{} + \frac{1}{\mu_{\mp}}+\frac{2}{1-s_{\mp}}+\frac{3}{s_{\mp}-\mu_{\mp}}+\frac{1}{\mu_{\mp}+1} +\frac{1}{s_{\mp}-\mu_{\mp}-1} ) ,
\end{eqnarray*}
and convergence of the integrals in Eq.(\ref{sgh}) is conditional upon 
$0\leq s_{\pm}\leq 2$.

Numerical solution of Eq.(\ref{2sol}) reveals the number of roots in the
interval $0\leq s^{\pm}\leq 2$ to be gauge-dependent. In the
Landau gauge there are two such roots and therefore in a general gauge
we wish to consider only the pair of roots which is continuously connected
to this pair via the change of gauge parameter.
Criticality corresponds to a choice of coupling constants $c_{\pm}$ for
which the two roots become equal.
The results of such a computation in the Feynman gauge $(G=1)$ are shown in 
Table 5.1.
\begin{table}[htb]
\begin{center}
\begin{tabular}{cccccccc}
\hline\hline
$c_{\pm}$ & $c_{\mp}$ & $\nu_{\pm}$ & $\nu_{\mp}$ & $s_{\pm}$ & $s_{\mp}$ &$\gamma^{\pm}_{m}$ & $\gamma^{\mp}_{m}$\\
\hline 
2.000 & 5.802 & 0.025  & 0.192& 0.462 & 0.466 & 1.076& 1.068\\
3.000 & 3.867 & 0.055  & 0.090& 0.467 & 0.466 & 1.066& 1.068\\ 
3.407 & 3.407 & 0.071  & 0.071& 0.466 & 0.466 & 1.068& 1.068\\ 
4.000 & 2.901 & 0.096  & 0.052& 0.456 & 0.468 & 1.088& 1.065\\  
5.000 & 2.324 & 0.147  & 0.034& 0.466 & 0.462 & 1.068& 1.076  \\
\hline \hline
\end{tabular}
\caption{Critical coupling values and scaling exponents 
in Feynman gauge.}
\end{center}
\protect \label{ad}
\end{table}
The first observation to be made about these results is that the
critical curve is again a parabola in the two-dimensional coupling constant
space $(c_{-},c_{+})$ as shown in Fig. \ref{cur2}.
In the  vector limit the critical coupling is $c_{+}=c_{-}\simeq 3.4065$
corresponding to a value $\alpha_{C}\simeq 0.9234$, identical 
with the value obtained for Feynman gauge QED \cite{Atkinson1994}.
In fact from Table 5.1 the equation of the critical parabola is given by
\begin{equation}
c_{+}c_{-}\simeq 11.61 \simeq 0.9234 (4 \pi). \label{eeq}
\end{equation}
There is an approximate symmetry between couplings $c_{+}\iff c_{-}$
and the solutions $A_{+}\iff A_{-}$, $B_{+} \iff B_{-}$ which may also
be pertinent for models of mirror matter.
\begin{figure}[htb]
\centering{
\rotatebox{0}{\resizebox{10cm}{6cm}{\includegraphics{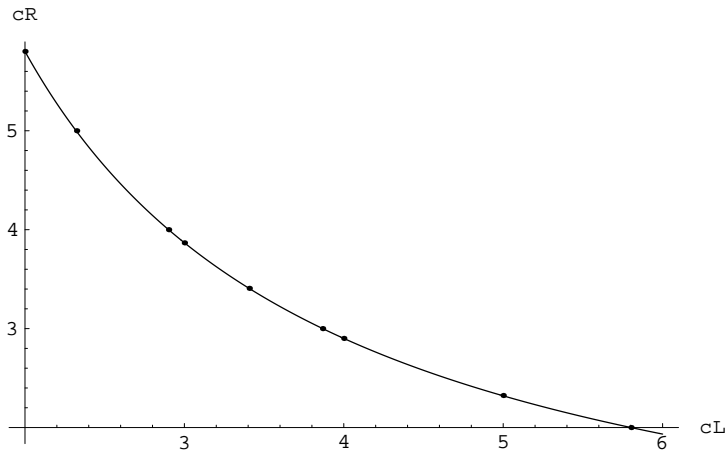}}}
}
 \caption{Critical curve Eq.(\ref{eeq}) (solid) and numerical points
obtained in Feynman gauge.}
 \protect \label{cur2}
\end{figure}
It is seen that at the critical point, both mass terms $M_{\pm}$ have
a large anomalous dimension 
\begin{equation}
\gamma_{m}^{\pm}=2(1-s^{\pm}),
\end{equation}
and become marginal 
operators.
Over the coupling range considered, the mismatch between the anomalous 
dimensions of these operators is typically less than $3\%$. Therefore
in the neighbourhood of the fixed point, the gaps due to each chiral 
condensate are approximately (i.e to $O(M^{2}_{\pm})$) degenerate.

It is also found for a large range of values for $z$, the scaling
exponents and critical couplings are quite robust. See, for example,
the values obtained in Table 5.2. In this study, motivated by 
\cite{BallChiu1980}, \cite{CurtisPennington1993},
we have proposed a vertex with longitudinal part satisfying the non-anomalous 
WTI and the transverse part constrained
to be (naively, the Goldstone-fermion coupling has been ignored) 
renormalisable to 1-loop. 
In addition to relatively gauge-insensitive results it was found that
{\it both} left and right fermion pairings acquired large anomalous dimensions
in the neighborhood of the phase transition with approximately equal
exponents.
\begin{table}[htb]
\begin{center}
\begin{tabular}{ccccc}
\hline\hline
$G$ & $c_{\pm}$ & $c_{\mp}$ & $s_{\pm}$ & $s_{\mp}$ \\
\hline 
1 & 3.407 &  3.407 & 0.466 & 0.466 \\ 
2 & 3.402 &  3.402 & 0.462 & 0.452 \\ 
5 & 3.445 &  3.445 & 0.455 & 0.444 \\ 
10 & 3.589 & 3.589 & 0.429 & 0.427  \\
\hline\hline
\end{tabular}
\caption{Gauge dependence of critical couplings and scaling exponents.}
\end{center}
\protect\label{gnv}
\end{table}

\section{Role of composite scalars} \label{sec:gol}
We conclude this section with another correction which needs to be noted.
The quenched, rainbow approximation is meant to use the perturbative
boson propagator and vertex function, however as Gribov \cite{Gribovgol} and 
others have pointed out, in order to construct a low-energy unitary perturbation theory 
in the presence of dynamically-broken, global chiral symmetry, the gauge 
boson-Goldstone mixing needs to be included. For the quenched rainbow hypercharge approximation this amounts 
to including diagrams where internal gauge boson lines are replaced by Goldstone propagators, with the bare vertex 
\begin{eqnarray*}
\Gamma^{\mu}(q,p)_{\eta}=(c_{+}\chi_{+}-c_{-}\chi_{-})\eta(q,p)\frac{(q-p)^{\mu}}{(q-p)^{2}}. 
\end{eqnarray*}
The effective ``Goldstone''-fermion coupling, required
to balance the left- and right-hand sides of Eq.(\ref{wti}), 
is given by Eq.(\ref{gb}) and the inverted commas refer to the fact 
that the bound state is a chiral object.
The fermion self-energy Eq.(\ref{int2}) has now two contributions, the first 
given by Eqs(\ref{acp}, \ref{b2cp}) above, while the second corresponds to the
``Goldstone'' loop. The kernel of Eq.(\ref{fse2}) contains the extra term
\begin{eqnarray*}
\frac{\delta_{\mu\nu}}{k^{2}}(c_{+}\chi_{+}-c_{-}\chi_{-})\eta(p,p+k) \frac{k^{\mu}}{k^{2}} S(k+p)
(c_{+}\chi_{+}-c_{-}\chi_{-})\eta(k+p,p) \frac{k^{\nu}}{k^{2}}\gamma^{5},
\end{eqnarray*}
which gives the form factors
\begin{eqnarray}
A^{'}_{\pm}(y)& = & \frac{c_{+}c_{-}}{16\pi^{2}}\int^{\Lambda^{2}}_{0} 
\frac{dx}{D_{\mp}(x)}\left(
A_{\mp}\eta(x,y)\eta(y,x)\frac{1}{x-y}(\frac{x^{2}}{y^{2}}\theta_{+}-\theta_{-})\right), \nonumber \\
 & & \\
B^{'}_{\pm}(y)& = & \frac{c^{2}_{\mp}}{16\pi^{2}}\int^{\Lambda^{2}}_{0} 
\frac{dx}{D_{\pm}(x)}\left(
B_{\pm}(x)\eta(x,y)\eta(y,x)\frac{1}{x-y}(\frac{x}{y}\theta_{+}-\theta_{-})\right). \nonumber \\
\end{eqnarray}
That is, we may write the full expressions as
\begin{eqnarray}
A_{\pm}(p) & = & A_{\pm}(p,c^{2}_{\mp})+A^{'}_{\pm}(p,c_{+}c_{-}),
\label{lga} \\
B_{\pm}(p) & = & B_{\pm}(p,c_{+}c_{-})+B^{'}_{\pm}(p,c^{2}_{\pm}). \label{lgb2}
\end{eqnarray}
At this stage inspection of Eq.(\ref{lgb2}) reveals the possibility for generation of a fermion mass satisfying
Eq.(\ref{stab}). While the fermion self-energy for $B_{\pm}$
due to gauge terms, e.g. Eq.(\ref{b2cp}), contained only 
terms with $B_{\mp}$ in the kernel, the Goldstone contributions are seen to be $B_{\pm}$. 
The equations for $B_{\pm}$ no longer depend solely on
$B_{\mp}$, condensation of one chirality, e.g., right if the SM couplings are used, as in \cite{BassThomas1996})
is now partially offset by formation of a condensate of the other at the same scale, similar to the anomaly-catalysed collapse of the left-fermion sector
\cite{BassThomas1996}. 

An opportunity for obtaining three physical masses at three separate scales now exists:
At small momentum scales with Eq.(\ref{lgb2}) in differential form
there would be no nontrivial real solutions compatible with the UV boundary 
condition, analogous to quenched QED. When the critical right-right coupling
is reached, the ``Goldstone'' term $B^{'}_{+}$ admits a nonzero solution. 
Due to the coupling of the $B_{\pm}$ equations both form factors acquire a non trivial 
real solution, hence a conventional ``mass'', that is, one
involving both left and right-handed condensates
could be generated.
The remaining two terms, ($B_{\pm}^{L}$ and $B_{-}^{'}$) would
be ``switched on'' at the right-left and left-left critical points 
respectively, leading to three distinct mass terms.
The two key steps, the appearance of three separate contributions to
the equations $B_{\pm}$ and the mixing of these equations, required
to generate mass terms are made possible 
here by the
Goldstone-gauge boson mixing. With pure gauge-boson contributions
the equations for $B_{\pm}$ only contain $B_{\mp}$ terms in the kernel and always couple $\sim c_{+}c_{-}$. 

\subsection{Anomalous vertex} \label{sec:av}
The last piece of the vertex WTI to be considered is the ABJ anomaly,
the motivation being to see whether fermion mass generation 
can be dynamically generated in conjunction with the chiral symmetry breaking
described previously in \ref{sec:nunon}. 
In the proposal of
\cite{BassThomas1996} the mismatch between the chiral sectors
(i.e right-condensed vs left-dynamical fermions) was suggested to be
removed via the ABJ anomaly. In the model under consideration the
discrepancy is manifested as the breakdown of Eq. (\ref{stab}), 
specifically $B_{+}(y)\neq B_{-}(y)$.
The anomalous term $\bar{F}$ omitted from the identity 
(Eq.\ref{wti})
is given by \cite{IZ}
\begin{eqnarray}
\bar{F}(q,p)&=&(c_{+}-c_{-})\frac{i(c^{2}_{+}-c_{-}^{2})}{4 \pi^{2}}\int d^{4}xd^{4}y e^{i(q.y -p.x)}
<\psi(x)\bar{\psi}(y)\epsilon_{\mu\nu\rho\sigma}F^{\mu\nu}F^{\rho\sigma}(0)>. \nonumber \\
& & \label{ann1}
\end{eqnarray}
We shall approximate this here by obtaining an effective anomalous
vertex from integrating over the pseudoscalar-2 boson anomalous 
diagram shown in Fig. \ref{avert}.
\begin{figure}[htb]
\centering{
\rotatebox{270}{\resizebox{4cm}{8cm}{\includegraphics{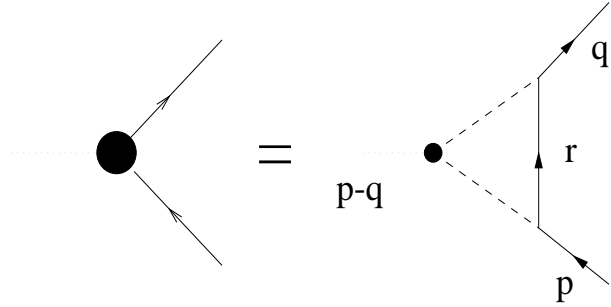}}}
}
 \caption{1-loop anomalous vertex.}
 \protect \label{avert}
\end{figure}
Using a momentum-cutoff regulation and with soft external momenta
a possible form for the 3-boson contact term is see \cite{sonoda}
in Eq.(\ref{ann1}) is
\begin{equation}
<-(k+l)Z_{\alpha}(k)Z_{\beta}(l)\frac{g}{m}\phi>\simeq 
\frac{i}{16\pi^{2}} \frac{g}{m}\epsilon_{\alpha\beta\mu\nu}k^{\mu}l^{\nu}, \label{son}
\end{equation}
where $\phi$ is a massless composite pseudoscalar, $m$ the fermion mass
and $g$ the Yukawa coupling.
Inserting this expression into the loop diagram Fig. \ref{avert} one finds 
the effective vertex
\begin{equation}
\Gamma^{F}_{\pm}(q,p)=\mp ic_{\pm}(c^{2}_{+}-c^{2}_{-})\frac{c_{+}c_{-}}{64 \pi^{3}}\epsilon_{\alpha\beta\mu\nu}p^{\mu}q^{\nu}\gamma^{\alpha}\gamma^{\beta}H(p,q;m)\chi_{\pm}, \label{aneff}
\end{equation}
where $H$ is a finite scalar function to be constrained by the physical 
properties of the anomaly: the anomalous vertex function must have the 
opposite CP properties to the gauge-boson vertex
\begin{eqnarray*}
\Gamma^{F}_{\pm}(p,q)= \Gamma^{F}_{\pm}(q,p), 
\end{eqnarray*}
i.e., $H$ must be symmetric in $p$ and $q$ and it must not vanish
for massless fermions. With the term
\begin{equation}
\frac{\delta^{\mu\nu}}{k^{2}}
\Gamma^{F}(p,k+p)\frac{k^{\mu}}{k^{2}}S(q)\Gamma^{F}(k+p,p)\frac{k^{\nu}}{k^{2}},
\end{equation}
added to the kernel of Eq.(\ref{fse2}), computation of these contributions
gives
\begin{eqnarray}
A^{F}_{\pm}(y) & = & -\frac{\alpha^{\pm}_{F}}{8\pi}\int_{0}^{\Lambda^{2}}\frac{dx}{D(x)}A_{\mp}(x)H(x,y)H(y,x)
\frac{xy}{x-y}(-\theta_{-}+\frac{x^{2}}{y^{2}}\theta_{+}), \nonumber \\
& & \label{Aanom}\\
B^{F}_{\pm}(y) & = & 
\frac{\alpha^{\pm}_{F}}{8\pi}\int_{0}^{\Lambda^{2}}\frac{dx}{D(x)}
B_{\mp}(x)H(x,y)H(y,x)\frac{xy}{x-y}
\nonumber (-\theta_{-}+\frac{x}{y}\theta_{+}), \nonumber \\
& & \label{Banom}
\end{eqnarray}
where we have defined the anomalous coupling
\begin{eqnarray*}
\alpha^{\pm}_{F}=\frac{1}{4\pi}(c^{\pm}\left(c^{2}_{+}-c^{2}_{-})\frac{c_{+}c_{-}}{64 \pi^{3}}\right)^{2} .
\end{eqnarray*}
Note that for the naive estimate $H$ these contributions would diverge
quadratically. In order to obtain the correct power counting behaviour it is 
clear that $H$ must have mass dimension $-1$. In fact in deriving the
vertex function (\ref{aneff}) from Eq.(\ref{son}), such a factor, $g/m$ has been
effectively absorbed into $H$.
A convenient form consistent with the above requirements and the 
momentum-squared rescaling $x \to \Lambda^{2} x$ is then just
\begin{equation}
H(p,q)= \frac{1}{\Lambda}\left(\frac{A_{+}(p)+A_{+}(q)}{2}\chi_{+}+
\frac{A_{-}(p)+A_{-}(q)}{2}\chi_{-}\right). \label{aitch}
\end{equation}
The form factor equations to be solved are now given by 
Eqs.(\ref{lga},\ref{lgb2}, \ref{Aanom}, \ref{Banom})
\begin{eqnarray}
A_{\pm}(y)&=&A^{L}_{\pm}(y)+A^{G}_{\pm}(y)+A^{F}_{\pm}(y), \label{aful}\\
B_{\pm}(y)&=&B^{L}_{\pm}(y)+B^{G}_{\pm}(y)+B^{F}_{\pm}(y). \label{bful}
\end{eqnarray}

For simplicity we numerically investigate Goldstone and anomaly
modifications to the rainbow approximation Eqs.(\ref{impA},\ref{impB}).
Upon performing the momentum scaling $x\to \Lambda^{2} x$, these equations
are solved for $A_{\pm}$ and $B_{\pm}(y)/\Lambda$ as before. The system of equations
(\ref{aful}, \ref{bful}) is highly sensitive to the value of $\alpha^{\pm}_{X}$: the 
anomalous terms are either negligible, in which
case the above-described \ref{sec:rain} behaviour occurs, or rapidly
come to dominate the iterative solution process, in which case $B_{\pm}(y)\to 0$.
Thus as expected, the anomaly removes the discrepancy between the chiral sectors, viz the restoration
of Eq.(\ref{stab}), however unfortunately not in the manner envisaged in \cite{BassThomas1996}: the dynamical
mass vanishes altogether!

Naturally a number of objections may be made to the above approximation. 
Firstly Eq.(\ref{son}) is only strictly valid for ``soft'' photon legs. Once included in the loop shown 
in Fig. \ref{avert}, large-$q$ corrections in the fermion self energy would rapidly become dominant.
Secondly if gauge-boson-Goldstone mixing is to be incorporated, it is necessary to move beyond the
quenched approximation, in which case $Z_{3}\neq 1$
and there are several diagrams (perturbatively, at least) more significant than the anomalous correction to the fermion self-energy. 

Indeed, if an anomaly-matching condition, such as the sum over all fermions within a SM generation, is imposed
the above discussion is rendered academic and the desirable features of the non-anomalous approximations
may be retained.
Nonetheless, it represents the first, to our knowledge, attempt to quantitatively understand a possible
role for the ABJ anomaly in the D$\chi$SB context. Such considerations are clearly necessary to evaluate 
whether the anomaly-catalysed vacuum decay proposed in \cite{BassThomas1996} can form the foundation
for self-consistent introduction of three fermion generations.

Finally we note that the Goldstone terms are at least quadratic in the
form factors $B_{\pm}$, therefore they will not contribute to the linearised
bifurcation analysis in section \ref{sec:nunon}. Moreover the chosen form 
of the anomalous vertex, being independent of $B_{\pm}$ also will not
affect the lowest order result. 

\chapter{Conclusion and outlook} \label{chap:concl}
In this thesis we have introduced two models in order to investigate
a proposal \cite{BassThomas1996} that dynamical chiral symmetry breaking in 
the hypercharge sector of the SM can give rise not only to fermion masses
but also the three-generation structure and attendant CP violation.

\section{Chiral-breaking model}
The first theory, a toy fermion-pairing model, was demonstrated in chapter 
\ref{chap:4fer} to exhibit the qualitative features of the SM which remain poorly 
understood and therefore are desirable for the proposal \cite{BassThomas1996}. 
The key feature was the observation that when both chiral-symmetric and -breaking
scalar 4-fermi operators are permitted
\begin{eqnarray*}
2\tilde{x}\bar{\psi}_{L}\psi_{R}\bar{\psi}_{R}\psi_{L}+\tilde{r}(\bar{\psi}_{L}\psi_{R})^{2}-
\tilde{r}^{*}(\bar{\psi}_{R}\psi_{L})^{2},
\end{eqnarray*}
the form of the mean-field approximation differs depending on whether one of, or 
both, such types of term are present.
In a 1-loop renormalisation group study of the running couplings this 
meant that there were three distinct renormalised flows depending upon whether
$\tilde{x}$ or $\tilde{r}$ were non-vanishing. The low-energy effective 4-fermi operators were 
thus found, in terms of the dimensionless couplings $x=\Lambda_{\chi}\tilde{x}$, $r=\Lambda_{\chi}\tilde{r}$ to be
\begin{eqnarray*}
O_{1} & = & (\bar{\psi}_{L}\psi_{R})^{2}+(\bar{\psi}_{R}\psi_{L})^{2};
\hspace{0.1cm}x=0, r-r^{*}=0,\\
O_{2} & = & \bar{\psi}_{L}\psi_{R}\bar{\psi}_{R}\psi_{L};
\hspace{0.1cm}r=0=r^{*}, x\neq 0,\\
O_{3} & = & (\bar{\psi}_{L}\psi_{R}+\bar{\psi}_{R}\psi_{L})^{2};
\hspace{0.1cm}x+r+r^{*}=0 .
\end{eqnarray*}
In each of these three cases a two-phase structure was suggested via solution
of the analogues of the NJL gap equations, with the critical coupling 
strengths given by
\begin{eqnarray*}
r& > & 4\pi, \\
x& > & 4\pi,\\
\ell &\equiv & \frac{x}{1+\frac{rr^{*}}{x^{2}}}\frac{1}{1-\textrm{Re}(r)(1+rr^{*}/x^{2})}> 4\pi,
\end{eqnarray*}
for the operators $O_{1-3}$ respectively. In particular it was found that,
while $O_{1,2}$ had trivial IR limits, the couplings $x$, $r$ flowed to
finite IR values for $O_{3}$, making it a suitable ``high energy'' effective
theory. In all cases  a fixed point in the ultraviolet region was also found.

The next step was to introduce a ``fundamental'' fermion with three components
which, while identical with respect to gauge interactions, participated
in either symmetric, breaking or both types of fermion pairing, thereby
living in distinct subspaces corresponding to the 4-fermi operators $O_{1-3}$.
As each of the couplings $x,r,\ell$ attains the critical strength at separate 
scales the theory has three distinct phase transitions, each associated with
dynamical mass generation for a component of the fundamental fermion.

This model was found to be similar to the 3-Higgs model of Kiselev\cite{Kiselev}
where the fermion families arise as a consequence of a $Z_{3}$-symmetric vacuum.
The latter \cite{Kiselev} has been found to give good tree-level estimates
of the CKM matrix elements. 

The main qualitative difference is that here the auxilary fields, which play the 
roles of the Higgs VEVs, are in general complex, chiral objects. 
Therefore in order to be left with the single CP-violating internal rotational 
uncertainty of \cite{Kiselev} one requires the existence of a discrete
$Z_{3} \otimes Z_{2}$ symmetry which may, as pointed out in \ref{sec:ol} be
interesting in terms of left-right-symmetric matter-only models. Otherwise
the model contains three extra (2 complex, 1 chiral) CP-violating phases.

While the proposed model was shown to have the required qualitative features
several important issues remain to be resolved. 
Firstly, while the presence of chiral condensates is anticipated in the
proposal \cite{BassThomas1996} it is not immediately clear how such
terms can be derived by integrating out the gauge-bosons from the hypercharge
Lagrangian:
\begin{equation}
{\cal L}_{eff}\sim (c_{L}\psi_{L}\gamma^{\mu}\psi_{L}+c_{R}\psi_{R}\gamma^{\mu}\psi_{R})
\Delta_{\mu\nu}(c_{L}\psi_{L}\gamma^{\nu}\psi_{L}+c_{R}\psi_{R}\gamma^{\nu}\psi_{R}). \label{leff}
\end{equation}
One possibility is to include the vacuum polarisation in the boson propagator,
$\Delta_{\mu\nu}$, where mismatches between the right-left and left-right
cross terms prevent the cancellation of parity-odd scalar terms in the subsequent
Fierz re-ordering.

Secondly the question of vector 4-fermi interactions has been ignored.
It has been demonstrated that for quenched QED \cite{Aoki1}, such terms remain 
irrelevant, even close to the fixed point. Here this is correlated with
the fact that the 1-loop RGE equation for the coupling $g_{V}$ does not contain
a quadratic term $\sim g^{2}_{V}$. Similarly the corresponding terms are seen to
be missing from the hypercharge equations for $r^{2}$, $\ell^{2}$ thus one
can tentatively anticipate similar behaviour.
While chiral symmetry guarantees the QED Wilson potential has next-to-leading 
order terms of degree eight in the fermion fields, in this model 6-fermi terms cannot 
be discounted. It is important to investigate the possible effect of these 
higher-order terms. 

Finally while the explicit-breaking terms will break chiral symmetry at
any scale the mean-field approximation suggests they only impart a 
fermion mass above a critical coupling strength. This result also
requires an explanation, which may be outlined as follows. From high-temperature 
superconductivity the mean-field approximation is known to be deficient. In 
particular there exists an intermediate phase where violent phase fluctuations 
mean that while fermion pairing occurs, condensate formation is inhibited.
Recently such a ``pseudogap'' has been demonstrated in the Gross-Neveu
model \cite{Babaev} and is conjectured to exist in other relativistic fermionic
theories.

\section{Quenched hypercharge}
The second model considered in this thesis is the hypercharge gauge interaction
itself. Analysis of dynamical chiral symmetry breaking in quenched QED
has an extensive literature. In particular we proceed by generalising the 
Schwinger-Dyson equation for the fermion self-energy.

In the simplest approximation with the quenched boson propagator and
bare vertex it is found that for an arbitrary covariant gauge separate
``gaps'' for the left- and right-chirality fermions appeared, evidence for the 
chiral pairings considered in the 4-fermi model of chapter \ref{chap:4fer}.
Moreover they were found to arise at separate scales, as shown in Figure 
\ref{run2}, suggesting a multiple-phase structure.

The exception to this behaviour was the Landau gauge, whereupon the gaps
and critical scales became degenerate. We therefore attempted to generalise the 
successful procedure \cite{BallChiu1980}, \cite{CurtisPennington1993} developed 
for construction of a QED vertex with greatly-reduced sensitivity to gauge choice.
 
The breakdown of the chiral Ward-Takahashi identity Eq.(\ref{wti}) due to the ABJ 
anomaly and dynamically-generated Golstone bosons meant that the procedure was not 
completely successful. 
In particular we constructed a vertex with longitudinal part satisfying
the (non-anomalous) WTI Eq.(\ref{wti}) and transverse term ensuring multiplicative
renormalisability of the fermion SDE Eq.(\ref{fse2}) to one loop. 
Diagrams due to the Goldstones and their anomalous mixing with the gauge-boson
vertex were of at least quadratic order in the scalar form factors $B_{\pm}$ 
in Eqs.(\ref{aful}, \ref{bful}) and therefore the proposed vertex was found to be sufficient for 
a linearised  bifurcation analysis, similar to that undertaken for QED 
\cite{Atkinson1994}. 
Here it was found that the seperation of chiral condensates persisted in all 
gauges, however with a difference in scaling exponents of $\sim 3\%$ the
effect was greatly reduced. In particular in the neighbourhood of a
critical coupling {\it both} mass terms acquired large anomalous 
dimensions, suggesting the splitting between the left- and right- gaps
is significantly less than that obtained in rainbow approximation.
These results were shown to be robust for a large range of covariant gauge 
parameter values. 

Finally the effect of the composite scalars was included in the quenched rainbow 
approximation to investigate the effect upon the dynamical chiral symetry 
breaking: the Goldstones are necessary to construct a unitary low-energy effective
perturbation theory, while it was hypothesised that the anomaly enables
the generation of physical fermion mass, by coupling left- (right-) fermions
with the right- (left-) condensates. 
Inclusion of the bare Goldstone propagator and effective bare scalar-fermion
couplings led to the appearance of mass terms depending upon all three
couplings Eqs.(\ref{lga}, \ref{lgb2}). Unfortunately the {\it ansatz} for the anomalous
vertex term was found to dominate the fermion SDEs (\ref{aful},\ref{bful})
causing vanishing of all mass functions. There are numerous reasons
for rejecting the {\it ansatz} Eq.(\ref{aitch}), discussed in section \ref{sec:anom}.
In particular a correct implementation would involve unquenching the boson
propagator. 

In summary we have shown in sections \ref{sec:rain}, \ref{sec:nunon} that
a fermion propagating in a background of hypercharge gauge fields 
appears to couple differently to chiral scalar condensates, which
become critical at different scales. These are necessary prerequisites
for the  dynamical generations hypothesis \cite{BassThomas1996}, however
its viability depends upon whether an improved treatment of the anomaly
can translate into physical fermion mass, with subsequent family
structure and mixing behavour.

\section{Outlook}
\label{sec:ol}
Until this stage, an appealing feature of the model is that no higher
physics is required to qualitatively produce the SM features outlined
in \cite{BassThomas1996}. Once this requirement is lifted, one can 
speculate that the non-perturbative effects embodied in this model 
facilitate a range of new theories with the Standard Model as a low energy 
limit. The fields of condensed matter and quantum field theory have a long history
of cross-pollination. Recently Volovik \cite{vvkbk} has written extensively on 
similarities between the electroweak sector of the SM and the superfluid $^{3}$HeB and it
seems logical to seek QFT analogues of recent developments in condensed matter. 
Phenomena such as chiral surface states in unconventional superconductors 
Sr$_{2}$RuO$_{4}$ \cite{Luke},\cite{Morinari} and 
Bi$_{1-x}$Ca$_{x}$MnO$_{3}$ \cite{Yoon}, 
chiral superfluidity in $^{3}$HeA \cite{vvkbk}, \cite{Goryo1} or spin-charge 
separation (SCS) \cite{Andrei},\cite{voit} may well have relevance to the
non-perturbative sector of the Standard Model.

We begin by noting the 4-fermi terms obtained upon quenching the inverse 
hypercharge boson propagator $\Delta_{\mu\nu}\sim g_{\mu\nu}$ in Eq.(\ref{leff}) resemble a higher-dimensional analogue of the Luttinger liquid Lagrangian \cite{voit},\cite{Haldane}, 
\begin{equation}
{\cal L}_{Lut}=\bar{\psi} \partial^{\mu}\gamma_{\mu} \psi + g_{2}\bar{\psi_{L}}\lambda^{\alpha}\gamma^{\mu}\psi_{L}
\bar{\psi}_{R}\lambda^{\alpha}\gamma_{\mu}\psi_{R}
+g_{4}((\bar{\psi}_{L}\lambda^{\alpha}\gamma^{\mu}\psi_{L})^{2}+
(\bar{\psi}_{R}\lambda^{\alpha}\gamma^{\mu}\psi_{R})^{2}), \label{lut1}
\end{equation}
if one identifies the terms $\bar{\psi}_{L,R}\lambda^{a}\gamma^{\mu}\psi_{L,R}$ with 
left-and right-moving charge density operators, which in the Luttinger
liquid are of the form $\psi_{L,R}\psi_{L,R}$, i.e., fermion-fermion condensates.

Model Eq.(\ref{lut1}) has, in $1+1$ dimensions a number of remarkable features:
the cross-term $g_{2}$ modifies the pole structure of the fermion propagator
(indeed the Fierz-ordering of this term in 4 dimensions leads to the scalar 
term breaking chiral symmetry).
The $g_{4}$ term lifts any residual degeneracies, similar to a hopping 
matrix element between spin chains \cite{voit}, and leads in 1+1 dimensions 
to SCS. Here, the effect is signalled by the appearance of two poles in 
the fermion propagator. An attempt to inject a free fermion into the second 
unoccupied energy level above the Fermi surface causes a hole excitation. 
The resulting  hole-electron pair (in the lowest free energy level) decomposes into spin and charge 
fluctuations which propagate through the medium with different velocities.

This, too reminds of the problem encountered in the SDE analysis of the
previous chapter with the appearance of new poles related to each condensate. 
We then see a possible interpretation of this feature is for some kind
of recombination of fermionic degrees of freedom.
A qualitative argument, based on the Dirac sea picture of the anomaly
\cite{Jackiw} runs as follows.  
The two-dimensional (for simplicity) hypercharge theory has Lagrangian:
\begin{equation}
{\cal L}_{2D}=\bar{\psi}\gamma^{\mu}(i \partial_{\mu}-(c_{L}\chi_{L}+c_{R}\chi_{R})Z_{\mu}\psi. \label{2d}
\end{equation}
In the Dirac sea picture, second quantisation corresponds to filling all
negative energy eigenmodes while leaving positive ones empty.
Setting $Z_{0}=0$ and the potential $Z_{1}=Z$ a space-time constant, the
eigenmodes satisfy the 2-D Dirac equation
\begin{eqnarray*}
E=-\gamma^{\mu}(p_{\mu}-(c_{L}\chi_{L}-c_{R}\chi_{R})Z_{\mu}.
\end{eqnarray*}
For $Z=0$ the energy-momentum dispersion relation 
\begin{eqnarray*}
E=\pm p,
\end{eqnarray*}
is shown in the
upper left of Figure \ref{disp}. The left- and right-hand branches correspond
to the separate fermion chiralities. If $Z$ is adiabatically changed to
a small (positive) value the relation is that of the upper right diagram,
\begin{eqnarray*}
E=\left\{ \begin{array}{c} 
p+c_{R}\delta Z,\\
-(p+c_{L} \delta Z), \end{array}\right.
\end{eqnarray*}
where we have assumed both $c_{L}$ and $c_{R}$ are positive.
Gauge transformations in this case cause a nett production of right 
antiparticles and left particles. While the total number of states
is conserved, the separate left and right numbers are not.
\begin{figure}[htb]
\centering{
\rotatebox{270}{\resizebox{7.5cm}{8cm}{\includegraphics{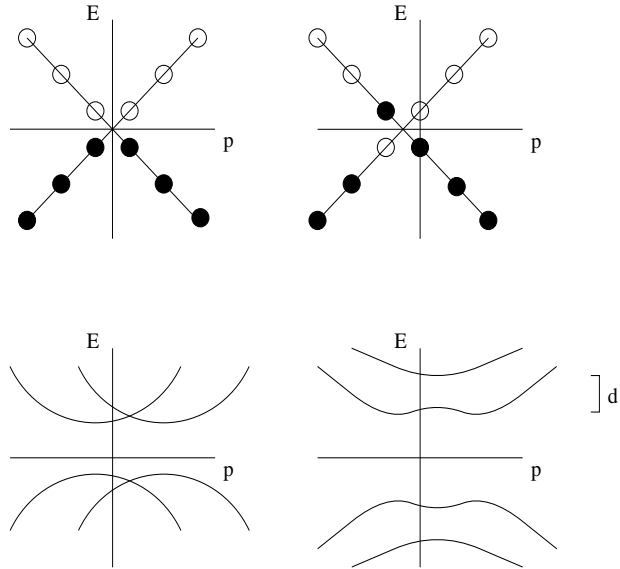}}}
}
 \caption{Dispersion relations for eigenmodes of Eq. (\ref{2d}). 
Black and white circles represent filled and empty states respectively.
}
 \protect \label{disp}
 \end{figure}
Introducing now a mass term $m\bar{\psi}\psi$ to the Lagrangian Eq.(\ref{2d})
would lead to the gapped lower left diagram in
Fig \ref{disp}. 
\begin{eqnarray*}
E=\pm \left\{\begin{array}{c} 
\sqrt{(p+c_{R}\delta Z)^{2}+m^{2}},\\
\sqrt{(p+c_{L} \delta Z)^{2}+m^{2}}. \end{array}\right. 
\end{eqnarray*}
Increasing the value $\delta Z$ has the effect of shifting the parabolas
upwards and to the left for positive couplings $c_{R}$, $c_{L}$, 
resulting in a nett production of right-handed particles at the expense of left
ones, as expected in \cite{BassThomas1996}.
Using this analogy the necessary anomaly-induced collapse of the left 
vacuum appears to break down if $c_{L}=-c_{R}$ whereby 
the left and right parabolas shift in opposite directions, such that there
is no net change in chirality. 
This happens for the hypercharge couplings
for the electron, up- and down-type quarks at the values
$\sin^{2} \theta_{W}=1/4$, $3/8$ and $3/4$ respectively.
In this case an alternative scenario, such as the SCS type transition
advocated here would be required to produce the fermion generations
in a manner outlined below.

Of course these dispersion relations do not correspond to the 
eigenmodes of a full Dirac fermion; each branch contains only half
the required number of degrees of freedom.
Up until now we have considered a single fermion, however SCS is a many-body
phenomenon. If we consider a superposition of a fermion and antifermion 
to make up the requisite number of degrees of freedom it is apparent that 
this diagram could also correspond to dispersion relations for two 
distinct bosonic objects, as shown in the lower right picture, in the 
limit of a vanishing gap $d$. These relations, of the form
\begin{eqnarray*}
E^{2}_{1}& = & p^{4} + m^{2}_{1}, \\
E^{2}_{2}& = & (p^{2}+ g S)^{2}+(m^{2}_{1}-d),
\end{eqnarray*}
have a natural interpretation as those of composite bosons. $E_{1}$ and
$E_{2}$ would then represent fluctuations in charge- and spin-type 
degrees of freedom respectively.

The hypothesised effect in 4-dimensions is that the SCS in the vector
interactions results in condensates containing ``charge'' and ``spin'' 
degrees of freedom. 
Consider now a free fermion with global $U(2f)_{L}\otimes U(2f)_{R}$
``isoflavour'' symmetry. 
Addition of charge density interactions,such as those in Eq.(\ref{lut1}), 
breaks the symmetry 
\begin{equation}
U(2f)_{L}\otimes U(2f)_{R}\supset (SU(2f)_{L}\otimes U(1))_{L}\otimes (SU(2f)_{R}\otimes U(1).
\end{equation}
For a single isoflavour, $f=1$, the $U(2)$ fermions decouple into commuting 
$SU(2)\otimes U(1)$ sectors reminiscent of a left-right-symmetric electroweak theory.
Upon breaking this to the QED scale, the degeneracy in the (iso)``spin''
condensates is lifted, leading to fermion mass terms of the form 
\begin{eqnarray*}
m=m_{\uparrow}(1+\tau_{3})/2 + m_{\downarrow}(1-\tau_{3})/2.
\end{eqnarray*}
Note that a mass term of this form is used in Gribov's calculation of the $W$ 
and $Z$ masses.
The dominant contribution (from the heaviest fermion generation) to vacuum 
polarisation reproduces good approximations to both boson masses and the 
expected Higgs VEV.

Alternatively if instead of Abelian densities of left- and 
right-movers, as in the Luttinger lagrangian Eq.(\ref{lut1}), interactions between isospin 
densities 
\begin{equation}
\sigma^{a}=\bar{\psi}\tau^{a}\psi, \hspace{0.1cm} a=1,\ldots 3,
\end{equation}
where $\tau^{a}$ are the $SU(2)$ isospin matrices,
the relevant decomposition into decoupled theories is \cite{Andrei}
\begin{equation}
U(2f) \otimes U(2f)\supset [SU(f)_{2} \otimes SU(2)_{f} \otimes U(1)]_{L}\otimes SU(f)_{2}
\otimes SU(2)_{f} \otimes U(1)]_{R} \label{fscs}
\end{equation}
Here the integer subscript denotes the fact that the interaction is in fact 
described by a chiral Wess-Zumino-Witten \cite{Witten} model, the value 
referring to the central charges.

If we identify $f$ with the number of known fermion generations $f=3$, then 
the model contains not only (iso)spin and (hyper)charge interactions but an 
$SU(3)$ ``flavour'' sector also. 
In $1+1$ dimensions \cite{Andrei} the analogous model contains a non-trivial
fixed point which generates a mass gap for the fermion propagators.
The bosonic spin fluctuations also acquire mass while the charge and flavour 
excitations remain gapless, strongly reminiscent of photons and gluons
in the SM.

In this context we note the dualised standard model 
(DSM) \cite{DSM} also associates the number of fermion generations with that of the fermionic colour
degrees of freedom. This colour "dual", analogous to the duality of 
electrodynamics under exchange of charge and magnetism, represents the 
{\it same} gauge symmetry as $SU(3)$, differing only by parity.
The question of whether the decomposition (\ref{fscs}) is equivalent to 
the DSM written in ``left-right'' rather than ``vector-axial'' notation certainly warrants 
further investigation.

The fact that the centre of the group $SU(3)$ is $Z_{3}$ also serves as
a motivation for the self-consistent introduction of a $Z_{3}$ symmetric 
``fundamental'' fermion in the generational model of Kiselev \cite{Kiselev}.
The  $Z_{3}$ would then be interpreted as a relic of the broken dual $SU(3)$.

The model has a natural three-step decomposition, from the ``Luttinger'' phase,
through the left-right symmetric SM, step (\ref{st1}), down to the Standard Model in stage (\ref{st2}) before, finally,
the conventional chiral symmetry breaking reproduces the familiar low-energy physics of stage (\ref{st3}):
\begin{eqnarray}
U(6)_{L}\otimes U(6)_{R} & \supset & [SU(3)_{2}\otimes SU(2) \otimes U(1)]_{L}\otimes
[SU(3)_{2}\otimes SU(2) \otimes U(1)]_{R} \nonumber \\
 & &  \label{st1} \\
& \supset &  SU(3)_{2} \otimes SU(2)_{L} \otimes U(1)_{H}\label{st2} \\
 & \supset & SU(3)_{c} \otimes U(1)_{QED}. \label{st3}
\end{eqnarray}
This symmetry breaking pattern is consistent with the hierarchy of the
three critical chiral scales shown to exist in the quenched hypercharge
theory of the previous chapter.

In conclusion we observe that certain recent aspects of condensed matter, in 
particular the behaviour of chiral fluids, are potentially fertile ground 
for understanding how the behaviour of the scalar sector of the SM 
gives rise to fermion mass, generation number and flavour mixing.
Regarding higher gauge theories, only one coupling and two six-dimensional 
fundamental fields (one ``quark'' and one ``lepton'') would be required as 
input at the $U(6)\otimes U(6)$ level.

We argue that it appears logical and compelling to investigate whether a 
technicolour version of the left-right symmetric SM might be analogous to
 certain forms of SCS.

\appendix
\chapter{Fierz transformations} \label{chap:fierz}
Consider an orthogonal basis $\Gamma_{a}$ for the space of $n\times n$
matrices where
\begin{eqnarray*}
\textrm{tr}(\Gamma^{a}\Gamma^{b})= \delta_{ab} /c .
\end{eqnarray*}
An arbitrary $n\times n$ matrix $X$ has the expansion
\begin{eqnarray*}
X=x_{a}\Gamma^{a}= c \Gamma^{a} \textrm{tr}(X\Gamma_{a}),
\end{eqnarray*}
and from the completeness of $\Gamma_{a}$ one obtains the general
Fierz identity for any pair of matrices:
\begin{eqnarray*}
X_{ij}Y_{kl} = c \textrm{tr}(X \Gamma_{a}Y)_{il}\Gamma^{a}_{jk}.
\end{eqnarray*}
In the case of the 4-dimensional Lorentz group ($c=1/4$) the identity
is frequently used for manipulating 4-fermion scalar interactions. 
For the 16 Dirac matrices we define
\begin{eqnarray*}
\rho^{a}_{bc}=\frac{1}{4}\textrm{tr}(\Gamma^{a}\Gamma_{b}\Gamma_{c}),
\end{eqnarray*}
where $a$, $b$ and $c$ range over $S=I$, $V=\gamma^{\nu}$, 
$T=-i \sigma^{\mu\nu} /\sqrt{2}$, $A=\gamma^{5}\gamma^{\nu}$ and $P=\gamma^{5}$. With the notation
\begin{eqnarray*}
(X.Y)_{42;31} \equiv
\bar{\psi}_{4}X^{42}\psi_{2}\bar{\psi}_{3}Y^{31}\psi_{1},
\end{eqnarray*}
and using the identity
\begin{eqnarray*}
\psi_{2}\psi_{1}=\delta_{2\bar{2}}\delta_{1\bar{1}}\psi_{\bar{2}}
\psi_{\bar{1}}=\frac{1}{4} \Gamma_{a2\bar{1}}\Gamma^{a}_{1\bar{2}}\psi_{\bar{2}}\psi_{\bar{1}},
\end{eqnarray*}
we arrive at the general result
\begin{equation}
(\Gamma^{a}.\Gamma_{b})_{42;31} = \frac{1}{4}\rho^{a}_{cd}\rho_{b}^{ce}
(\Gamma^{d}.\Gamma^{e})_{41;32}. \label{master}
\end{equation}
In chapter \ref{chap:4fer} we encounter four-fermion interactions 
of the form 
\begin{eqnarray*}
((V \pm A).(V \pm A))_{42;31}, & & ((V-A).(V+A))_{42;31}.
\end{eqnarray*}
The results for Lorentz scalars are commonplace in texts, for example, \cite{IZ},\cite{Miransky}) or are readily computed from the identity (\ref{master}):
\begin{eqnarray}
(V.V)_{42;31}& = & 
\rho^{V}_{VS}\rho_{V}^{VS}(S.S)_{41;32}+ (\rho^{V}_{SV}\rho_{V}^{SV}+\rho^{V}_{TV}\rho_{V}^{TV})(V.V)_{41;32} \nonumber \\
& & \mbox{} + (\rho^{V}_{PA}\rho_{V}^{PA}+\rho^{V}_{TA}\rho_{V}^{TA})(A.A)_{41;32}+ \rho^{V}_{AP}\rho_{V}^{AP}(P.P)_{41;32} \nonumber \\
& \equiv & 
-(\bar{\psi}\psi)^{2}+(\bar{\psi}i \gamma^{5}\psi)^{2}
+\frac{1}{2}((\bar{\psi}\gamma^{\mu}\psi)^{2}+(\bar{\psi}\gamma^{5}\gamma^{\mu}\psi)^{2}), \label{vv}\\
(A.A)_{42;31}& = & 
\rho^{A}_{AS}\rho_{A}^{AS}(S.S)_{41;32}+ (\rho^{A}_{PV}\rho_{A}^{PV}+\rho^{A}_{TV}\rho_{A}^{TV})(V.V)_{41;32} \nonumber \\
& & \mbox{} + (\rho^{A}_{SA}\rho_{A}^{SA}+\rho^{A}_{TA}\rho_{A}^{TA})(A.A)_{41;32}+ \rho^{A}_{VP}\rho_{A}^{VP}(P.P)_{41;32} \nonumber \\
& \equiv & 
(\bar{\psi}\psi)^{2}-(\bar{\psi}i \gamma^{5}\psi)^{2}
+\frac{1}{2}((\bar{\psi}\gamma^{\mu}\psi)^{2}+(\bar{\psi}\gamma^{5}\gamma^{\mu}\psi)^{2}). \label{aa}
\end{eqnarray}
The parity-violating, chiral symmetric terms, which cancel in \ref{chap:4fer},
are included for completeness.
The non-zero contributions from Eq.(\ref{master}) are
\begin{eqnarray}
(V.A)_{42;31} & = & 
(\rho^{V}_{SV}\rho_{A}^{SA}+\rho^{V}_{TV}\rho_{A}^{TA})(V.A)_{41;32} +
(\rho^{V}_{TA}\rho_{A}^{TV}+\rho^{V}_{PA}\rho_{A}^{PV})(A.V)_{41;32} \nonumber \\
 & & \mbox{} + \rho^{V}_{VS}\rho_{A}^{VP}(S.P)_{41;32}+\rho^{V}_{AP}\rho_{A}^{AS}(P.S)_{41;32}  \nonumber \\
 & \equiv & \frac{1}{2}(\bar{\psi}\gamma_{\mu}\psi \bar{\psi}\gamma^{5}\gamma^{\mu}\psi+\bar{\psi}\gamma_{\mu}\gamma^{5}\psi \bar{\psi}\gamma^{\mu}\psi)+
(\bar{\psi}\psi \bar{\psi}\gamma^{5}\psi+
\bar{\psi}\gamma^{5}\psi \bar{\psi}\psi),\nonumber \\
& & \label{va}\\
(A.V)_{42;31} & = & 
\frac{1}{2}(\bar{\psi}\gamma_{\mu}\psi \bar{\psi}\gamma^{5}\gamma^{\mu}\psi+\bar{\psi}\gamma_{\mu}\gamma^{5}\psi \bar{\psi}\gamma^{\mu}\psi)-
(\bar{\psi}\psi \bar{\psi}\gamma^{5}\psi+
\bar{\psi}\gamma^{5}\psi \bar{\psi}\psi). \nonumber\\
& & \label{av}
\end{eqnarray}
Now the full expansion of 4-fermion terms in Eq.(\ref{imedt1}) is
\begin{eqnarray*}
& &(c_{R}^{2}+c_{L}^{2}+c_{X}^{2})(V.V)_{42;31}+
(c_{R}^{2}+c_{L}^{2}-c_{X}^{2})(A.A)_{42;31} \nonumber \\
& & +(c_{R}^{2}-c_{L}^{2})(V.A+ A.V)_{42;31} +c_{X}^{2}(VA-AV+AV-VA)_{42;31}.
\end{eqnarray*}
Upon substituting the expressions (\ref{vv}-\ref{av}), 
one arrives at the claimed form, Eq.(\ref{genform}), of the Lagrangian.
\vspace{4cm}

\pagebreak
\chapter{1-loop running couplings}\label{chap:1loop}
\label{sec:NPRG}
\begin{figure}[htb]
\centering{
\rotatebox{270}{\resizebox{6cm}{7cm}{\includegraphics{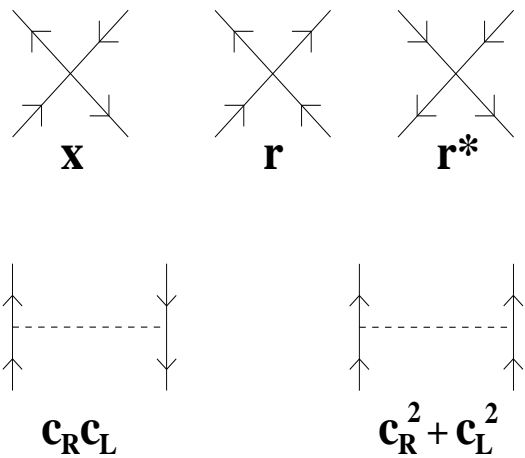}}}
}
 \caption{4-fermion and fermion-boson vertices in the gauged model.}
 \protect \label{vert}
 \end{figure}

\begin{figure}[htb]
\centering{
\rotatebox{270}{\resizebox{7.5cm}{8cm}{\includegraphics{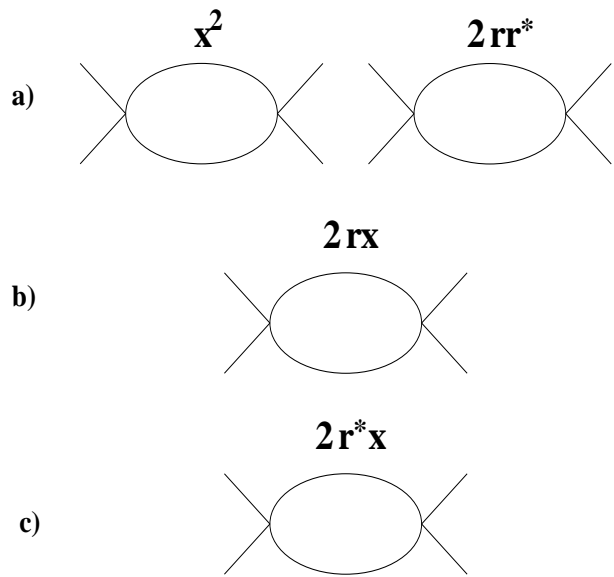}}}
}
 \caption{1-loop diagrams in the toy 4-fermion model contributing to the
running of a) $x$, b) $r$ and c) $r^{*}$ respectively.}
 \protect \label{4fer}
 \end{figure}

\chapter{Transverse boson propagator}\label{chap:huh}
In this appendix we compute the 1-loop hypercharge boson propagator.
Following Gribov \cite{Gribovgol}  it is demonstrated that modifying the 
fermion-boson vertex to include a Goldstone-fermion coupling cancels the 
non-transverse contributions of massive fermion loops to the vacuum 
polarisation. 

From the hypercharge Lagrangian
\begin{eqnarray*}
{\cal L}=\bar{\psi} \gamma^{\mu}(i \partial_{\mu} -m + (c_{R}\chi_{R}+c_{L}\chi_{L}))Z_{\mu} )\psi +{\cal F}_{\mu\nu}{\cal F}^{\mu\nu},
\end{eqnarray*}
the 1-loop vacuum polarisation is computed to be
\begin{eqnarray}
\Pi^{\mu\nu}&=&-\int \frac{d^{4}p}{(2\pi)^{4}} Tr \{
\Gamma^{\mu}S(p)g_{j}\Gamma^{\nu}S(p-k)\} \label{barec}\\
& =& (c_{R}-c_{L})^{2} m^{2}\delta^{\mu\nu}\int\frac{d^{4}p}{(2\pi)^{4}}X(k^{2},p^{2}) \nonumber \\
& & + (c_{L}+c_{R})^{2}(m^{2} \int\frac{d^{4}p}{(2\pi)^{4}}X(k^{2},p^{2})\delta^{\mu\nu}+Y^{\mu\nu})+i(c^{2}_{R}-c^{2}_{L}) \epsilon^{\mu\rho\nu\sigma} Z_{\rho \sigma}, \nonumber
\end{eqnarray}
where
\begin{eqnarray*}
X(k^{2},p^{2})&=&\frac{1}{p^{2}-m^{2}}\frac{1}{(p-k)^{2} -m^{2}},\\
Y^{\mu\nu}&=&- \int\frac{d^{4}p}{(2\pi)^{4}} X(k^{2},p^{2})(p^{\mu}(p-k)^{\nu} + p^{\nu}(p-k)^{\mu} -\delta^{\mu\nu}p(p-k)),\\
Z_{\rho\sigma}&=&- \int\frac{d^{4}p}{(2\pi)^{4}}X(k^{2},p^{2})p_{\rho}(p-k)_{\sigma}.
\end{eqnarray*}
For $c_{L}=c_{R}$ only the second term survives and 
yields the familiar QED result: 
\begin{equation}
-\frac{i}{4 \pi^{2}}(\delta^{\mu\nu}k^{2}-k^{\mu}k^{\nu})\Pi(k^{2}),\label{qedr}
\end{equation}
where $\Pi(k^{2})$ is the usual $U(1)$ scalar polarisation.
More generally, following Gribov\cite{Gribovgol} we include the 
Goldstone contribution in each vertex in order to restore transversality to 
$\Pi^{\mu\nu}$:
\begin{eqnarray*}
\Gamma^{\mu}\to \Gamma^{\mu} - F.\eta \frac{k^{\mu}}{k^{2}},
\end{eqnarray*}
where $F$ is the boson-Goldstone transition amplitude and the 
Goldstone-fermion coupling $\eta$ is given by the Ward-Takahashi
identity:
\begin{eqnarray*}
k_{\mu}(\Gamma^{\mu}-F.\eta\frac{k^{\mu}}{k^{2}})=c_{R}\chi_{R}
S^{-1}(p) - S^{-1}(p-k) c_{L}\chi_{L}.
\label{WGT}
\end{eqnarray*}
For the free fermion propagator it is just
\begin{eqnarray*}
F.\eta = 2 m (c_{L}-c_{R}) \gamma{5}.
\end{eqnarray*}
Now Eq.(\ref{barec}) is replaced by 4 terms:
\begin{eqnarray*}
\Pi^{\mu\nu}&=&-\int \frac{d^{4}p}{(2\pi)^{4}} Tr \{
(\Gamma^{\mu}-F.\eta\frac{k^{\mu}}{k^{2}})S(p)(\Gamma^{\nu}-F.\eta\frac{k^{\nu}}{k^{2}})S(p-k)\} \\
&\equiv& P_{1}+P_{2}+P_{3}+P_{4}.
\label{nubit1}
\end{eqnarray*}
At the one-loop level, the contraction $\epsilon_{\mu\nu\rho\sigma}Z^{\rho\sigma}$ vanishes.
$P_{1}$ is just the bare result, Eq.(\ref{qedr}) above. The second and 
third terms are equal (and vanish in this case) while
\begin{eqnarray*}
P_{4}&=&\int \frac{d^{4}p}{(2\pi)^{4}} Tr \{F.\eta
\frac{k^{\mu}}{k^{2}}S(p)F.\eta\frac{k^{\nu}}{k^{2}}S(p-k)\} \\
&=&2 m^{2} (\alpha_{S}-\alpha_{D}) \frac{k^{\mu}k^{\nu}}{k^{2}}\int\frac{d^{4}p}{(2\pi)^{4}} X(k^{2},p^{2})\frac{2}{k^{2}}(m^{2}-p.(p-k)).
\end{eqnarray*}
Combining this with Eqs.(\ref{barec}) we see
\begin{eqnarray*}
\Pi^{\mu\nu}&=&(\alpha_{S}-\alpha_{D}) m^{2}(\delta^{\mu\nu}-\frac{k^{\mu}k^{\nu}}{k^{2}}
(-\frac{4 m^{2}}{k^{2}})) X(k^{2}) \nonumber \\
& & \mbox{} + \alpha_{D}(\delta^{\mu\nu}k^{2}-k^{\mu}k^{\nu})\Pi(k^{2}),
\end{eqnarray*}
which is transverse when the Goldstone self-energy 
\begin{equation}
\Sigma_{G}=k^{2}_{G}=-4m^{2},
\end{equation}
a natural condition for composite bosons.


\begin{thebibliography}{499}
\bibitem{Pich94} A. Pich, 
Lectures at the CERN Academic Training (Geneva, November, 1993), hep-ph/9412274.
\bibitem{BassThomas1996} S.D. Bass and A.W. Thomas, Mod. Phys. Lett. {\bf A11},(1996) 339.
\bibitem{Pich97} A. Pich, 
Lectures at the Cargese '96 School --Masses of Fundamental Particles-- Cargese, Corsica, 5-17 August 1996, 
hep-ph/9701263.
\bibitem{Renton} P.B. Renton, Eur. Phys. J. {\bf C8} (1998) 585.
\bibitem{bklt} D.E. Groom {\it et al.}, Eur. Phys. J. { \bf C15} (2000) 1.
\bibitem{FrogNil1996} C.D. Froggatt, H.B. Nielsen and D.J. Smith, Z.Phys. {\bf C73} (1997) 333. 
\bibitem{Seiberg} H. Harari and N. Seiberg, Phys. Lett. {\bf B102} (1981) 263.
\bibitem{Adlerpre} S.L. Adler, hep-th/9711393 (1997).
\bibitem{Volkas} A. Davidson, T. Schwartz and R.R. Volkas,  J.Phys. {\bf G25} (1999) 1571.
\bibitem{Nardi} E. Nardi, Talk given at 3rd Latin American Symposium on High Energy Physics (SILAFAE III), Cartagena de Indias, Colombia, April 2-8, 2000, hep-ph/0009329.
\bibitem{Ibanez1} L.E. Ib\'{a}\~{n}ez, Phys. Lett. {\bf B303} (1993) 55.
\bibitem{Ibanez2} L.E. Ib\'{a}\~{n}ez and G.G. Ross, Phys. Lett. {\bf B332} (1994) 100.
\bibitem{Ramond1} 
P. Binetruy, C. Deffayet, E. Dudas and P. Ramond,
Phys.Lett. {\bf B441} (1998) 163.
\bibitem{GSW} M. Green and J. Schwartz, Phys. Lett. {\bf B149} (1984) 117.
\bibitem{NJL} Y. Nambu and G. Jona-Lasinio, Phys. Rev. {\bf 127} (1962) 962.
\bibitem{MaskawaNakajima} T. Maskawa and H. Nakajima, Prog. Theor. Phys. {\bf 52}, (1974) 1326.
\bibitem{Miransky} V.A. Miransky, {\it Dynamical Symmetry Breaking in Quantum 
Field Theories} (World Scientific 1993).
\bibitem{RobertsWilliams1994} C.D. Roberts and A.G. Williams, in {\it Progress in Particle and Nuclear Physics} {\bf 33}, (Ed. Amand Faessler, Pergamon Press 1994).

\bibitem{Adlerdisc} S.L. Adler, Phys.Rev. {\bf}D59 (1999) 015012; {\it erratum-ibid.} {\bf D59} (1999) 099902; Phys.Rev. {\bf D60} (1999) 015002.
\bibitem{ChanTsou} H. Chan and S.T. Tsou, Int. J. Mod. Phys. {\bf A14} (1999) 2139.
\bibitem{Kiselev} V.V. Kiselev, hep-ph/9806523 (1998).
\bibitem{Visnjic} V. Visnjic, hep-ph/0002067 (2000).
\bibitem{K2} V.V. Kiselev, hep-ph/9909545 (1998).
\bibitem{Bardeen1989} W.A. Bardeen, C.N. Leung and S.T. Love, Nucl. Phys.
{\bf B323}, (1989) 493.
\bibitem{Beg1989a} M.A.B. B\'{e}g, Phys. Rev. {\bf D39} (1989) 2373.
\bibitem{G1} M. G\"{o}ckeler, R. Horsely, V. Linke, P. Rakow, G. Schierholz 
and H. St\"{u}ben, Phys. Rev. Lett. {\bf 80}, (1998) 4119.
\bibitem{G2} V. Bornyakov, A. Hoferichter and G. Schierholz, 
Nucl. Phys. Proc. Suppl. 94 (2001) 773-776.
\bibitem{Reenders} M. Reenders, {\it The gauged Nambu-Jona Lasinio model},
(Ph.D thesis, University of Groningen 1999).
\bibitem{Adler} S.L. Adler, Phys. Rev. {\bf 177}, 2426 (1969).
\bibitem{BJ} J.S. Bell and R. Jackiw, Nuovo Cim. {\bf 60}, 47 (1969).
\bibitem{Gribovgol} V.N. Gribov, Phys. Lett. {\bf B336}, (1994) 243.
\bibitem{Li1} B-A. Li, hep-ph/0012051 (2000). 
\bibitem{Li2} B-A. Li, Nucl. Phys {\bf B} Proc. Suppl., (1999) 76,263.
\bibitem{Wilson} K. Wilson, Phys Rep C {\bf 12}, (1974) 75. 
\bibitem{WK}  K. Wilson and J. Kogut, Rev. Mod. Phys. {\bf 47}, (1975) 773.
\bibitem{BallChiu1980} J.S. Ball and T.W. Chiu, Phys. Rev. {\bf D22}, 2542 (1980).

\bibitem{IZ} C. Itzykson and J-B. Zuber, {\it Quantum Field Theory},
(McGraw-Hill 1985).
\bibitem{Muta} T. Muta, {\it Foundations of Quantum Chromodynamics}
(World Scientific 1987).
\bibitem{Fukuda} H. Fukuda and Y. Miyamoto, Prog. Theor. Phys. {\bf 4}, 49 (1949).
\bibitem{Jackiw} R. Jackiw, 
Dirac Prize Lecture delivered at Trieste, Italy, in March 1999, hep-th/9903255.
\bibitem{Gribbk} V.N. Gribov, in Proceedings of non-perturbative Methods
in Quantum Field Theories, Z. Horvath, L. Palla and A. Patk\'{o}s (Eds.),
(World Scientific 1987) 65.
\bibitem{Dass} N.D. Hari Dass Int. J. Mod. Phys. {\bf B14} (2000) 1989-2010
\bibitem{Fuji1} K. Fujikawa, Phys. Rev {\bf D21}, 2848 (1980).
\bibitem{Swanson} M Swanson, {\it Path Integrals and Quantum Processes}, (Academic Press 1992.
\bibitem{Bardeena} W.A. Bardeen, Phys. Rev. {\bf 184}, 1848 (1969).
\bibitem{Kieu1} T.D. Kieu, Chin. J. Phys. 32 (1994) 1099-1108
\bibitem{Kieu2} T.D. Kieu, Mod. Phys. Lett. A {\bf 11} (1996) 2601-2610
\bibitem{cas1} R. Casana and S.A. Dias, Int. J. Mod. Phys. {\bf A15} 4603, (2000).
\bibitem{cas2} R. Casana and S.A. Dias, J. Phys. {\bf G27} (2001) 1501-1518.
\bibitem{WZ} J. Wess and B. Zumino, Phys. Lett. {\bf 37B} (1971), 95.
\bibitem{Witten} E. Witten, Nucl. Phys. {\bf B223}, (1983) 422.
\bibitem{ColeWein72} E. Coleman and S. Weinberg, Phys. Rev. D {\bf 7}, 1888 (1973).
\bibitem{salcedo} L.L Salcedo, 
Eur. Phys. J. {\bf C20} (2001) 161-184.
\bibitem{Landau} L.D. Landau, in {\it Niels Bohr and the Development of Physics},
(McGraw-Hill, 1955).
\bibitem{WH} F. Wegner and J. Houghton, Phys. Rev. {\bf A8}, 401 (1973).
\bibitem{Aoki2} K-I. Aoki, K. Morikawa, J-I. Sumi, H. Terao and M. Tomoyose, 
Prog. Theor. Phys. {\bf 102} 1151 (1999).
\bibitem{Aitchison} I.J.R. Aitchison and A.J.G. Hey, {\it Gauge theories in 
particle physics : a practical introduction}, (Hilger 1982)
\bibitem{Glashow} S.L. Glashow, Nucl. Phys.{\bf 22}, 579 (1961).
\bibitem{Weinberg} S. Weinberg, Phys. Rev. Lett. 19 1264 (1967).
\bibitem{Salam} A. Salam, in {\it Elementary Particle Theory}, Ed. N. 
Svartholm (Almquist and Wiksells 1969) 367.
\bibitem{Doff} A. Doff and F. Pisano, 
Mod. Phys. Lett. {\bf A15} (2000) 1471-1480
\bibitem{Higgs} P.W. Higgs, Phys. Rev. {\bf 145}, 1156 (1966).
\bibitem{Kibble} T.W.B. Kibble, Phys. Rev. {\bf 155}, 1554 (1967).
\bibitem{higgs} M. Krawczyk, Acta Phys. Polon. {\bf B29}, 3543 (1998). 
\bibitem{Nicholson} A.F. Nicholson and D.C. Kennedy, 
Int. J. Mod Phys. {\bf A15} (2000) 1497.
\bibitem{Leblanc} D. Caenepeel and M. LeBlanc, hep-th/9404088 (1994).
\bibitem{Weinberg2} S. Weinberg, Phys. Rev. {\bf D19}, 1277 (1979).
\bibitem{Susskind} L. Susskind, Phys. Rev. {\bf D20}, 2619 (1979).
\bibitem{topcon} W.A. Bardeen, C.T. Hill and M.Lindner, Phys. Rev. {\bf D41},
1647 (1990).
\bibitem{topcon2} V.A. Miransky, M. Tanabashi and K. Yamawaki, Phys. Lett. {\bf 221} (1989) 177.
\bibitem{Kobayashi} M. Kobayashi and T. Maskawa, Prog. Theor. Phys {\bf 49},
652 (1973).
\bibitem{kaons} J.H. Christenson, J.W. Cronin, V.L. Fitch and R. Turlay,
Phys. Rev. Lett. {\bf 13} 138 (1964).
\bibitem{superK} Y. Fukuda {\it et al.} (Super-Kamiokande collaboration),
Phys. Lett. {\bf B436}, 33 (1998).
\bibitem{superK2} Y. Fukuda {\it et al.} (Super-Kamiokande collaboration),
Phys. Rev. Lett. {\bf 81}, 1562 (1998).
\bibitem{Rosner} J.L. Rosner, 
invited talk presented at 2nd Tropical Workshop in Particle Physics and Cosmology, San Juan, Puerto Rico, May 1-6, 2000, hep-ph/0005258.
\bibitem{Espriu} D. Espriu and J. Manzano,Phys. Rev. {\bf D63} (2001) 073008.
\bibitem{Einhorn} M.B. Einhorn and J. Wudka, Phys. Rev. {\bf D63} (2001) 073008.
\bibitem{Hall} L. Hall and S. Weinberg, Phys. Rev. {\bf D48} (1993) 979-983.
\bibitem{Zarikas} V. Zarikas, 
Phys.Lett. {\bf B384} (1996) 180-184.
\bibitem{DSM} H. Chan, 
Int. J. Mod. Phys. {\bf A16} (2001) 163-178.
\bibitem{transmut} J. Bordes, H. Chan, and S.T. Tsou, 
Phys. Rev. {\bf D65} (2002) 093006
\bibitem{FominMiransky1976} P.I. Fomin and V.A. Miransky, Phys. Lett.
{\bf B64}, (1976) 166.
\bibitem{FGM1978} P.I. Fomin, V.P. Gusynin and V.A. Miransky, Phys. Lett.
{\bf B78}, (1978) 136.
\bibitem{FukudaKugo} R. Fukuda and T. Kugo, Nucl. Phys. {\bf B117}, 250 (1976).
\bibitem{bif} D. Atkinson and P.W. Johnson, J. Math. Phys. {\bf 28} (1987) 248.
\bibitem{Kondobif} K. Kondo, H. Mino and K. Yamawaki, Phys. Rev. {\bf D39}
(1989) 2430.
\bibitem{FGM1981} P.I. Fomin, V.P. Gusynin and V.A. Miransky, Phys. Lett.
{\bf B105}, (1981) 387.
\bibitem{Miransky1985} V.A. Miransky, Nuovo Cimento {\bf 90A}, (1985) 149.
\bibitem{BurdenRoberts1993} C.J. Burden and C.D. Roberts, Phys. Rev. {\bf D47} (1993) 5581.
\bibitem{lk} L.D. Landau and I.M. Khalatnikov, Sov. Phys. JETP {\bf 2} (1956) 69. 
\bibitem{CurtisPennington1993} D.C. Curtis and M.R. Pennington, Phys. Rev. {\bf D42}, 4165 (1990); Phys. Rev. {\bf D48}, (1993) 4933.
\bibitem{Ayse1995} A. Kizilers\"{u}, M. Reenders and M.R. Pennington, Phys. Rev. {\bf D52} (1995) 1242.
\bibitem{Ayse1997} A. Bashir, A. Kizilers\"{u} and M.R. Pennington, Phys. Rev. {\bf D57} (1998) 1242.
\bibitem{Takahashi} Y. Takahashi, in {\it "Quantum Field Theory"}, F. Mancini (Ed.), (Elsevier Science, 1986)
\bibitem{Kondo1996} K. Kondo, 
Int. J. Mod. Phys. {\bf A12} (1997) 5651-5686.
\bibitem{DelbourgoT} R. Delbourgo and G. Thompson, J. Phys. {\bf G8}, L185.
\bibitem{Bernstein} J. Bernstein, {\it Elementary particles and their currents}, p 64 ( Freeman, 1968).
\bibitem{Atkinson1993} D. Atkinson, V.P. Gusynin and P. Maris, Phys. Lett {\bf 303}, 157,1993.
\bibitem{Atkinson1994} D. Atkinson, J.C.R. Bloch, V.P. Gusynin, M.R. Pennington and M. Reenders, Phys. Lett. {\bf B329}, (1994) 117.
\bibitem{OliensisJohnson1990} J. Oliensis and P.W. Johnson, Phys. Rev. {\bf D42}, (1990) 656.
\bibitem{Gusynin1990} V. Gusynin, Mod. Phys. Lett. {\bf A5} (1990) 133.
\bibitem{Kondo1992} K. Kondo and H. Nakatani, Prog. Theor. Phys. {\bf 88} (1992) 737.
\bibitem{Rakow} P.E.L. Rakow, Nucl. Phys. {\bf B356}, (1991) 27.

\bibitem{Gasser} J. Gasser and H. Leutwyler, Phys. Ann. {\bf 158}, 142 (1984).
\bibitem{bender} C. Bender, K. Milton and V Savage, Phys. Rev. {\bf D62} (2000) 085001.
\bibitem{Aoki1} K-I. Aoki, K. Morikawa, J-I. Sumi, H. Terao and M. Tomoyose,
Prog.Theor.Phys. {\bf 97} 479 (1997).
\bibitem{Branchina} V. Branchina, 
Phys. Lett. {\bf B549} (2002) 260-266.
\bibitem{Babaev} E. Babaev, 
Int. J. Mod. Phys. {\bf A16} (2001) 1175-1197.
\bibitem{joyce} M. Joyce, K. Kainulainen and T. Prokopec,
JHEP 0010 (2000) 29
\bibitem{Matsuda} K. Matsuda, N. Takeda, T. Fukuyama and H, Nishiura,
Phys. Rev. {\bf D62}, 093001 (2000).
\bibitem{pisarski} R. Pisarski and D. Rishcke, Phys.Rev. {\bf D60} (1999) 094013.
\bibitem{sonoda} H. Sonoda, hep-th/0005188 (2000).

\bibitem{vvkbk} G.E. Volovik, {\it Exotic properties of superfluid} $^{3}$He,
(World Scientific 1992).
\bibitem{Luke} G.M. Luke {\it et al.}, Nature {\bf 394}, (1998) 558.
\bibitem{Morinari} T. Morinari and M. Sigirist, J. Phys. Soc. Jpn. 69, 2411 (2000).
\bibitem{Yoon} S. Yoon {\it et al.}, cond-mat/0003250 (2000).
\bibitem{Goryo1} J. Goryo, 
J. Phys. Soc. Jap. 69 (2000) 3501-3504.
\bibitem{Andrei} N. Andrei, M.R. Douglas and A. Jerez, cond-mat/9803134 (1998).
\bibitem{voit} J. Voit, cond-mat/9510014 (2000).
\bibitem{Haldane} F.D.M. Haldane, J. Phys. {\bf C14}, 2585 (1981).


\end{thebibliography}
\end{document}